\documentclass[structabstract]{aa}

\usepackage{graphicx}
\usepackage{subfigure}

\usepackage{txfonts}

\usepackage{natbib} 
\bibpunct{(}{)}{;}{a}{}{,} 

\begin{document}

\newcommand{\red}{\textcolor{red}}
\newcommand{\green}{\textcolor{green}}
\newcommand{\blue}{\textcolor{blue}}
\newcommand{\countss}{{\rm counts}\ {\rm s}^{-1}}
\newcommand{\ergs}{{\rm erg}\ {\rm s}^{-1}}
\newcommand{\ergcms}{{\rm erg}\ {\rm cm}^{-2}\ {\rm s}^{-1}}
\newcommand{\ergcmssr}{{\rm erg}\ {\rm cm}^{-2}\ {\rm s}^{-1}\ {\rm sr}^{-1}}

\title{
White dwarf masses in intermediate polars \\
observed with the Suzaku satellite
}

\author{
Takayuki Yuasa\inst{1}, 
Kazuhiro Nakazawa\inst{1},
Kazuo Makishima\inst{1,2},
Kei Saitou\inst{3,4}, \\
Manabu Ishida\inst{3,5}, 
Ken Ebisawa\inst{3,4},
Hideyuki Mori\inst{3,6} and
Shin'ya Yamada\inst{1}
}

\institute{
Department of Physics, School of Science,
The University of Tokyo,
7-3-1 Hongo, Bunkyo, Tokyo 113-0033, Japan\\
\email{yuasa@juno.phys.s.u-tokyo.ac.jp}
\and
Cosmic Radiation Laboratory,
The Institute of Physical and Chemical Research (RIKEN),
2-1 Hirosawa, Wako, Saitama 351-0198, Japan
\and
Japan Aerospace Exploration Agency,
Institute of Space and Astronautical Science,
3-1-1 Yoshino-dai, Chuo-ku, Sagamihara, Kanagawa 252-5210, Japan
\and
Department of Astronomy, School of Science,
The University of Tokyo,
7-3-1 Hongo, Bunkyo, Tokyo 113-0033, Japan
\and
Department of Physics, Tokyo Metropolitan University,
1-1 Minami-Osawa, Hachioji, Tokyo 192-0397, Japan
\and
Department of Astrophysics, School of Science,
Nagoya University,
Furo-cho, Chikusa-ku, Nagoya 464-8602, Japan
}

\date{Received March 31, 2010; accepted June 7, 2010}

\authorrunning{T. Yuasa et al.}
\titlerunning{White dwarf masses in intermediate polars observed with \it Suzaku \rm}

\abstract
{White dwarfs (WDs) in cataclysmic variables (CVs) are important experimental laboratories where the electron degeneracy is taking place on a macroscopic scale. Magnetic CVs increase in number especially in the hard X-ray band ($\gtrsim 10$~keV) thanks to sensitive hard X-ray missions.}
{From X-ray spectroscopy, we estimate the masses of nearby WDs in moderately-magnetized CVs, or Intermediate Polars (IPs).}
{Using the {\it Suzaku\rm} satellite, we aquired wide-band spectra of 17 IPs, covering $3-50$~keV. 
An accretion column model of Suleimanov et al. (2005) and an optically-thin thermal emission code were used to construct a spectral emission model of IPs with resolved Fe emission lines.
By simultaneously fitting the Fe line complex and the hard X-ray continuum of individual spectra, the shock temperature and the WD mass were determined with a better accuracy than in previous studies. }
{We determined the WD masses of the 17 IPs with statistical fitting errors of $\lesssim0.1-0.2~M_\odot$ in many cases.
The WD mass of a recently-found IP, IGR J17195-4100, was also estimated for the first time ($1.03^{+0.24}_{-0.22}~M_\odot$).
The average WD mass of the sample is  $0.88\pm0.25~M_\odot$. When our results were compared with previous X-ray mass determinations, we found significant deviation in a few systems although the reason of this is unclear. The iron abundance of the accreting gas was also estimated, and confirmed the previously reported sub-solar tendency in all sources with better accuracy. }
{}


\maketitle

\section{Introduction}\label{section:introduction}
White dwarfs (WDs) are attractive experimental laboratories where the electron degeneracy plays an important role for supporting the stars against gravity. An accreting WD is of great importance in the universe because some of them probably cause Type Ia supernovae as the WD mass reaches the Chandrasekhar limit, and deliver synthesized heavy elements out to the surrounding space. 
The mass of a WD is thus one of the critical physical parameters which determines the fate of the WD. 

Forming a subgroup of cataclysmic variables (CVs), a magnetic cataclysmic variable (mCV) involves a magnetized WD as a compact star. mCVs are further classified into polars and intermediate polars (IPs), according to whether their magnetic field strengths are $10^{7-9}$~G or $10^{5-7}$~G, respectively. Polars are so named after their strong polarization in the optical and infrared wavelengths, and sometimes referred to as AM Her type following this prototype \citep[for a review of polars, see][]{cropper1990review}. The magnetic field of a polar is strong enough for the orbital and spin periods of its WD to be synchronized ($P_{\rm{orb}}=P_{\rm{spin}}$). 
In IPs, the two periods are asynchronous, typically $P_{\rm{spin}}\sim0.1~P_{\rm{orb}}$ \citep[for reviews of IPs, see ][]{patterson1994,hellier1996ip}, and polarization is generally weak, although there are some recent measurements of polarization (e.g. \citealt{buttersetal2009}).
In both types of these binary systems, mass accretion is taking place from non-degenerate companions, probably low-mass stars, via Roche-lobe overflow.
Polars do not have an accretion disk. Most IPs have accreting disks, while in the inner region, the disk is truncated, and gas almost freely falls onto a WD channeled along its magnetic field.
Near the WD surface, a stationary shock stands to convert the kinematic energy of the bulk gas motion into thermal energy.
Because the temperature of the shock-heated gas is typically $>10$~keV, and the density is low, hard X-rays are emitted via optically thin thermal emission.
Then, the heated gas forms a post shock region (PSR) with a temperature gradient, wherein the gas descends while it cools via the X-ray emission \citep[e.g. ][]{aizu1973, franketal1992accretionpower, wu1994, wuetal1994}. As a result, the total spectrum from the PSR is observed as a sum of multi-temperature emission components \citep[e.g.][]{ishidaetal1994exhya}.

Although mCVs account for a small fraction ($\sim10-25\%$) of all CVs, they dominate in the hard X-ray band ($>10$~keV), and most ($\sim80\%$) of CVs detected in recent hard X-ray surveys with {\it INTEGRAL}/IBIS and {\it Swift} are magnetized \citep{barlowetal2006,landietal2009,birdetal2010integral4thcatalogue}, including both known \citep[e.g.][]{revnivtsevetal2004v1223sgr} and new objects \citep{ajelloetal2006,bonnet-bidaudetal2007,buttersetal2007,evansetal2008,sazonovetal2008integralchandrafollowup,landietal2009}.

According to recent X-ray studies \citep[][]{sazonovetal2006luminosityfunction,revnivtsev2008cvproperties}, mCVs form an important X-ray source population in the lower luminosity range ($L_{\rm{X}}\sim10^{31-34}~\ergs$), while low-mass X-ray binaries with neutron stars and black hole binaries dominate in the high-luminosity range.
It has also been suggested that the Galactic ridge X-ray emission (GRXE; \citealt{bleachetal1972ridge,worralletal1982ridge,iwanetal1982ridge,koyamaetal1986ridge67kev,kanedaetal1997,valiniaetal1998ridge,ebisawaetal2001ridge,tanaka2002ascaridge,ebisawaetal2005ridgeir,koyamaetal2007suzakugc,revnivtsevandsazonov2007chandraridge,revnivtsevetal2009nature}) and the Galactic Center X-ray emission \citep{koyamaetal1989gc67kev,yamauchietal1990gc67kev,koyamaetal1986ascagc,munoetal2004gcdiffuse,koyamaetal2007suzakugc,revnivtsevetal2007chandragc} mostly consist of numerous dim point sources \citep[][and references therein]{revnivtsevetal2009nature} including RS CVn systems, dwarf novae, and mCVs, each contributing to different X-ray energy regimes.
Especially in the hard X-ray band, mCVs are attractive candidates because of their spectral resemblance to the GRXE \citep{revnivtsevetal2006,krivonosetal2007}, including both continua and iron-K lines. 
If we obtain a reliable ``averaged mCV spectrum" in the hard X-ray band, we can directly compare it with that of the GRXE.
For this purpose, the construction of a wide-band spectral model of mCVs and estimation of WD masses are both mandatory, because mCV spectra can be well expressed as a function of the WD mass as described in Sect. \ref{section:model}.

Although the mass of a WD is a basic physical quantity, its determination is generally not easy, since an uncertainty of the inclination angle of the orbital plane degrades the accuracy of estimation except for eclipsing systems. However, especially in mCVs, X-ray spectroscopy provides an independent method; the X-ray spectrum is primarily determined by the shock temperature, which in turn reflects the $M/R$ ratio, where $M$ and $R$ are the mass and radius of a WD, respectively. Combining the X-ray estimate on $M/R$ with a theoretical WD mass-radius relation, for example one by \citet{nauenberg1972},
\begin{equation}
R=
7.8\times10^8\left[ \left( \frac{1.44M_\odot}{M} \right)^{2/3} -
 \left( \frac{M}{1.44M_\odot} \right)^{2/3} \right]^{1/2}~\rm{cm}\label{equation:mrrelation},
\end{equation}
we can determine $M$.

The X-ray spectroscopic approach is in practice subdivided into two methods, the continuum method and the emission line method. The former requires a sensitive hard X-ray detector since the relevant spectral feature, namely exponential cutoff which reflects the shock temperature, appears in the hard X-ray range (energy $>10$~keV). The latter relies on the resolved emission lines (e.g. from Fe), hence requiring a detector with a relatively high energy resolution such as an X-ray CCD (charge coupled device) camera. The continuum method has been applied to the data taken with \it Ginga \rm \citep{ishida1991, cropperetal1998}, \it RXTE \rm \citep{ramsay2000,suleimanovetal2005,suleimanovetal2008}, and \it Swift \rm \citep{brunschweigeretal2009,landietal2009}. \cite{fujimotoishida1997} and \cite{ezukaishida1999} applied the emission line method to the {\it ASCA} data, deriving the metal abundances as well.

\it Suzaku \rm \citep{mitsudaetal2007}, which is the fifth Japanese X-ray astrophysical observatory, has a wide-band (0.2-600~keV) energy coverage with a high sensitivity, realized by the X-ray Imaging Spectrometer (XIS; \citealt{koyamaetal2007xis}), which uses CCDs and the Hard X-ray Detector (HXD; \citealt{takahashietal2007}) which consists of semiconductor detectors and scintillator crystals.
The hard X-ray response of the HXD allows us to directly measure the thermal cutoff of an IP spectrum, and hence estimate its shock temperature. At the same time, we can use the XIS to resolve emission lines, especially Fe-K$\alpha$ lines, into three typical components; from almost neutral, He-like, and H-like species. Thus, \it Suzaku \rm is expected to provide significantly advanced information on the masses of mCVs.

In the present paper, we report masses and iron abundances of IPs estimated by analyzing wide-band \it Suzaku \rm spectra of 17 objects. Compared to IPs, hard X-ray emission from polars is weaker because cyclotron emission induced by the stronger magnetic field reduces their plasma temperatures. 
It is also difficult to incorporate the effect of strong magnetic field into a spectral model. Therefore, we only analyzed IPs.

In Sect. \ref{section:model}, we describe a numerically calculated emission model which is used to derive the mass estimations. Target information and data reduction procedure are given in Sect. \ref{section:observationanddatareduction}. Section \ref{section:result} deals with the WD mass estimation. In Sect. \ref{section:discussion}, we compare our results with previous reports.

\section{Modeling of X-ray emission from an accretion column}
\label{section:model}
\subsection{Previous works}
A number of works have been performed to model X-ray spectra emitted from PSRs of accretion columns in mCVs, and to estimate the PSR temperatures by comparing the model spectrum with observations. The temperatures were then translated into the $M/R$ ratio of their WDs. As mentioned in Sect. \ref{section:introduction}, these works can be divided into continuum and emission line methods.

In the continuum method, the vertical structure of the accretion column (e.g., profiles of bulk velocity, temperature, and density) is calculated based on plausible physical assumptions, following which an X-ray spectral model can be constructed by convolving thin thermal emission models with the plasma profile. The $M/R$ ratio is a free parameter of the synthesized model spectrum, and should be determined through its comparison with an observed spectrum.
\citet{aizu1973} was the very first who explored this theoretical method.
By neglecting the gravitational attraction inside accretion columns, \citet{wu1994} and \citet{wuetal1994} presented profiles of plasma parameters in one-dimensional plane-parallel PSR in closed-integral formulae by considering plasma cooling via bremsstrahlung and cyclotron emission.

\citet{ishida1991} analyzed {\it Ginga} data of several IPs using an isothermal assumption to a PSR, and reported WD mass estimations (or lower limits) for the first time by fitting spectra with an isothermal emission model.
By comparing the isothermal emission model with their more realistic spectral calculations, \citet{wuetal1995} gave correction factors to the WD masses estimated by \citet{ishida1991}.
\citet{cropperetal1998} and \citet{ramsayetal1998} also used the model of \citet{wu1994} and fitted data obtained with {\it Ginga}, {\it ASCA}, and {\it RXTE}. 

To construct a more physically realistic model, \citet{cropperetal1999}, \citet{ramsay2000} and \citet{suleimanovetal2005} calculated the accretion column structure by taking into account the gravitational attraction, and found that the inclusion of the gravity slightly reduces the WD mass estimates especially in systems with higher WD masses (e.g. $>1.0~M_{\odot}$; \citealt{ramsay2000}).
\citet{brunschweigeretal2009} also applied the model of \citet{suleimanovetal2005} to the data of 22 IP systems observed with the {\it Swift}/BAT, and gave mass estimates. 
Some authors considered more sophisticated emission models;
for example, \citet{suleimanovetal2008} considered the effect of Compton up-scattering of photons by post-shock plasmas although it was found to change the WD mass estimates only slightly. \citet{canalleetal2005} calculated PSR temperature and density profiles by taking into account the dipole magnetic field structure, and showed that this increases estimated WD masses mainly in high mass (e.g. $\sim1~M_\odot$) systems.

As the first application of the emission line method, \citet{ishidafujimoto1995linediagnostics} and \citet{fujimotoishida1997} analyzed the {\it ASCA}/SIS data of EX Hya. They thus estimated the plasma temperature behind the shock by comparing observed line-intensity ratios with those theoretically predicted. Employing the PSR structure of \citet{aizu1973}, they obtained a WD mass estimate of $0.48^{+0.10}_{-0.06}M_{\odot}$, which agreed with that of $0.49\pm0.03M_\odot$ derived with optical spectroscopic observations by \citet{hellier1996ip}.
Using the same line-ratio model, \citet{hellier1996aopsc}, \citet{fujimoto1996}, and \citet{ezukaishida1999} investigated WD masses in other systems, giving comparable results with those obtained with the X-ray continuum methods.

\subsection{The constructed model}
\label{section:the_model_in_this_study}
To make the best use of the capability of {\it Suzaku} (Sect. 1), we constructed a spectrum model by improving the PSR model of \citet{suleimanovetal2005} in two ways. One is to include the Fe-K emission lines, while the other is to improve the plasma cooling function, which is used when solving hydrodynamical equations in the high-temperature regime ($>10^{8.5}$~K; see the following section). Below we describe how we constructed our model.

\begin{figure}
\centering
	\resizebox{!}{6cm}{\includegraphics{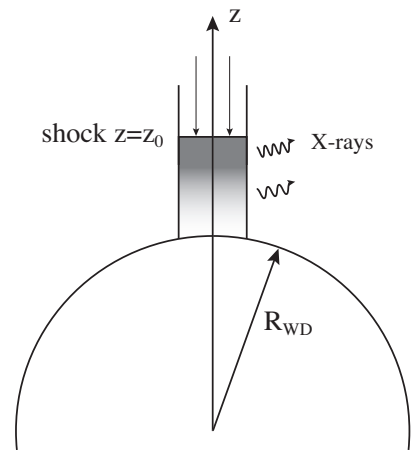}}
	\caption{
		Assumed geometry of the accretion column of an IP.
	}
	\label{figure:geometry}
\end{figure}

As illustrated in Fig. \ref{figure:geometry}, a one-dimensional cylindrical accretion column, or PSR, is assumed. Channeled by the magnetic field, gas is considered to fall into the accretion column with its supersonic free-fall velocity. At a certain height, a strong shock converts some of the bulk motion of the gas into thermal energy in the PSR. 
While the shock-heated gas falls subsonically down along the magnetic field lines, it cools by emitting X-rays. Finally, the gas is assumed to softly land onto the atmosphere of the WD.
Because the magnetic fields of IPs are not as strong as those of polars, the dominant cooling process of the heated gas is optically thin thermal X-ray emission (free-free plus free-bound plus bound-bound), and the cyclotron cooling can be considered negligible (e.g. Eq.(2) of \citealt{wuetal1995}).
Including these, the assumptions made in the present model are summarized as
\begin{itemize}
\item 1-dimensional cylindrical accretion\\
(curvature of magnetic field neglected)
\item temperature before the shock to be zero
\item gas to be ideal
\item a strong shock
\item cooling via free-free, free-bound, and bound-bound process\\
(cyclotron cooling to be negligible)
\item gravitational force in the PSR taken into account
\item soft landing of cooled gas onto the WD surface
\end{itemize}
These are the same assumptions as those used in \citet{suleimanovetal2005} and the negligible magnetic field case of \citet{cropperetal1999}.

\subsection{Calculation of the model}
\label{section:formulation_and_numerical_implementation}
With these assumptions, the accreting plasma flow can be described with hydrodynamical equations; conservations of the mass, the momentum, and the energy. Since these equations and initial conditions assumed in the numerical calculation have been already published by several authors (e.g. \citealt{suleimanovetal2005}; \citealt{cropperetal1999}), we do not recall them here, and follow the same formalization and notation as \citet{suleimanovetal2005}.
Below, we describe major changes and some attentions related to the model calculation. 

In the model used in \citet{suleimanovetal2005} and \citet{brunschweigeretal2009}, the plasma cooling function $\Lambda$ was taken from \citet{sd93}. Although we could integrate the optically-thin emission code (\verb|apec|; see below) to construct a cooling function, we also use the result of \citet{sd93} here. This is because spectra calculated with the emission code only cover  $E<50$~keV currently, and this would artificially reduce cooling rates in higher plasma  temperatures (e.g. $kT\gtrsim10$~keV).
However, for rather massive WDs, we need to specially treat a higher temperature regime which is not covered by \citet{sd93} (i.e. $T>10^{8.5}$~K). In this regime, we extended the cooling function using a thermally averaged relativistic free-free Gaunt factor tabulated by \citet{nakagawaetal1987}. We averaged the Gaunt factor over the frequency separately for hydrogen and helium, and then converted the averaged values into the cooling rate (or energy loss rate) in terms of the total power per unit volume emitted by thermal bremsstrahlung (see e.g. \citealt{rybickilightman1979}), 
\begin{equation}
\frac{dW}{dV dt}=1.426\times10^{-23}n_{\rm{e}}n_{\rm{i}}Z^2\bar{g_{\rm{B}}}~\rm{erg}~\rm{cm}^{-3}~\rm{s}^{-1},
\end{equation}
where $n_{\rm{e}}$, $n_{\rm{i}}$, $Z$, and $\bar{g_{\rm{B}}}$ are electron and ion number densities, charge of ion, and the averaged Gaunt factor, respectively. 
Then, by rescaling the obtained cooling rate so that it matches that of \citet{sd93} in $T<10^{8.5}$~K, we constructed the cooling rate up to $T<10^{9.0}$~K.
The extrapolated cooling function exceeds a simple linear extrapolation of the cooling function by \citet{sd93} by a factor of $\lesssim2$ . 

When a fraction $f$ of the PSR cross section to the WD surface area is set, 
the specific accretion rate $a$ can be related to the total accretion rate $\dot{M}$ via $\dot{M}=4\pi R^2_{\rm{WD}}af$. 
Since the luminosity of a PSR is determined by liberated gravitational energy ($\propto \dot{M}$), 
the product $af$ can be regarded as the model normalization. 
Although $a$ and $f$ can be free parameters in general (see e.g. \citealt{cropperetal1999}; \citealt{suleimanovetal2005}), and the numerical solution changes as $a$ varies, we fixed them at $a=1.0$~g~cm$^{-2}$~s$^{-1}$ and $f=0.001$ (0.1\% of the WD surface area) for the sake of an easy comparison with the previous reports.
To rescale the model spectrum, a normalization parameter,
which linearly adjusts the normalization of the model calculated with the fixed $a$ and $f$, was introduced. Effects of changing $a$ on our result are discussed in Sect. \ref{section:mdot_change}.

After setting $\Lambda$ and $a$, finding the solution of the three conservation equations is reduced to solving an initial value problem of ordinary differential equations. As explained for example in  \citet{cropperetal1999} and \citet{suleimanovetal2005}, this can be practiced with the so-called shooting algorithm. Each trial starts from an assumed shock height $z=z_0$, where the initial parameters are set according to the strong shock limit of the Rankine-Hugoniot relation (\citealt{suleimanovetal2005}).
The lower limit of the calculation, $z=R_{\rm{WD}}$, is specified by the WD mass-radius relation of Eq.(\ref{equation:mrrelation}). At the WD surface, another boundary condition of the bulk gas speed $v=0$ is imposed to fulfill the soft-landing assumption.
In the algorithm, we first solve the differential equations from a trial initial value of $z_0$ (for example, 1.01~$R_{\rm{WD}}$), and see how the final state of each quantity at $z=R_{\rm{WD}}$ deviates from the conditions assumed above. By taking into account the deviation as a feedback to a new initial $z_0$, we solve the equations again. Until the feedback becomes small enough, below $\sim10^{-4}$ of derived $z_0$, this procedure is repeated.
Practically, this method converged in several to a few tens of iterations at each value of $M_{\rm{WD}}$.

\begin{figure}
\centering
	\resizebox{\hsize}{!}{\includegraphics{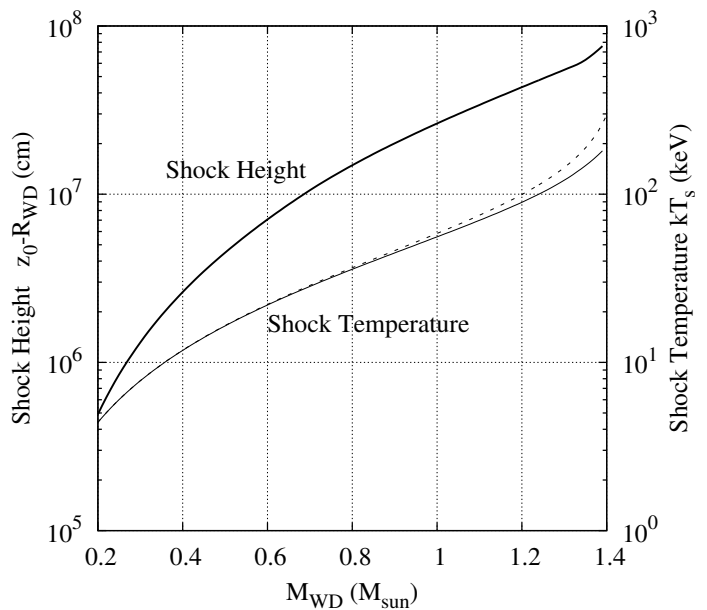}}
	\caption{
Results of the numerical solutions for the shock height from the WD surface (thick solid line) and the shock temperature (thin solid line), shown against the WD mass. For comparison, the dashed line shows the shock temperature calculated by assuming no-gravity in the PSR.}
	\label{figure:kts_and_zs_vs_mwd}
\end{figure}

We executed the calculation over a range of $0.20~M_\odot\le M_{\rm{WD}}\le 1.39~M_\odot$ with a step of $0.01~M_\odot$. 
Figure \ref{figure:kts_and_zs_vs_mwd} shows the dependency of the shock height and the shock temperature on the WD mass by interpolating data points with spline functions.
In the higher WD masses, the gas velocity at the shock height becomes higher, and therefore the gas density falls for a given accretion rate.
This causes a decrease in the plasma cooling rate and cooling time, yielding taller shock heights.

\citet{cropperetal1999} showed that an inclusion of the gravitational attraction changes WD mass estimations especially in high mass ($\sim1.0~M_{\odot}$) systems.
This can be confirmed with the comparison of the profiles of shock temperature calculated with and without gravitational attraction shown in Fig. \ref{figure:kts_and_zs_vs_mwd}. For example, at $M_{\rm{WD}}=1.2~M_\odot$, no-gravity profile overestimates a shock temperature and following $M_{\rm{WD}}$ by $\sim10\%$ compared with the present result. Note that in \citet{cropperetal1999}, they used an analytic free-free cooling rate \citep{wu1994}, which is described in their Appendix A. Therefore, the resulting shock temperatures (their Fig. 1) are slightly different from our values (Fig. \ref{figure:kts_and_zs_vs_mwd}).

Based on the calculation results, we convolved profiles of the temperature and the density along the accretion column with (single-temperature) spectra calculated by the optically-thin emission code \verb|apec| \citep{smithetal2001apec} to obtain a total multi-temperature emission spectrum.
Since we employed \verb|apec| spectra instead of the simple bremsstrahlung continuum which \citet{suleimanovetal2005} utilized, our model includes atomic (particularly Fe) emission lines, which can be an important signature of the plasma temperature in the present spectral analysis.
In this way, we prepared a series of model spectra for each $M_{\rm{WD}}$ by changing the metal abundance of \verb|apec| from 0.1~$Z_\odot$ to 2.0~$Z_\odot$ with a step of $0.1~Z_\odot$. 

Because the currently available version of \verb|apec| does not calculate the spectrum above $E>50$~keV, we can miss the exponential cutoff feature, which is crucial in the calculation of high temperature layers ($kT>20-30$~keV). Although this energy range is not directly employed in our data analysis, the model shapes in that range are considered important, because the HXD has the broad spectral redistribution function, and photons with $E>50$~keV sometimes produce signals in significantly lower energies.
Therefore, above $E>50$~keV, we substituted spectra estimated by \verb|meka| model (which still provides a spectrum in that energy range), with its normalization set 5.5\% smaller than that of the corresponding \verb|apec| model. This renormalization is done to make the two spectra match smoothly at $E=50$~keV under high temperatures where this substitution is of particular importance.
In this manner, the model was calculated over the $0.1-100$~keV band. Note however that we only used it only above 3~keV in spectral fitting to avoid modeling of complex absorption in lower energies, which is less important for WD mass estimations (see Sect. \ref{section:widebandfitting}).

Near the WD surface, where the density of a PSR plasma rapidly increases, the optically-thin assumption will be invalid, and optically-thick emission will emerge. Although this can alter a PSR spectrum in the lower energy range (e.g. $E\lesssim0.1$~keV), we simply used \verb|apec| because our analysis (Sect. \ref{section:observationanddatareduction}) utilizes spectra in higher energies ($E>3$~keV).

Finally, a set of total spectra calculated in this way were converted to a so-called ``local model'' file of the spectral analysis software \verb|Xspec| \citep{arnaud1996}.
The free parameters of the model are the WD mass, the metal abundance, and the model normalization. In this local model, spectra for masses and abundances that are in between the calculation grid points (see above) are generated by interpolating spectra of the surrounding four grid points.

\begin{figure}
\centering
	\resizebox{\hsize}{!}{\includegraphics{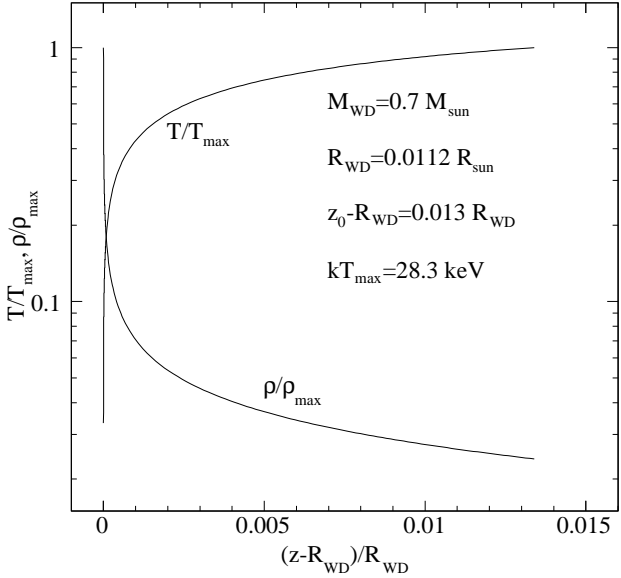}}
	\caption{
		Temperature and density profiles of a PSR calculated for an IP of $M_{\rm{WD}}=0.7~M_\odot$. For easy comparison with \citet{suleimanovetal2005}, the plot employs the same parameter values and the same style as their Fig. 2. Labels give the WD radius calculated using the mass-radius relation of Eq.(\ref{equation:mrrelation}), as well as the shock height ($z_0-R_{\rm{WD}}$) and the shock temperature ($kT_{\rm{max}}$) from Fig. \ref{figure:kts_and_zs_vs_mwd}.
	}
	\label{figure:profilesof0.7Msun}
\end{figure}

In Fig. \ref{figure:profilesof0.7Msun} we present normalized profiles of the temperature and density for an IP with a WD mass of 0.7~$M_\odot$, for comparison with the results of \citet{suleimanovetal2005}. 
The shock height of 0.013~$R_{\rm{WD}}$, which we derived, is somewhat different from that of \citet{suleimanovetal2005}, 0.018~$R_{\rm{WD}}$. This difference might arise from difference of implementation of the algorithm.
In the present study, we do not investigate this difference since the shock temperatures in the two cases differ only by $\sim0.5\%$.

\begin{figure}
\centering
	\resizebox{\hsize}{!}{\includegraphics{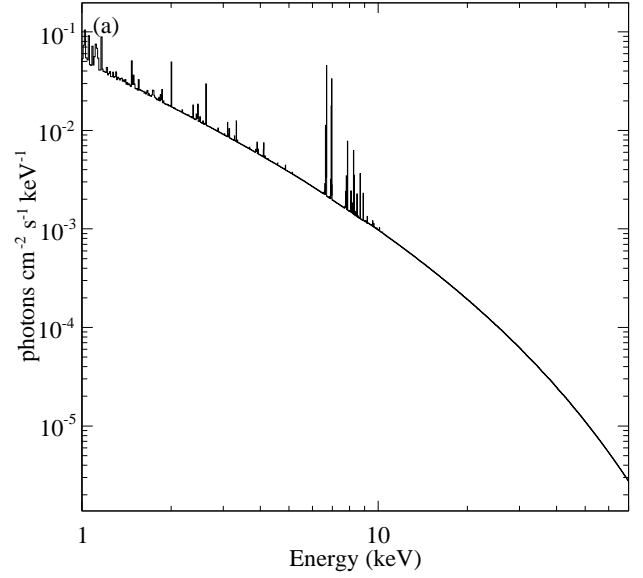}}
	\resizebox{\hsize}{!}{\includegraphics{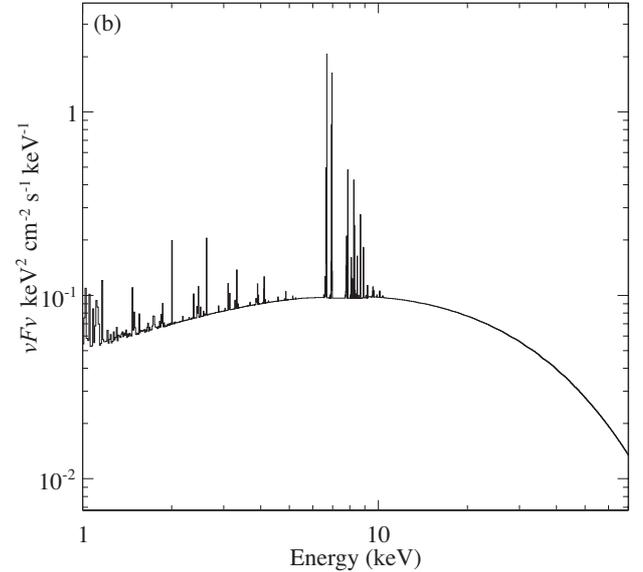}}
	\caption{
		A calculated total spectrum from a PSR of an IP with a WD mass of 0.7~$M_\odot$ and a gas abundance of $1.0~Z_\odot$, plotted in units of (a) photons~cm$^{-2}$~s$^{-1}$~keV$^{-1}$ and (b) keV$^2$~cm$^{-2}$~s$^{-1}$~keV$^{-1}$. The calculation assumed a distance of 100~pc to the WD.
	}
	\label{figure:modelspectrum}
\end{figure}

\section{Observations and data reduction}
\label{section:observationanddatareduction}
We assembled our IP sample of 17 systems observed with \it Suzaku\rm. Table \ref{table:observation_info} lists information of their \it Suzaku \rm observations together with their background-subtracted count rates from each detector. 

The XIS consists of four X-ray imaging charge coupled device (CCD) cameras as a focal plane detector of the X-ray Telescope (XRT; \citealt{serlemitsosetal2007}). Three (XIS0, 2, and 3) use front-side illuminated (FI) CCD chips, while the other is back-side illuminated (BI; XIS1). The energy coverage is nominally 0.2-12~keV \citep{koyamaetal2007xis}.
Since 2006 November, one of the FI sensors, XIS2, has not been operational, and therefore, we utilize data taken with the other three XIS sensors for observations performed after that date. 

The HXD is a non-imaging collimated X-ray detector that consists of two main detection parts; silicon $p$-intrinsic-$n$ diodes (hereafter PIN) and gadolinium silicate (GSO) crystals. They cover the energy range of $10-70$~keV (PIN) and $40-600$~keV. In the present analysis, we only utilize PIN data because signals from IPs were hardly detected by GSO.

Since the sample has been constructed by combining different observations, it is heterogeneous and incomplete, although it can be regarded as a flux-limited sample in the hard X-ray band. To perform a detailed spectral analysis, we only selected targets with enough statistics in the hard X-ray band;
sources with background-corrected HXD count rates higher than 0.04 counts~s$^{-1}$ are used in the present analysis. This rate is equivalent to 10~\% of the typical HXD background count rate \citep{kokubunetal2007,fukazawaetal2009}, or $\sim1$~mCrab in terms of flux. With these criteria, AE Aqr, GK Per, and the IP-like object SAX J1748.2-2808 \citep{nobukawaetal2009iplike} were excluded from the sample, although the XIS clearly detects their spectra.

The typical energy bands utilized in the present study are $3-12$~keV and $12-50$~keV for the XIS and the HXD (Table \ref{table:observation_info}).
We limited ourselves to use the XIS data above 3~keV since lower energy spectral shape are affected by the intrinsic multi-column absorption; the time-averaged spectrum is integrated over multiple azimuthal angles, and thus including spectra absorbed by different column density.
The higher floor bound of the XIS in some sources are explained in the section below.
In some observations, especially in that of FO Aqr, we applied higher floor bound for the HXD data since an enhanced electrical noise contaminated the data therein. The HXD data above 40~keV are discarded when source signals were below 5~\% of the background.

All observations were performed in the normal operation mode of the XIS and HXD. 
Obtained data were processed with the \it Suzaku \rm off-line processing software version 2. We utilized the analysis software package HEASOFT of version 6.6.2 together with a detector calibration database (CALDB) version 2009-05-11.
We excluded the XIS and HXD/PIN data taken when the target is occulted by the Earth, or the spacecraft was in the South Atlantic Anomaly. To reduce the contamination from scattered Solar X-rays, we also removed the XIS data taken while the target elevation angle above the Sun-lit Earth rim was less than $20^\circ$. We applied the same event filtering criteria as used for creating nominal cleaned event files, as described at the \it Suzaku \rm website\footnote{http://www.astro.isas.jaxa.jp/suzaku/process/v2changes/criteria\_xis.html}. 

In the XIS data analysis, we extracted time-averaged spectra of the targets using circular regions with radii of $2'$. Background spectra were also extracted from the same observation data, excluding apparent point sources if any. This background spectrum contains both non-X-ray background (NXB) and diffuse X-ray background (DXB).
Because the signal-to-noise ratio is sufficiently high for all targets, the background subtraction ignored the vignetting effect of the XRT \citep{serlemitsosetal2007}. We generated redistribution matrices and auxiliary response files of the XIS/XRT using xisrmfgen version 2009-02-28 and xissimarfgen \citep{ishisakietal2007} version 2009-01-08, respectively (both contained in HEASOFT). In the spectral analysis, we summed the spectra taken by the FI sensors, and used a detector response averaged over them.

We accumulated a time-averaged PIN spectrum of each target, and subtracted the non X-ray background events extracted from the NXB event file, which in turn was produced from a modeling method by \citet{fukazawaetal2009} and distributed via the PIN NXB release website\footnote{http://www.astro.isas.ac.jp/suzaku/analysis/hxd/pinnxb/}. The version keyword of the NXB file we used is METHOD=LCFITDT, i.e. so-called tuned background. 
We subtracted the DXB component simulated after \citet{boldt1987}, and subtracted it from the observed spectrum.
This procedure is summarized in the website of NASA Goddard Space Flight Center\footnote{http://heasarc.gsfc.nasa.gov/docs/suzaku/analysis/pin$\_$cxb.html}. Since the present sample objects are located off the Galactic plane, we neglected contributions from the GRXE.
When fitting a PIN spectrum, we employed an appropriate redistribution matrix file contained in CALDB, depending on the pointing mode and the observation date.
Because the observation of XY Ari was performed at an off-axis position of the HXD, the effective area of PIN decreased by $\sim20\%$; we took this effect into account by applying an additional auxiliary file calculated with hxdarfgen.

In some observations, XIS images contained weak point sources other than target IPs. These sources could contaminate the data of PIN because PIN has a wider field of view ($1^\circ\times1^\circ$ bottom to bottom). However, their contributions in the PIN energy band are estimated to be $\lesssim2-3$\% in XY Ari, and much lower in other observations. Therefore, we simply neglected them in the present analysis. 

\begin{table*}
\centering
\caption{
\it Suzaku \rm observations of intermediate polars analyzed in the present study.
}
\label{table:observation_info}
\begin{tabular}{lllllllllllll}
\hline
\hline
System             & \multicolumn{2}{c}{Coordinate$^{\rm{a}}$}     & 
\multicolumn{1}{c}{Start Time} & Exp.$^{\rm{b}}$
& \multicolumn{2}{c}{Energy Band$^{\rm{c}}$} & \multicolumn{2}{c}{Rate$^{\rm{d}}$}  & Obs. ID \\
                   & RA       &  Dec &    & ks &  XIS (keV) & HXD (keV) & XIS & HXD  \\
\hline
FO Aquarii         &334.481&-8.351& 2009-06-05 08:14 &33.4&$3.0-12.0$&$17.0-40.0$&0.95& 0.08 &404032010&\\ 
XY Arietis         &44.036&19.442& 2006-02-03 23:02 &93.6&$3.0-12.0$&$12.0-50.0$&0.65& 0.07 &500015010&\\ 
MU Camelopadalis   &96.318&73.578& 2008-04-14 00:55 &50.1&$3.0-12.0$&$12.0-40.0$&0.27& 0.06 &403004010&\\ 
BG Canis Minoris   &112.871&9.94& 2009-04-11 12:11 &45.0&$3.0-12.0$&$12.0-40.0$&0.51& 0.10 &404029010&\\ 
V709 Cassiopeiae   &7.204&59.289& 2008-06-20 10:24 &33.3&$3.0-12.0$&$12.0-50.0$&1.10&0.16&403025010&\\ 
TV Columbae        &82.356&-32.818& 2008-04-17 18:00 &30.1&$3.0-12.0$&$12.0-50.0$&1.21&0.20&403023010&\\ 
TX Columbae        &85.834&-41.032& 2009-05-12 16:19 &51.1&$3.0-12.0$&$12.0-40.0$&0.29& 0.04 &404031010&\\ 
YY Draconis        &175.896&71.703& 2008-06-15 18:37 &27.4&$3.0-12.0$&$12.0-40.0$&0.85& 0.10 &403022010&\\ 
PQ Geminorum       &117.822&14.74& 2009-04-12 13:46 &43.2&$3.0-12.0$&$13.0-50.0$&0.56& 0.09 &404030010&\\ 
EX Hydrae          &193.107&-29.249& 2007-07-18 21:23 &91.0&$3.0-12.0$&$12.0-33.0$&2.12&0.14&402001010&\\ 
NY Lupi            &237.061&-45.478& 2007-02-01 15:17 &86.8&$3.5-12.0$&$12.0-50.0$&0.79&0.19&401037010&\\ 
V2400 Ophiuchi &258.146&-24.247& 2009-02-27 11:42 &110&$3.0-12.0$&$12.0-50.0$&2.60&0.21&403021010&\\ 
AO Piscium         &343.825&-3.178& 2009-06-22 11:50 &35.6&$3.0-12.0$&$12.0-50.0$&1.21&0.15&404033010&\\ 
V1223 Saggitarii   &283.759&-31.163& 2007-04-13 11:31 &46.2&$4.0-12.0$&$13.0-50.0$&2.38&0.52&402002010&\\ 
RX J2133.7+5107    &323.432&51.124& 2006-04-29 06:50 &62.8&$4.0-12.0$&$12.0-50.0$&0.83&0.15&401038010&\\ 
IGR J17303-0601    &262.59&-5.993& 2009-02-16 10:09 &27.7&$3.0-12.0$&$13.0-50.0$&0.56&0.16&403026010&\\ 
IGR J17195-4100    &259.898&-41.015& 2009-02-18 11:03 &26.9&$3.0-12.0$&$12.0-50.0$&1.02&0.15&403028010&\\ 
\hline
\end{tabular}
\begin{list}{}{}
\item[]$^{\mathrm{a}}$ In units of degree (J2000 equinox).
$^{\mathrm{b}}$ Net exposure.
$^{\mathrm{c}}$ Energy band used in the spectral analysis (\S\ref{section:result}).
$^{\mathrm{d}}$ Background-corrected count rates in units of counts s$^{-1}$ integrated over the noted energy bands.
\end{list}
\end{table*}

\section{Spectral analysis and results}
\label{section:result}

\begin{figure*}[htb]
\centering
\subfigure{
\includegraphics[height=6cm]{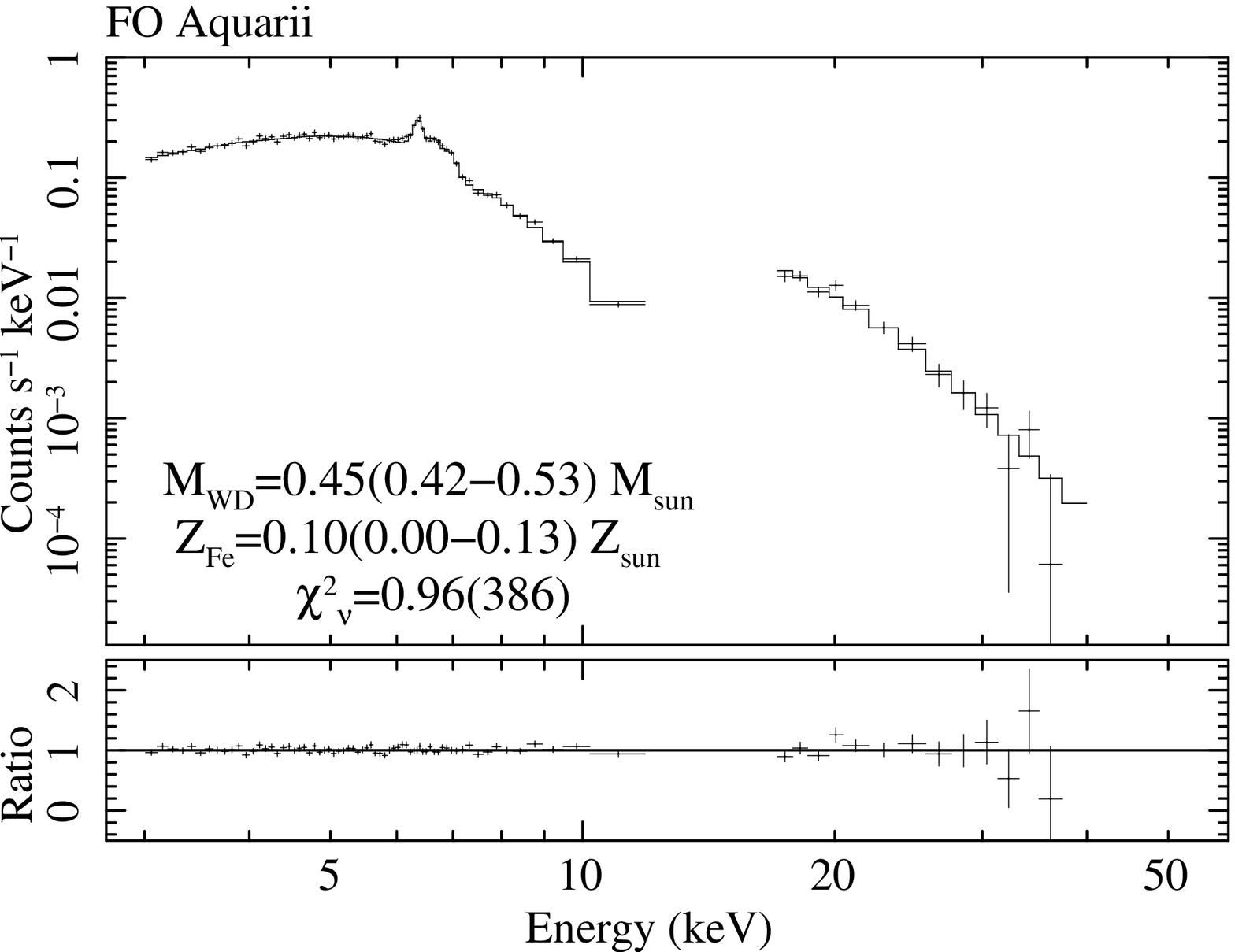}
\includegraphics[height=6cm]{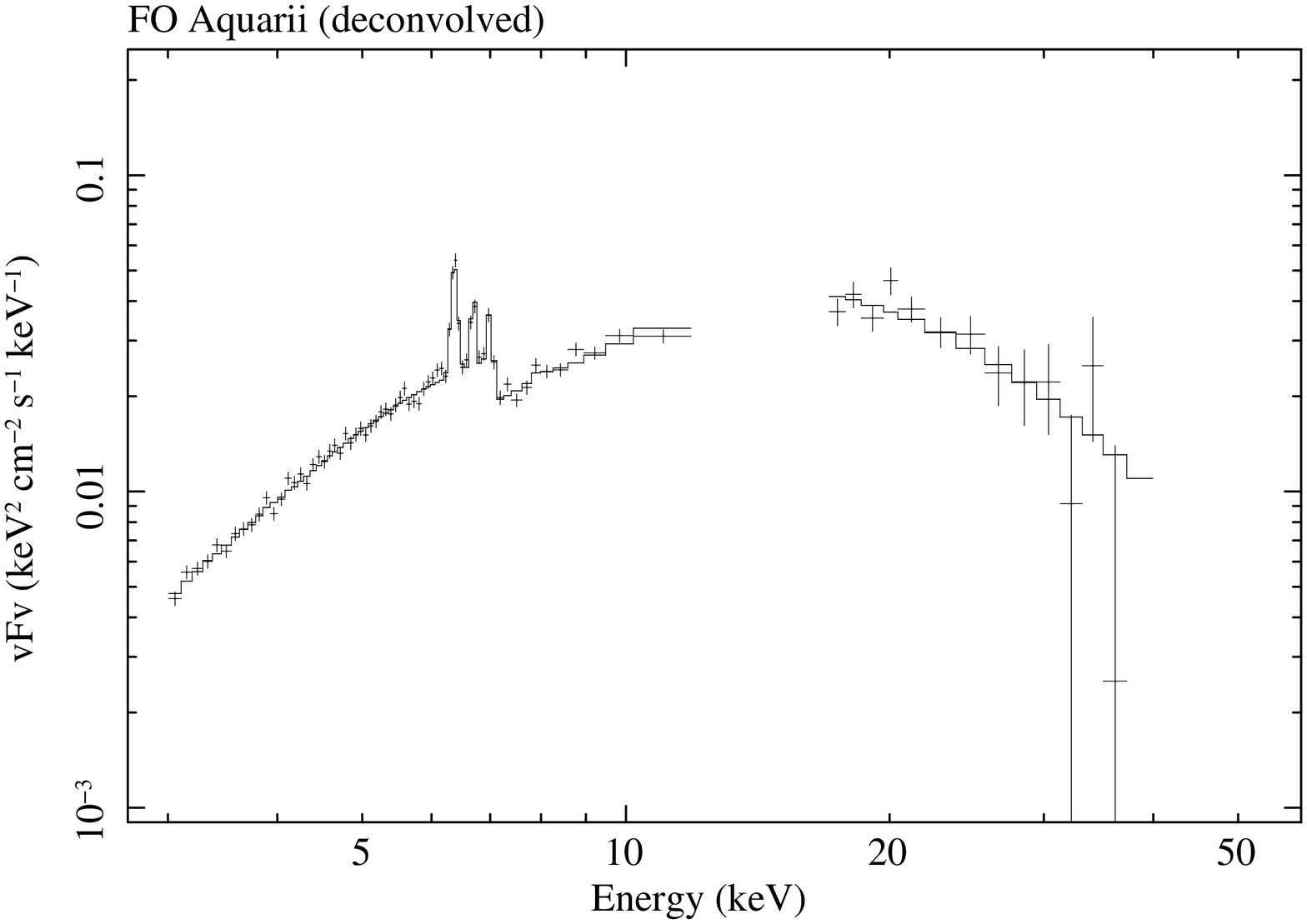}
}
\subfigure{
\includegraphics[height=6cm]{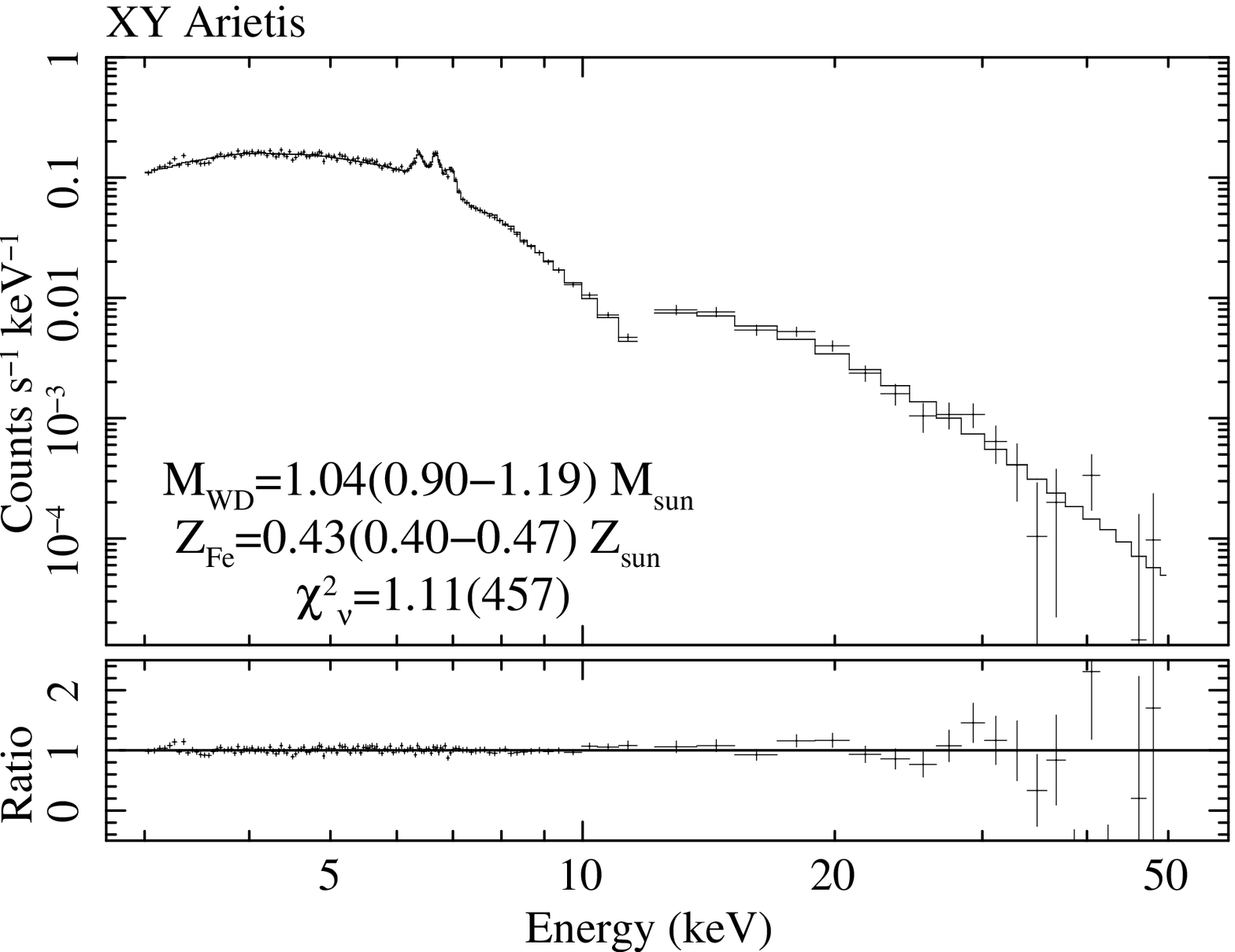}
\includegraphics[height=6cm]{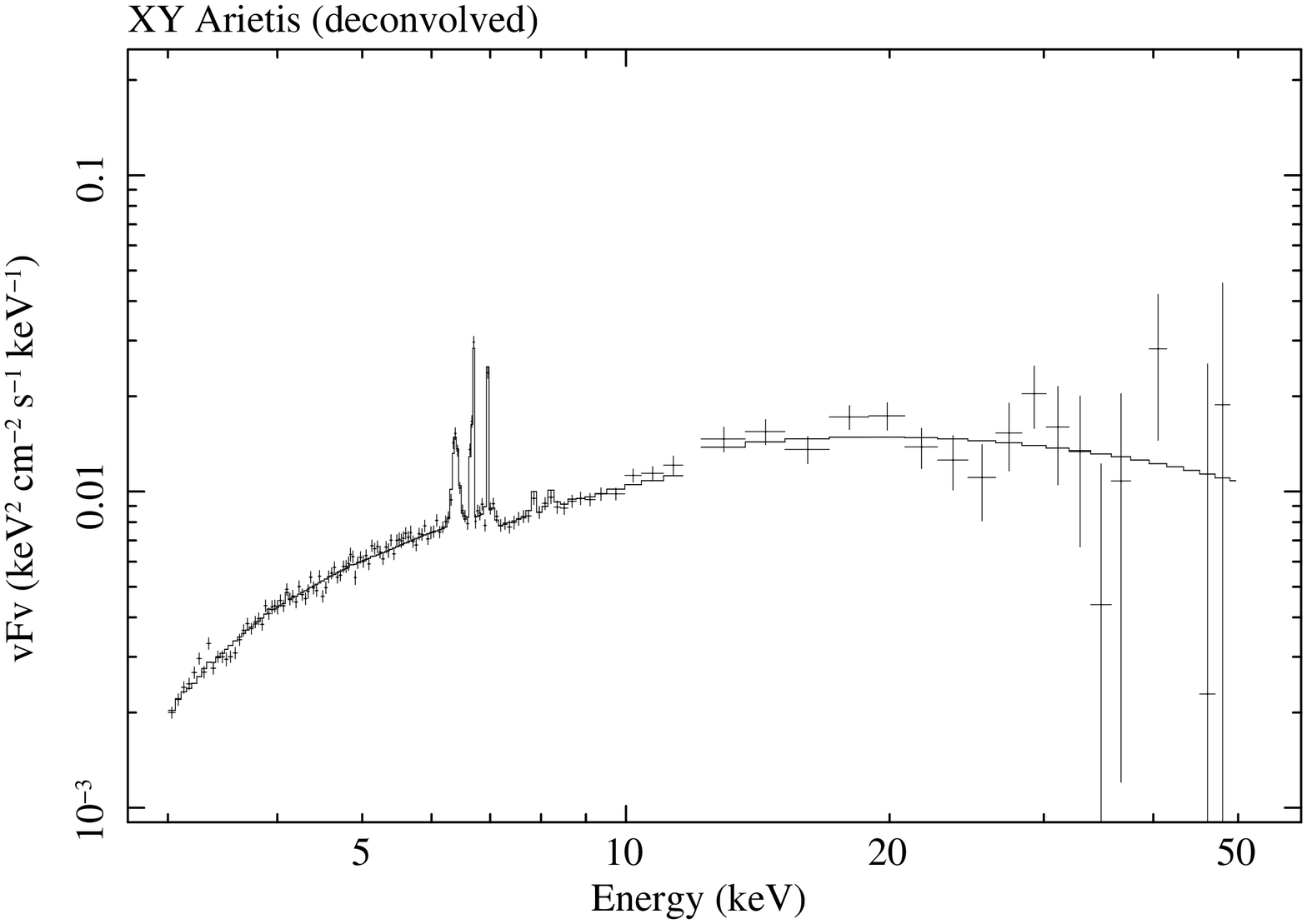}
}
\caption{
Observed XIS and HXD spectra (black crosses) of individual IPs compared with the best-fit model (solid line) . 
{\it Left panels}: Raw spectra, wherein the detector responses are not removed. The lower panels show the data-to-model ratio.
{\it Right panels}: The same spectra, shown after removing of the detector responses and multiplied by $E^2$. Equivalent to the $\nu F_\nu$ presentation. The discontinuity seen in the model curve at 12~keV is due to the renormalization factor, which was introduced to adjust relative effective areas of the XIS and PIN (see text).
{\it Remaining figures are shown in Fig. A.1 of Appendix A.}
}
\label{figure:pc_result}
\end{figure*}

\subsection{Fitting to the XIS/HXD spectra with the partial covering model}
\label{section:widebandfitting}
Using the tabulated model (Sect. \ref{section:model}; hereafter \verb|ip_psr|), we performed chi-square fitting to the spectra of individual IPs.
The lower energy band ($<3$~keV) was ignored because analyses in this band are hampered by multi-column and ionized absorption caused by pre-shock gas \citep[e.g.,][]{cropperetal1999,demartinoetal2001v709cas}. 
To account for the intrinsic multi-column absorption, we utilized a model with a partial-covering (PC) absorber expressed by \verb|vphabs| (available in \verb|Xspec|). 
A Gaussian was added to the model to express the neutral (or low ionized) Fe K$\alpha$ emission line at $\sim6.4$~keV. In addition, the entire model spectrum was subjected to interstellar absorption using \verb|phabs|. The total model expression is thus
\begin{eqnarray}
&&\verb|constant| \times \verb|phabs|\times\nonumber\\ 
&&(C_{\rm{PC}}\cdot\verb|vphabs|\cdot\verb|ip_psr|\nonumber\\
&&+(1-C_{\rm{PC}})\cdot\verb|ip_psr|+\verb|gaussian|),\nonumber
\end{eqnarray}
where \verb|constant| is a renormalization factor for the two detectors (see \it Suzaku \rm Memo 2008-06\footnote{ftp://legacy.gsfc.nasa.gov/suzaku/doc/xrt/suzakumemo-2008-06.pdf}); fixed at 1 for the XIS, and 1.16 (XIS-nominal attitude) or 1.18 (HXD-nominal attitude) for PIN. $C_{\rm{PC}}$ denotes a partial covering fraction. The two \verb|ip_psr| components are constrained to have a common set of parameters.
The Fe abundance of \verb|vphabs| was equalized to that of \verb|ip_psr|, considering that the X-ray absorbing and emitting gas must have the same chemical composition.

First, we fitted the wide-band spectra using all XIS data starting from 3~keV. The HXD spectra were fitted over the energy ranges listed in Table \ref{table:observation_info}. 
The table also gives the shock temperatures, which are related to $M_{\rm{WD}}$ through Fig. \ref{figure:kts_and_zs_vs_mwd}.
Hereafter, the quoted errors are at 90\% confidence levels unless otherwise noted.
Note that the WD mass of IGR J17195-4100 was estimated for the first time as far as we know.

In the present fitting, we considered fits with a null-hypothesis probability higher than 1\% as acceptable. 
Based on this criterion, acceptable fits were obtained with this partial covering model in 11 systems; FO Aqr, EX Hya, NY Lup, V2400 Oph, V1223 Sgr, and RX J2133 are remainders. 
The best-fit parameters are listed in Table \ref{table:pc_result} and
Figs. \ref{figure:pc_result} and \ref{figure:pc_result:b} show the spectra and the best-fit models of individual IPs. For the remaining six objects, we listed results obtained after prescriptions described in the paragraphs below.
The average WD mass and the mean iron abundance associated with their 1$\sigma$ standard deviations were $0.88\pm0.25~M_\odot$ and $0.38\pm0.14~Z_\odot$, respectively.

When fitted with the partial covering model, EX Hya showed large bipolar residuals in the energy band slightly below the H-like Fe K$\alpha$ line. As reported by \citet{hellierandmukai2004fek} with {\it Chandra}/HETG data, the highly ionized Fe lines from EX Hya have complex shapes with considerable contribution from unresolved dielectric satellite lines. Although this might be a cause of the residual, we leave this speculation as a future task because it is difficult to diagnose the line structure with the energy resolution of the XIS. In the present study, we therefore put the original fitting result obtained with the partial covering model in Table \ref{table:pc_result}. The estimated WD mass did not change within errors even if the H-like Fe emission line was ignored in the fitting since the thermal cutoff is clearly detected in the PIN spectrum, and it strongly constrains the shock temperature (and hence a WD mass).

In NY Lup, V1223 Sgr, and RX J2133, positive residuals were observed near the lower bound of the XIS data.
We suspect that these residuals reflect the complex absorption structure in the lower energy spectra, which  are not as so important when determining shock temperatures and WD masses.
Therefore, we gradually changed the lower bound of the XIS spectrum, and found that 
fits become acceptable when the lower bound energies listed in Table \ref{table:observation_info}, 3.5~keV for NY Lup, and 4~keV for V1223 Sgr and RXS J2133, were employed.
These changes did not significantly affect the best-fit parameters, and therefore we simply regarded the resultant parameters of these fittings with the narrower energy bands as the current best estimations.

In FO Aqr and V2400 Oph, the partial covering model did not give an acceptable fit even when the employed energy ranges were changed.
Instead, the fits improved significantly to an acceptable level when an additional absorbed \verb|ip_psr| component was incorporated. Since these sources are supposed to have high accretion rates \citep{suleimanovetal2005}, we consider that taking this additional absorbed component into account is reasonable. Therefore, we estimated WD masses in these sources using this modified model.
The column densities of the additional absorbed component are also listed below those of the primary PC model in Table \ref{table:pc_result}.

In systems with special accreting and viewing geometries, the Fe emission lines (6.7~keV) could be artificially enhanced or suppressed by ``resonance trapping'' of the Fe-K line photons \citep{teradaetal2001}, because the optical depth of the process is higher than 1 \citep[e.g. 25 by ][]{doneetal1995}. This could affect our abundance measurements, and in turn, those of $M_{\rm{WD}}$.
However, this effect is only of a importance when the column is observed either pole-on (the Fe-K line enhanced) or side-on (suppressed) through a large fraction of the spin phase \citep{teradaetal2001}.

Because our 17 IPs are more or less pulsating at their spin periods, they are not likely to be either pole-on or side-on, and the enhancement or suppression, if any, can be averaged out by the WD spin (changing viewing angle).
We therefore consider that this effect is negligible for the present study, and that the derived sub-solar abundances reflect a true property of the accreting gases.

\begin{table*}
\centering
\caption{
The best-fit spectral parameters obtained with the partial-covering absorber model.
The errors are at 90\% confidence levels.
}
\label{table:pc_result}
\begin{tabular}{llllllllll}
\hline
\hline
System   &   $Z_{\rm{Fe}}$   & $M_{\rm{WD}}$     & $kT_{\rm{s}}$  & $n\rm{H}$$^{\mathrm{a}}$   & $n\rm{H}_{\rm{PC}}$$^{\mathrm{b}}$   & $C_{\rm{PC}}$$^{\mathrm{c}}$ & $\chi^2_\nu(\nu)$ & $F_{2,10}$$^{\mathrm{d}}$ & $F_{12,40}$$^{\mathrm{e}}$ \\
         &   $(Z_\odot)$                    & $(M_\odot)$       &  (keV)         & $10^{22}~\rm{cm}^{-2}$     & $10^{22}~\rm{cm}^{-2}$\\
\hline
FO Aqr & $0.10^{+0.03}_{-0.09}$ & $0.45^{+0.08}_{-0.03}$ & $14.0^{+4.0}_{-1.4}$ & $8.20^{+2.50}_{-2.67}$ & $373^{+133}_{-78}$ &  $0.79^{+0.05}_{-0.16}$& $0.96(386)$ & $4.27$ & $6.00$ \\
  &  &  &  &  & $38.4^{+13.4}_{-7.5}$$^{\mathrm{f}}$ &  1.0 (fixed) &  &  &  \\
XY Ari & $0.43^{+0.04}_{-0.03}$ & $1.04^{+0.15}_{-0.14}$ & $61.0^{+26.2}_{-16.2}$ & $8.97^{+0.44}_{-0.42}$ & $114^{+34}_{-25}$ &  $0.36^{+0.06}_{-0.04}$& $1.11(457)$ & $1.59$ & $2.73$ \\
MU Cam & $0.59^{+0.08}_{-0.08}$ & $0.95^{+0.36}_{-0.13}$ & $49.9^{+75.6}_{-12.5}$ & $5.05^{+1.07}_{-1.32}$ & $91.0^{+33.2}_{-26.9}$ &  $0.56^{+0.06}_{-0.07}$& $1.02(291)$ & $1.03$ & $1.79$ \\
BG CMi & $0.22^{+0.05}_{-0.05}$ & $1.14^{+0.09}_{-0.17}$ & $76.9^{+20.3}_{-24.7}$ & $4.08^{+3.09}_{-4.08}$ & $25.4^{+10.0}_{-6.7}$ &  $0.53^{+0.23}_{-0.22}$& $0.98(350)$ & $2.24$ & $3.18$ \\
V709 Cas & $0.22^{+0.05}_{-0.04}$ & $1.22^{+0.05}_{-0.20}$ & $94.5^{+15.0}_{-36.2}$ & $1.87^{+0.56}_{-0.65}$ & $90.7^{+31.2}_{-21.3}$ &  $0.38^{+0.03}_{-0.03}$& $0.99(382)$ & $4.12$ & $5.42$ \\
TV Col & $0.49^{+0.06}_{-0.06}$ & $0.91^{+0.14}_{-0.10}$ & $45.7^{+16.6}_{-9.1}$ & $3.38^{+1.30}_{-2.63}$ & $35.2^{+19.6}_{-14.4}$ &  $0.39^{+0.16}_{-0.08}$& $1.08(383)$ & $5.08$ & $5.92$ \\
TX Col & $0.47^{+0.12}_{-0.12}$ & $0.82^{+0.24}_{-0.16}$ & $37.4^{+26.3}_{-11.8}$ & $1.56^{+1.67}_{-1.56}$ & $32.9^{+27.8}_{-15.7}$ &  $0.38^{+0.13}_{-0.13}$& $1.18(341)$ & $1.23$ & $1.18$ \\
YY Dra & $0.50^{+0.19}_{-0.11}$ & $0.67^{+0.30}_{-0.14}$ & $26.3^{+25.9}_{-8.3}$ & $0.47^{+1.12}_{-0.47}$ & $90.7^{+77.5}_{-72.3}$ &  $0.32^{+0.20}_{-0.26}$& $1.07(340)$ & $3.77$ & $2.74$ \\
PQ Gem & $0.24^{+0.05}_{-0.05}$ & $1.15^{+0.10}_{-0.18}$ & $78.8^{+24.2}_{-26.6}$ & $1.94^{+1.60}_{-1.94}$ & $38.0^{+28.8}_{-14.2}$ &  $0.41^{+0.12}_{-0.10}$& $1.17(386)$ & $2.49$ & $3.11$ \\
EX Hya & $0.56^{+0.04}_{-0.03}$ & $0.42^{+0.02}_{-0.02}$ & $12.7^{+0.9}_{-0.9}$ & $0.69^{+0.23}_{-0.20}$ & $101^{+13}_{-12}$ &  $0.42^{+0.05}_{-0.04}$& $1.22(358)$ & $9.10$ & $3.56$ \\
NY Lup & $0.55^{+0.04}_{-0.04}$ & $1.15^{+0.08}_{-0.07}$ & $78.8^{+18.4}_{-12.0}$ & $6.13^{+0.60}_{-0.63}$ & $145^{+31}_{-22}$ &  $0.46^{+0.02}_{-0.02}$& $1.19(307)$ & $3.29$ & $6.03$ \\
V2400 Oph & $0.25^{+0.02}_{-0.04}$ & $0.62^{+0.06}_{-0.05}$ & $23.1^{+3.8}_{-2.9}$ & $1.69^{+0.89}_{-1.69}$ & $308^{+40}_{-26}$ &  $0.60^{+0.07}_{-0.09}$& $1.12(457)$ & $5.00$ & $5.99$ \\
  &  &  &  &  & $38.3^{+18.7}_{-15.0}$$^{\mathrm{f}}$ &  1.0 (fixed) &  &  &  \\
AO Psc & $0.44^{+0.08}_{-0.04}$ & $0.61^{+0.08}_{-0.04}$ & $22.5^{+5.1}_{-2.3}$ & $4.99^{+0.86}_{-2.41}$ & $54.4^{+23.4}_{-29.7}$ &  $0.38^{+0.08}_{-0.07}$& $1.03(595)$ & $4.96$ & $4.06$ \\
V1223 Sgr & $0.29^{+0.03}_{-0.02}$ & $0.75^{+0.05}_{-0.05}$ & $31.9^{+3.9}_{-3.6}$ & $9.45^{+0.85}_{-0.84}$ & $219^{+29.7}_{-25.7}$ &  $0.50^{+0.04}_{-0.02}$& $1.19(333)$ & $9.52$ & $15.57$ \\
RX J2133 & $0.38^{+0.04}_{-0.06}$ & $0.91^{+0.19}_{-0.17}$ & $45.7^{+24.2}_{-14.6}$ & $9.33^{+1.39}_{-1.04}$ & $275^{+63.3}_{-39.7}$ &  $0.58^{+0.08}_{-0.06}$& $1.16(331)$ & $2.10$ & $4.64$ \\
IGR J17303 & $0.29^{+0.05}_{-0.05}$ & $1.06^{+0.19}_{-0.14}$ & $63.8^{+39.2}_{-17.0}$ & $4.54^{+0.61}_{-0.66}$ & $240^{+37.0}_{-26.6}$ &  $0.69^{+0.03}_{-0.03}$& $1.08(364)$ & $2.07$ & $5.60$ \\
IGR J17195 & $0.38^{+0.05}_{-0.05}$ & $1.03^{+0.24}_{-0.22}$ & $59.6^{+49.9}_{-23.0}$ & $1.99^{+0.54}_{-1.06}$ & $79.6^{+47.3}_{-29.0}$ &  $0.36^{+0.10}_{-0.04}$& $0.95(373)$ & $3.99$ & $4.96$ \\
\hline
\end{tabular}\\
\begin{list}{}{}
\item[ ]$^{\mathrm{a}}$The absorption column density of fully-covering absorbent.
$^{\mathrm{b}}$ The absorption column density of the partial-covering absorbent (see text).
\item[]$^{\mathrm{c}}$The partial-covering fraction.
$^{\mathrm{d}}$$2-10$~keV flux in units of $10^{-11}~\ergcms$.
$^{\mathrm{e}}$$12-40$~keV flux in units of $10^{-11}~\ergcms$.
$^{\mathrm{f}}$The column density of the additional absorbed component (see text).
\end{list}
\end{table*}

\begin{table*}
\centering
\caption{Estimated WD masses compared with previous reports in the X-ray band. The mass ranges associated are all 90\% confidence intervals.}
\label{table:mwd_compare}
\begin{tabular}{lllllllll}
\hline
\hline
System   & {\it Suzaku}$^{\rm{a}}$         & {\it Swift}$^{\rm{b}}$   & {\it RXTE}$^{\rm{c}}$    & {\it RXTE}$^{\rm{d}}$             & {\it Ginga}$^{\rm{e}}$            & {\it ASCA}$^{\rm{f}}$ \\
         & XIS+HXD              & BAT           & PCA+HEXTE     & PCA                    & LAC                    & SIS\\
         & $M_{\rm{WD}}$        & $M_{\rm{WD}}$ & $M_{\rm{WD}}$ & $M_{\rm{WD}}$          & $M_{\rm{WD}}$          & $M_{\rm{WD}}$\\
\hline
FO Aqr &  $0.45^{+0.08}_{-0.03}$  & $0.61\pm0.05$ & $0.60\pm0.05$ & $0.88^{+0.07}_{-0.09}$ & $0.92^{+0.30}_{-0.34}$ & $1.05(>0.44)$\\
XY Ari &  $1.04^{+0.15}_{-0.14}$  & $0.96\pm0.12$ & $$            & $0.97^{+0.16}_{-0.17}$ & $$                     & $$\\
MU Cam &  $0.95^{+0.36}_{-0.13}$  & $0.74\pm0.13$ & $$            & $$                     & $$                     & $$\\
BG CMi &  $1.14^{+0.09}_{-0.17}$  & $0.67\pm0.19$ & $0.85\pm0.12$ & $1.15^{+0.06}_{-0.06}$ & $1.09(>0.94)$          & $$\\
V709 Cas &  $1.22^{+0.05}_{-0.20}$  & $0.96\pm0.05$ & $0.90\pm0.10$ & $1.08^{+0.05}_{-0.17}$ & $$                   & $$\\
TV Col &  $0.91^{+0.14}_{-0.10}$  & $0.78\pm0.06$ & $0.84\pm0.06$ & $0.97^{+0.02}_{-0.05}$ & $1.3(>0.9)$            & $0.51^{+0.41}_{-0.22}$\\
TX Col &  $0.82^{+0.24}_{-0.16}$  & $0.67\pm0.10$ & $0.70\pm0.30$ & $0.74^{+0.05}_{-0.05}$ & $0.48^{+0.06}_{-0.09}$ & $0.66^{+0.73}_{-0.42}$\\
YY Dra &  $0.67^{+0.30}_{-0.14}$  & $0.50\pm0.11$ & $0.75\pm0.05$ & $$                     & $$                     & $$\\
PQ Gem &  $1.15^{+0.10}_{-0.18}$  & $0.65\pm0.09$ & $0.65\pm0.20$ & $$                     & $1.29(>1.17)$          & $$\\
EX Hya &  $0.42^{+0.02}_{-0.02}$  & $0.66\pm0.17$ & $0.50\pm0.05$ & $0.44^{+0.03}_{-0.03}$ & $0.46^{+0.04}_{-0.04}$ & $0.48^{+0.10}_{-0.06}$\\
NY Lup &  $1.15^{+0.08}_{-0.07}$  & $1.09\pm0.07$ & $$            & $$                     & $$                     & $$\\
V2400 Oph &  $0.62^{+0.06}_{-0.05}$  & $0.81\pm0.10$ & $0.59\pm0.05$ & $0.71^{+0.07}_{-0.03}$ &  & $0.68^{+0.42}_{-0.24}$ \\
AO Psc &  $0.61^{+0.08}_{-0.04}$  & $0.55\pm0.06$ & $0.65\pm0.05$ & $0.60^{+0.03}_{-0.03}$ & $0.56^{+0.16}_{-0.20}$ & $0.40^{+0.13}_{-0.10}$\\
V1223 Sgr &  $0.75^{+0.05}_{-0.05}$  & $0.65\pm0.04$ & $0.95\pm0.05$ & $1.07^{+0.08}_{-0.09}$ & $$                  & $1.28(>0.84)$\\
RX J2133 &  $0.91^{+0.19}_{-0.17}$  & $0.91\pm0.06$ & $$          & $$                     & $$                     & $$\\
IGR J17303 &  $1.06^{+0.19}_{-0.14}$  & $1.08\pm0.07$ & $$        & $$                     & $$                     & $$\\
\hline
\end{tabular}\\
\begin{list}{}{}
\item[ ]$^{\mathrm{a}}$This work.
$^{\mathrm{b}}$\citet{brunschweigeretal2009}.
$^{\mathrm{c}}$\citet{suleimanovetal2005}.
$^{\mathrm{d}}$\citet{ramsay2000}.
$^{\mathrm{e}}$\citet{cropperetal1999}.
$^{\mathrm{f}}$\citet{ezukaishida1999}.
\end{list}
\end{table*}

\subsection{HXD/PIN-only fitting}
We also fitted the HXD/PIN spectra alone to confirm the result of the partial covering model at the energy band where even the strongest absorption little changes the spectral model. In the fitting, we fixed $C_{\rm{PC}}=0$.

\begin{figure}
\centering
\resizebox{8cm}{!}{\includegraphics{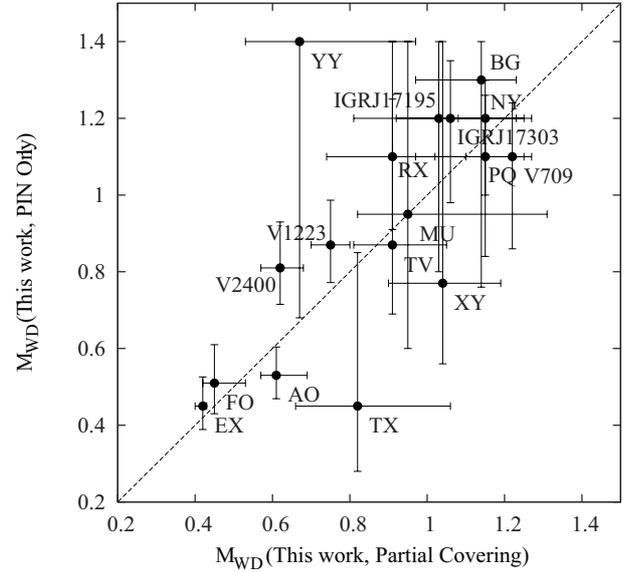}}\\
\caption{
Comparison of the WD masses estimated from the partial covering model using the XIS and HXD/PIN data, with those determined by the HXD/PIN data alone ignoring the partial covering.
Source names are abbreviated.
}
\label{figure:mwd_pin_compare}
\end{figure}

The results of this analysis are compared with those from the wide-band fitting (Sect. \ref{section:widebandfitting}) in Fig. \ref{figure:mwd_pin_compare}.
Large y-axis errors are simply due to lower statistics in the PIN data. 
The masses derived with the two methods thus agree with each other within errors.
In particular, we do not find any systematic differences between the two sets of estimates. This self-calibration ensures the consistency of the continuum and line methods.

\section{Discussion}
\label{section:discussion}

\subsection{Comparison with previous reports : WD masses}
\label{section:comparison_of_wd_masses}
\subsubsection{Based on X-ray data}
\begin{figure}
\centering
\resizebox{8cm}{!}{\includegraphics{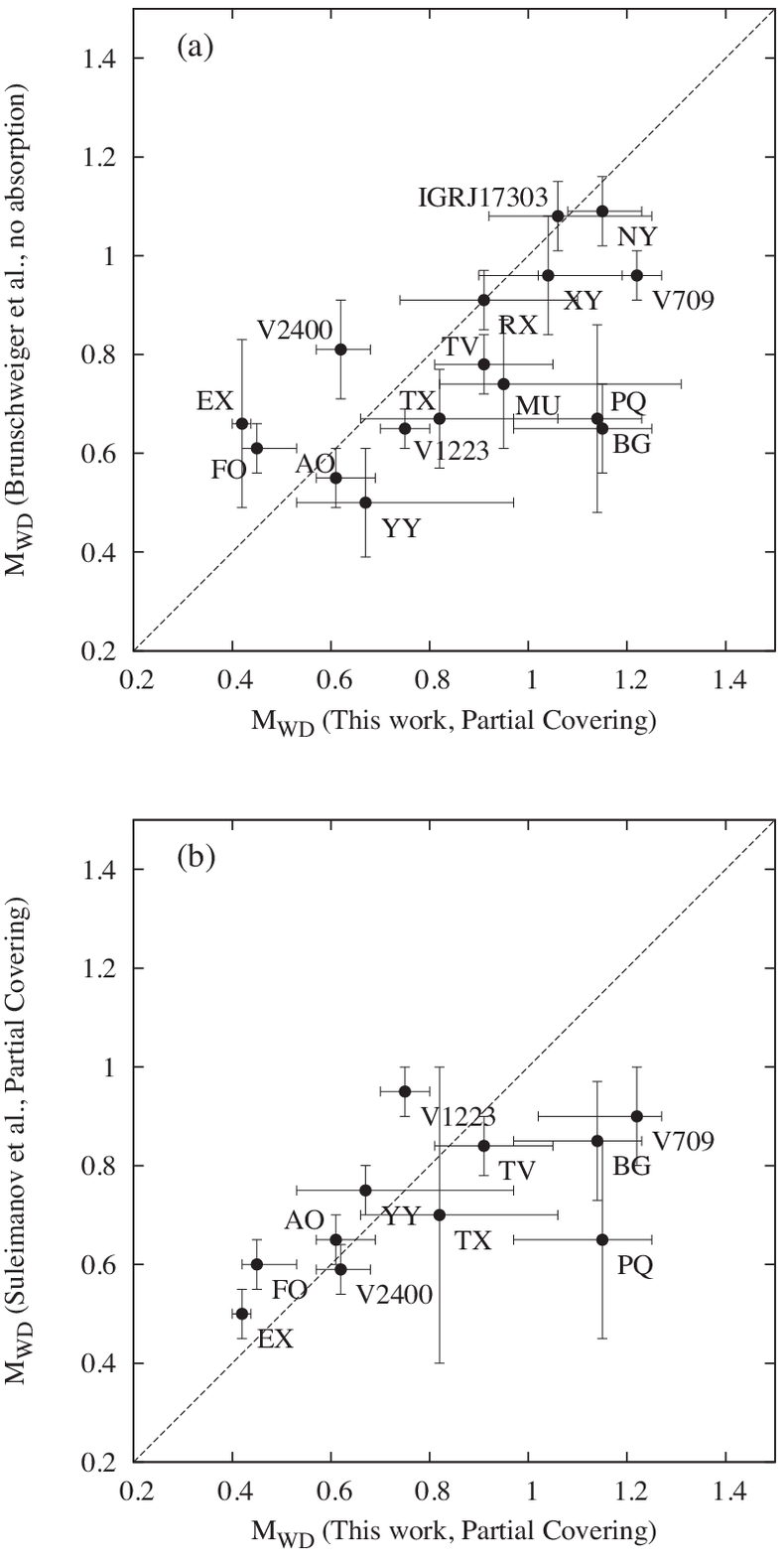}}
\caption{
Comparison of the WD masses estimated in the present study with those by (a) \citet{suleimanovetal2005} and (b) \citet{brunschweigeretal2009}.
}
\label{figure:mwd_compare}
\end{figure}

Table \ref{table:mwd_compare} compares the present {\it Suzaku} XIS+HXD results with the WD masses reported by the previous studies in the X-ray band. 
In Fig. \ref{figure:mwd_compare} we plot correlations between our WD masses and those from the {\it RXTE}/PCA+HEXTE data by \citet{suleimanovetal2005} and the {\it Swift}/BAT data by \citet{brunschweigeretal2009}; the spectral models and the energy band used in these three studies are very similar to one another. 

In Figs. \ref{figure:mwd_compare} (a) and (b), our results derive higher WD masses, especially for massive ($\gtrsim1~M_\odot$) systems, although rough correlations are seen among the three measurements.
For a few sources, for example V709 Cas, BG CMi, and PQ Gem, this discrepancy amounts to $\sim0.3-0.4~M_\odot$. Reasons for these differences are unclear at present.
Unfortunately, the WD masses of \citet{suleimanovetal2005} and \citet{brunschweigeretal2009} show the same level of disagreement as seen in these figures, although the two studies use the same spectral model calculated in the former. This might indicate that the results suffer from rather large systematic uncertainties, which could exceed the quoted statistical fitting errors. We are aware that our analysis is not free either of possible systematic uncertainties (Sect. \ref{section:limitations}) involved in the detectors and the spectral model. Yet our result may be less subject to various systematic effects, because we jointly utilized the Fe-K line spectroscopy and the hard-band continuum shape, and the wide-band fitting result is consistent with that from the PIN-only fitting (Fig. \ref{figure:mwd_pin_compare}).

From X-ray light curve analysis in the eclipsing IP XY Ari, \citet{hellier1997xyaritiming} estimated $M_{\rm{WD}}$ to be $0.91-1.17~M_\odot$. 
Our result, $M_{\rm{WD}}=1.14^{+0.15}_{-0.14}~M_\odot$, nicely agrees with this value, and 
this agreement provides a convincing calibration to the present methodology (see also \citealt{brunschweigeretal2009}).

\subsubsection{Other methods}
\begin{figure}
\centering
\resizebox{8cm}{!}{\includegraphics{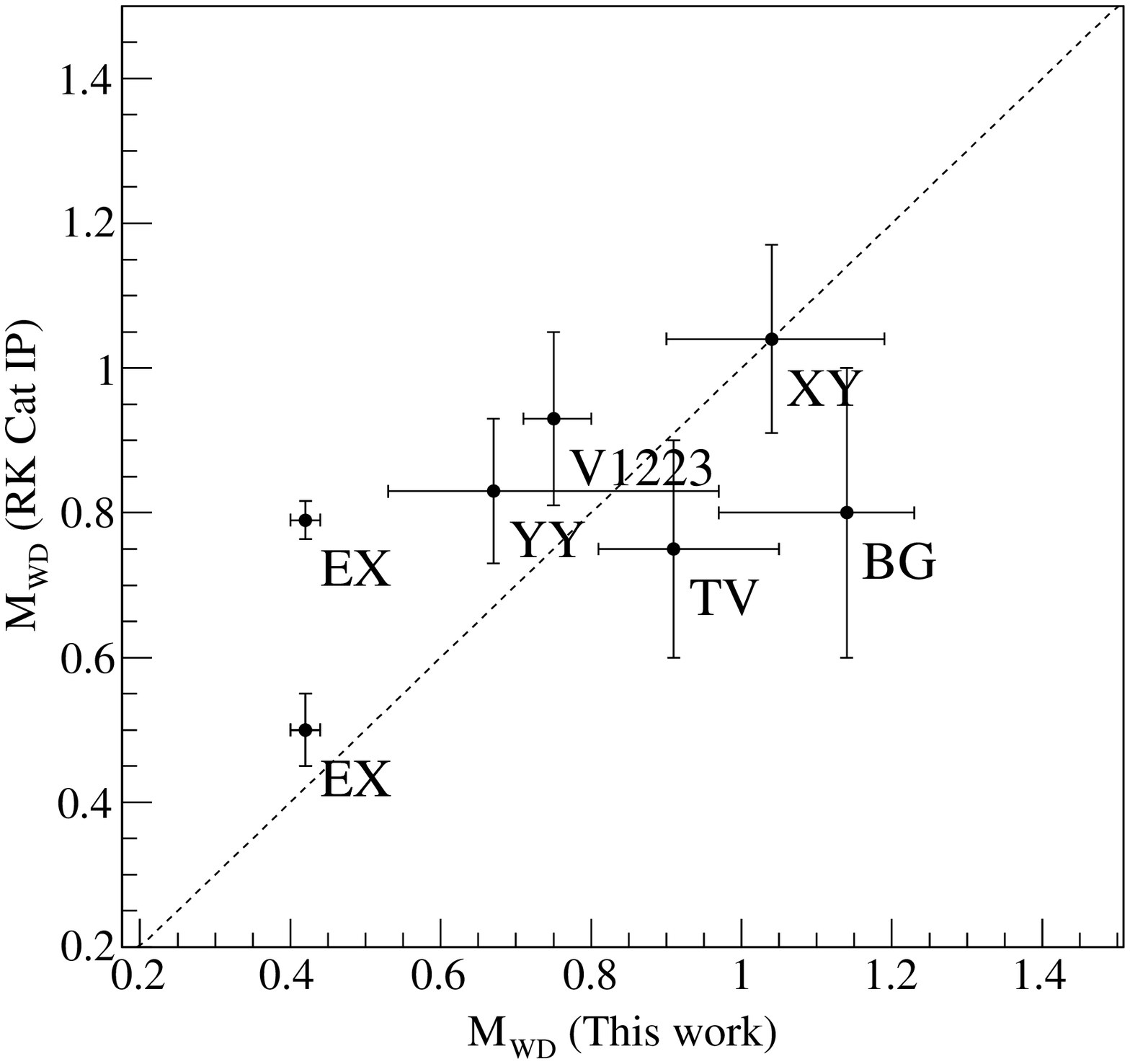}}
\caption{
Comparison of the WD masses estimated in the present study with those in the optical and infra-red wavelengths. References are \citet{hellier1997xyaritiming} for XY Ari, \citet{penning1985bgcmi} for BG CMi, \citet{hellier1993tvcol} for TV Col, \citet{haswelletal1997yydra} for YY Dra, and \citet{beuermannetal2004v1223sgr} for V1223 Sgr. For EX Hya, two different values are presented; $0.790\pm0.026~M_\odot$ and $0.50\pm0.05~M_\odot$ by \citet{beuermannandreinsch2008} and \citet{mhlahloetal2007}, respectively (see text). Quoted errors of the present result and the previous reports are at 90\% confidence levels.}
\label{figure:mwd_compare_optical_ir}
\end{figure}

It is also meaningful to compare the derived WD masses with those estimated from binary kinematics based mainly on optical/IR spectroscopy. Using references in the CV catalog by \citet{ritterkolb2003}, we collected reports on $M_{\rm{WD}}$ of IPs, and then plotted them as shown in Fig. \ref{figure:mwd_compare_optical_ir}. Note that most of the reports employed an assumption of a Roche-lobe filling Zero Age Main Sequence star to estimate radius and mass of the secondary (non-degenerate) stars, and this may be a cause of systematic uncertainties in their WD mass estimations.

All masses agree (with rather large errors) except for EX Hya. For this object, WD masses determined spectroscopically in optical/IR wavelength have been somewhat controversial.
The difference of the WD masses are mainly caused by different radial velocity estimations.
Based on NaI and CaII line profiles, \citet{beuermannandreinsch2008} estimated the amplitude of a radial velocity of the secondary (non-degenerate) star $K_2$ to be $432.4\pm4.8$~km~s$^{-1}$ by assuming that of the primary WD $K_1$ of $59\pm3$~km~s$^{-1}$ \citep{belleetal2003,hoogerwerfetal2004}.
From these values, they obtained $M_{\rm{WD}}=0.790\pm0.026~M_\odot$.
On the contrary, \citet{mhlahloetal2007} derived $K_1=74\pm2$~km~s$^{-1}$ from H$\alpha$, H$\beta$, and H$\gamma$ lines, and used $K_2$ of $360\pm35$~km~s$^{-1}$ reported by \cite{vandeputteetal2003}, leading to the WD mass $M_{\rm{WD}}=0.50\pm0.05~M_\odot$ which disagrees with the former value by more than four times the error range.

Independently of the radial velocities, the PIN spectrum of EX Hya clearly shows a cutoff in the $10-20$~ keV band (Sect. \ref{section:widebandfitting}), and because of this cutoff together with the Fe line structure, the lower WD mass is favored by the fitting.
The cutoff cannot be explained by a WD of 0.7~$M_{\rm{WD}}$ since 
our model implies a shock temperature of $30$~keV and a similar cutoff-energy, and even multiple absorption or reflection cannot fake a lower cutoff energy from a spectrum of the high shock temperature.
The inconsistency with the WD mass of \citet{beuermannandreinsch2008} might imply the necessity of some modifications to the assumptions of the geometry and the physical processes (Sect. \ref{section:the_model_in_this_study}) in this object.

\subsection{The WD mass spectrum and its average}
\label{section:wd_mass_spectrum_and_its_average}
\begin{figure}
\centering
\resizebox{7cm}{!}{\includegraphics{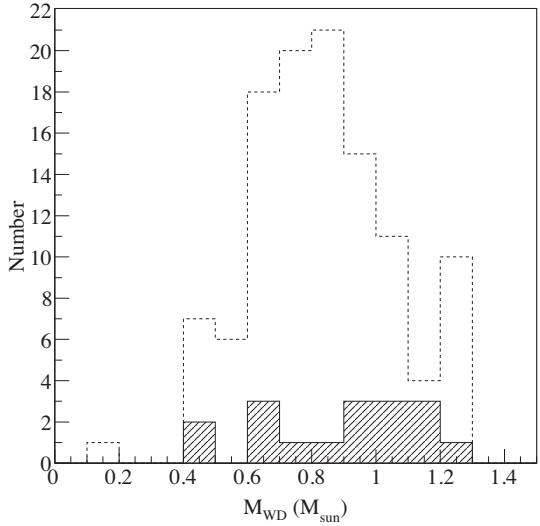}}\\
\caption{
WD mass spectrum of our sample IPs (shaded histogram), compared with that of 104 CVs (histogram with dashed line; taken from \citealt{ritterkolb2003}).
}
\label{figure:mwd_comparison_with_rkcat}
\end{figure}

Figure \ref{figure:mwd_comparison_with_rkcat} shows the distribution of the WD mass of our sample, compared with those of CVs determined kinematically, mainly based on optical/IR spectroscopy \citep{ritterkolb2003}. The two distributions give the mean masses of $0.88\pm0.25~M_\odot$ for the present study and $0.82\pm0.23~M_\odot$ for the optical sample (errors are 1$\sigma$ standard deviations), and the Kolmogorov-Smirnov probability is 0.39. Therefore it is inferred with this limited sample that the mass spectra of CVs and IPs are statistically not largely different. With this distribution, although the systematic uncertainties (Sect. \ref{section:limitations}) should be examined carefully, we consider that the WD masses of IPs are truly distributed, not extracted from a single common value with errors caused by the analysis method.

The mean WD mass of our IP sample, $0.88\pm0.25~M_\odot$, can be compared with the value of $\sim0.5~M_{\odot}$ inferred by \cite{krivonosetal2007} based on the analysis of Galactic Ridge X-ray Emission (GRXE) in the hard X-ray band with {\it INTEGRAL}/SPI. This comparison is intriguing because IPs are thought to have deep connection with the origin of the GRXE (especially in the hard X-ray band). However, before examining any discussion about the discrepancy of the two values, we should consider the incompleteness of our sample. The current sample only consists of IPs localized to the solar vicinity ($\lesssim500$~pc) because of their intrinsic low luminosity. Because their selection is usually flux-limited, it can be biased to IPs with higher WD masses which give higher fluxes in the hard X-ray band.
In addition, the GRXE spectrum may be decomposed into various types of X-ray sources 
over a wide energy range (e.g. \citealt{revnivtsevetal2006}). Among them, RS CVn systems and dwarf novae have softer spectra, and if their contribution is considered simultaneously with IPs,
the IP component in the GRXE spectrum may become harder, leading to a higher
WD mass. Then the difference of the two values may be reduced.

\subsection{Comparison of Fe abundances}
\citet{fujimotoishida1997}, \citet{ezukaishida1999}, and \citet{evansetal2004} estimated metal abundances of IPs by analyzing the resolved Fe line complex ($6-7$~keV) and X-ray continua with the multi-temperature PSR model.
\citet{fujimotoishida1997} obtained the Fe abundance of EX Hya as $0.6\pm0.2~Z_{\odot}$, which was later revised by 
\citet{ezukaishida1999} to $0.72^{+0.08}_{-0.12}~Z_\odot$ together with measurements of the Fe abundances of 12 other IPs. 
\citet{evansetal2004} gave the metal abundance of $0.121^{+0.028}_{-0.023}~Z_\odot$ for FO Aqr (whilst they did not take their result as evidence of genuine sub-solar abundance).

\begin{figure}
\centering
\resizebox{7.5cm}{!}{\includegraphics{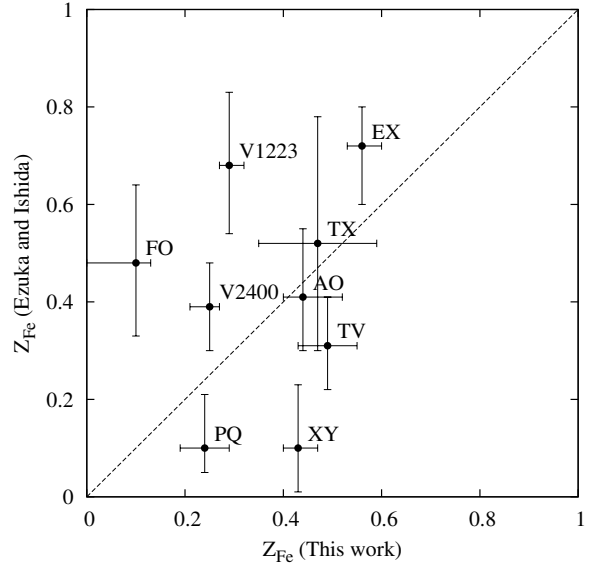}}\\
\caption{
Fe abundances determined in the present study, compared with the measurements by \citet{ezukaishida1999}. Source names are abbreviated.
}
\label{figure:zfe_compare}
\end{figure}

Since our IP sample is well covered by that of \citet{ezukaishida1999}, we compare our result with theirs in Table \ref{table:zfe_compare} and Fig. \ref{figure:zfe_compare}. 
Although the two results consistently indicate sub-solar abundances, they exhibit no particular correlation. Considering that the {\it Suzaku} data have a much wider spectral coverage, higher statistics, and a somewhat better energy resolution than those from {\it ASCA} which \citet{ezukaishida1999} used, our results may be seen as more updated. Note also that \citet{ezukaishida1999} did not take into account partial covering or the reflection component.

When our 17 objects are divided into seven objects with $M_{\rm{WD}}>1.0~M_\odot$ and the other 10 with $M_{\rm{WD}}\le1.0~M_\odot$, the former and latter sub-samples have averaged Fe abundances of $0.33\pm0.12~Z_\odot$ and $0.41\pm0.15~Z_\odot$, respectively. The lack of significant difference ensures that our abundance determinations are not strongly affected by the continuum shape.
We suppose that these different sub-solar values reflect various metal abundances of the accreting gas and the companion (non-degenerate) stars.

Previous studies of summed (unresolved) Fe lines also reported sub-solar Fe abundances in IPs. 
Using {\it Ginga}, \citet{beardmoreetall2000} reported an underabundance of metals by a factor of $\sim2$ relative to solar for V1223 Sgr. 
This roughly agrees with our result, $0.30-0.35~Z_\odot$.
\citet{demartinoetal2004} gave $0.33^{+0.07}_{-0.09}~Z_\odot$ for PQ Gem (RE0751$+$14), which is also comparable to ours ($0.19-0.32~Z_\odot$).

\begin{table}
\centering
\caption{Comparison of the Fe abundances with values reported by \citet{ezukaishida1999} associated with 90\% confidence intervals.}
\label{table:zfe_compare}
\begin{tabular}{llllll}
\hline
\hline
System & This work & Ezuka \& Ishida \\
& $Z_{\rm{Fe}} (Z_\odot)$ & $Z_{\rm{Fe}} (Z_\odot)$\\
\hline
FO Aqr & $0.10(0.01-0.13)$ & $0.48(0.33-0.64)$\\
XY Ari & $0.43(0.40-0.47)$ & $0.10(0.01-0.23)$\\
TV Col & $0.49(0.43-0.55)$ & $0.31(0.22-0.41)$\\
TX Col & $0.47(0.35-0.59)$ & $0.52(0.30-0.78)$\\
PQ Gem & $0.24(0.19-0.29)$ & $0.10(0.05-0.21)$\\
EX Hya & $0.56(0.53-0.60)$ & $0.72(0.60-0.80)$\\
V2400 Oph & $0.25(0.21-0.27)$ & $0.39(0.30-0.48)$\\
AO Psc & $0.44(0.40-0.52)$ & $0.41(0.30-0.55)$\\
V1223 Sgr & $0.29(0.27-0.32)$ & $0.68(0.54-0.83)$\\
\hline
\end{tabular}
\end{table}

\subsection{Limitations and systematic uncertainties of our assumptions}
\label{section:limitations}
In this section, we discuss the limitations in our assumptions of the present model calculation
and the WD mass estimation.
As a whole, these may cause systematic uncertainties, and
affect our WD mass results (Table \ref{table:pc_result}) to some extent.
Nevertheless, we regard the present results as meaningful,
because the estimated systematic effects are not significantly larger
than the statistical precision associated with our determinations of $M_{\rm WD}$.
More detailed spectral analyses, which are less subject to the limitations
will be possible with data taken with future high-resolution and 
wide-band instruments, such as the X-ray microcalorimeter and the hard X-ray imager
onboard the \it ASTRO-H \rm satellite \citep{takahashietal2008astroh}.

\subsubsection{Dynamics of accreting gas and system geometries}
As detailed in \citet{suleimanovetal2005}
and \citet{brunschweigeretal2009},
the assumption that the accreting gas falls freely
from an infinite distance is not necessarily warranted,
since the gas may well form an accretion disk
and radiate away part of its energy release there.
If such a disk is truncated, under plausible parameters,
at a radius of $\sim 3 \times 10^{9}$~cm in an IP
with a WD of $0.9~M_\odot$,
the shock temperature, and hence $M_{\rm WD}$,
would be underestimated by $\sim 20\%$.

Similarly, a fast spin of a WD would also
cause the present method to underestimate $M_{\rm WD}$
because of the centrifugal force.
If corrected for this effect,
the WD masses in Table \ref{table:pc_result} of two rapid rotators in our sample,
V709 Cas (with a spin period of 312.78 s) and YY Dra ($\sim530$~s),
could increase by a few tens of percent depending
on their accretion rates.

In addition to this, a fast spinning WD might obey
a different (equatorially elongated) mass-radius relation (e.g. \citealt{nauenberg1972};\citealt{geroyannisandpapasotiriou1997}). However, precise numerical treatment of this effect exceeds the scope of our present study, and we leave it untouched.

According to \citet{canalleetal2005}, the inclusion of the dipole curvature in the PSR calculation 
reduces the calculated shock temperature by $\sim10\%$ for $M_{\rm{WD}}=1.0~M_\odot$ under typical accreting colatitudes, and $<$2-3\% for $M_{\rm{WD}}\lesssim0.8~M_\odot$.
At present, it is hardly possible to observationally constrain the magnetic field structure near the WD magnetic poles. Therefore, we reluctantly neglected the curvature of the WD magnetic field when calculating the PSR structure (Sect. \ref{section:the_model_in_this_study}). In Table \ref{table:pc_result}, five systems have upper WD mass limits below 0.8~$M_\odot$, and we suppose this systematic effect can be safely neglected at least for these light systems.

\subsubsection{Accretion rate dependence}
\label{section:mdot_change}
In the spectral fitting,
we freely varied the normalization (i.e. accretion rate) of the PSR spectral model,
which was calculated assuming a fixed specific
accretion rate of 1.0~g~cm$^{-2}$~s$^{-1}$ (Sect. \ref{section:formulation_and_numerical_implementation}).
Strictly speaking, this means a self-inconsistency,
because the calculated shock height and the shock temperature should increase and decrease, respectively,
as the accretion rate decreases;
this, in turn, is because the free-free cooling rate is
proportional to the plasma density squared (e.g. \citealt{aizu1973}).

To examine the accretion-rate dependence of our numerical result,
we performed calculations with several different accretion rates as shown in Fig. \ref{figure:mdot_change_kts}.
As noted above, a lower value of $a$ gives a lower shock temperature at any $M_{\rm{WD}}$.
For higher $M_{\rm{WD}}$, difference of the resulting shock temperatures is more or less conspicuous.
At $M_{\rm{WD}}=0.8~M_\odot$, we see a difference of a factor of 1.13 between the results of $a=0.1$ and 1.0~g~cm$^{-2}$~s$^{-1}$, and a factor of 1.02 between $a=1.0$ and 10~g~cm$^{-2}$~s$^{-1}$. Correspondingly, at $M_{\rm{WD}}=1.2~M_\odot$, the difference increases to factors of 1.33 ($a=0.1$ to $a=1.0$) and 1.10 ($a=1.0$ to $a=10$).
In most systems we analyzed (except for V709 Cas), WD masses are below 1.2~$M_\odot$, 
and therefore we consider that this uncertainty affects the WD mass estimates only slightly, much less than $\sim30\%$.
Especially in lower $M_{\rm{WD}}$ ($\lesssim0.6~M_\odot$), 
the $a$ dependence is small, and therefore
we presume that the estimated shock temperatures and WD masses are considerably secure in this mass range.

\begin{figure}
\centering
\resizebox{8cm}{!}{\includegraphics{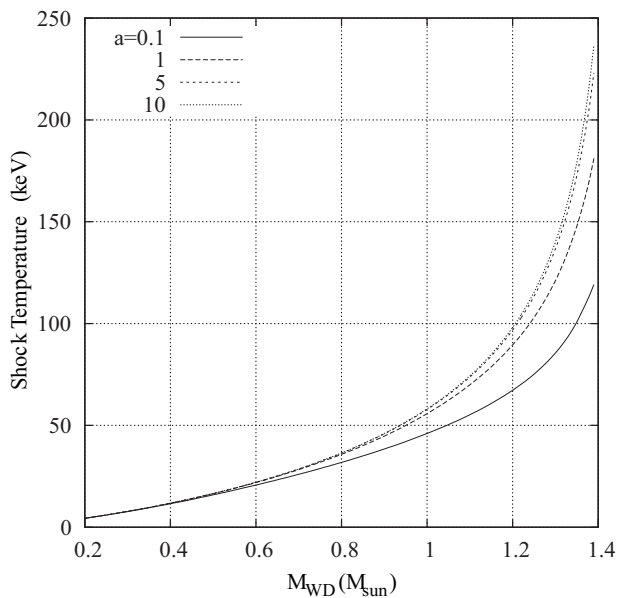}}
\caption{
Result of the numerical calculation of the PSR model when changing the accretion rate.
Shock temperatures are plotted with solid, long dashed, short dashed, and dotted lines
for accretion rates of 0.1, 1, 5, and 10~g~cm$^{-2}$~s$^{-1}$.
}
\label{figure:mdot_change_kts}
\end{figure}

\subsubsection{PSR base temperature}
In the present study, the PSR temperature was
assumed to fall to zero at its base although irradiating hard X-rays from the PSR may heat the WD photosphere (e.g. \citealt{cropperetal1998}).
Indeed, non-zero base temperatures have been suggested observationally by several authors (see. \citealt{ezukaishida1999}, and references therein).

This uncertainty could be avoided
by iteratively calculating the PSR emission model
with a finite base temperature until a model fit converges into a
self-consistent result. We did not perform this because of the limitation on
the computation time and the statistical quality of the data.
Furthermore, as noted in Sect. \ref{section:formulation_and_numerical_implementation},
emission will become thick near the WD surface, and therefore a more sophisticated spectral model which takes this effect into account is needed to be constructed for performing this analysis.

\subsubsection{Reflection component}
The spectra of the present IP sample show clear detections of the fluorescent Fe line at $\sim6.4$~keV. This strongly indicates that some of the incident X-rays are emitted from a PSR reflected by cold gas near the WD surface or pre-shock region (see \citealt{ezukaishida1999}, and references therein).
However, in the present spectral fitting (Sect. \ref{section:widebandfitting}) we did not take into account a reflection component from the WD surface explicitly because the computation power needed to precisely calculate the reflection is unaffordable at this moment. Qualitatively, its effect may be represented to some extent by the strongly absorbed component in the partial-covering model, which has a similar spectral shape to a reflection component, both contributing mainly in the PIN energy band. 

\citet{cropperetal1998} revealed that an inclusion of a reflection component changes the obtained WD masses only slightly from those of their partial covering fittings.
By analyzing a wide-band spectrum of V1223 Sgr with a reflection component, \citet{revnivtsevetal2004v1223sgr} derived a WD mass of $0.71\pm0.03~M_\odot$. Our result, $0.75\pm0.05~M_\odot$ well agrees with their value, and thus an effect of the reflection component can be considered not as very large.
A full treatment of a reflection component is left as our future work.

\section{Summary}
\label{section:summary}
We numerically solved the hydrostatic equations for PSRs of mCVs of \citet{suleimanovetal2005}, and updated it (Sect. \ref{section:the_model_in_this_study}) in terms of (1) the additional cooling function at higher temperatures and (2) the optically thin thermal emission code, which includes Fe emission lines. By constructing the spectral model file (Sect. \ref{section:the_model_in_this_study}), we fitted 17 IPs observed with {\it Suzaku}, and successfully reproduced the Fe line structures in the $6-7$~keV band and at the same time the thermal cutoff detected in the hard X-ray band (Sect. \ref{section:widebandfitting}).
From the best-fit models, we estimated the WD masses and the Fe abundances. 
The average WD mass for the present sample were $0.88\pm0.25~M_\odot$.
All sources exhibited sub-solar Fe abundances, and the average iron abundance was $0.38\pm0.14~Z_\odot$. The WD mass of the newly found IP, IGR J17195-4100, was estimated to be $1.03^{+0.24}_{-0.22}~M_\odot$ for the first time.

The derived WD masses are compared with previous reports (Sect. \ref{section:comparison_of_wd_masses}), and a rough correlation was confirmed, although results on a few systems differ significantly ($\sim0.3-0.4~M_\odot$). We also tentatively compared the mean WD mass with the value inferred from the analysis of the GRXE in the hard X-ray band (Sect. \ref{section:wd_mass_spectrum_and_its_average}), and found a difference by $\sim0.4~M_\odot$, although this may arise from various systematic biases.

\begin{acknowledgements}
T.Y. is financially supported by the Japan Society for the Promotion of Science.
This research has made use of data and softwares obtained from the Data Archive and Transmission System at JAXA/ISAS and the High Energy Astrophysics Science Archive Research Center at NASA Goddard Space Flight Center. 
\end{acknowledgements}


\appendix
\section{Spectral fitting result}
In Fig. \ref{figure:pc_result:b} we present the observed spectra overlaid with the best-fit model spectra (online version only).

\onlfig{1}{


\begin{figure*}[htb]
\centering
\subfigure{
\includegraphics[height=6cm]{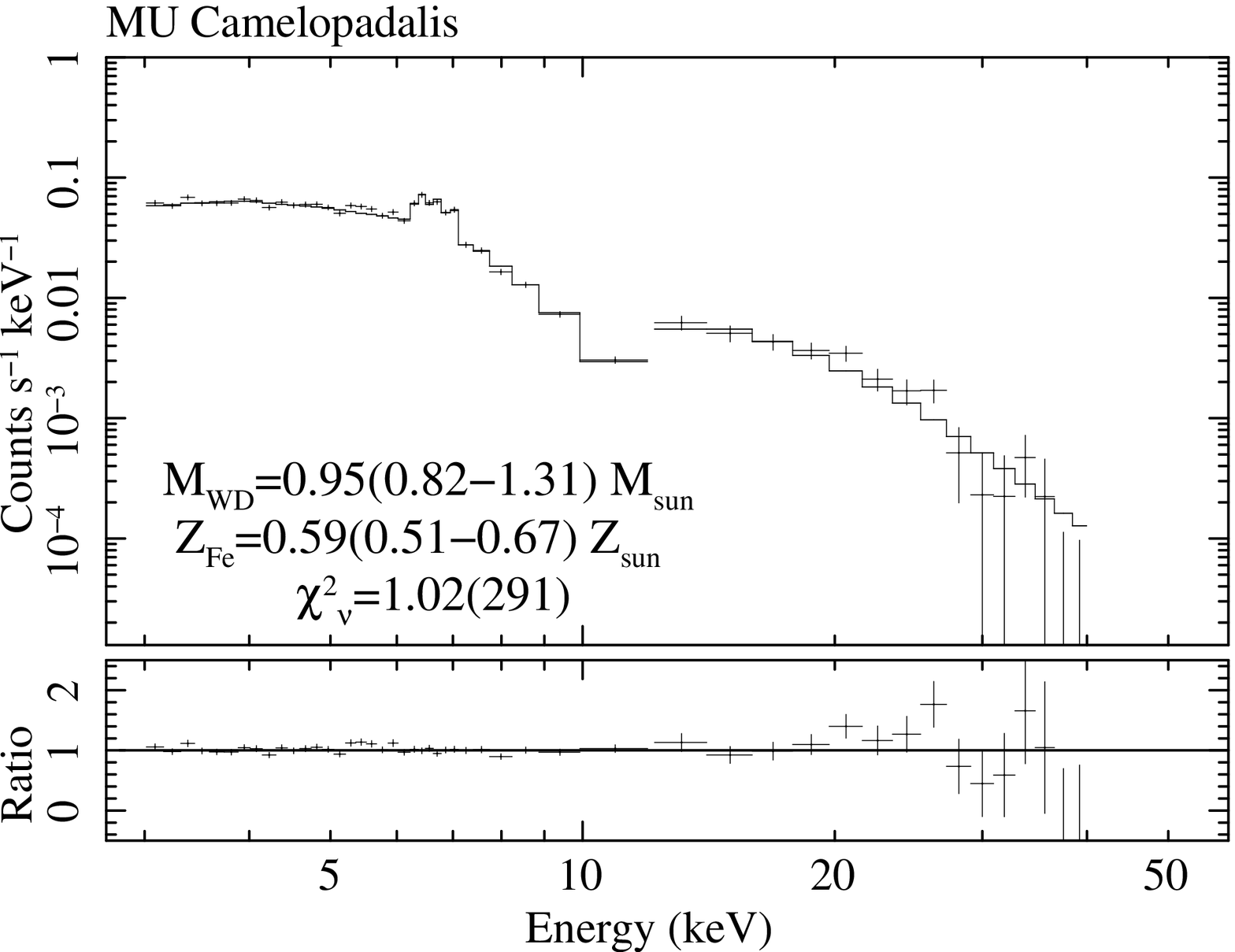}
\includegraphics[height=6cm]{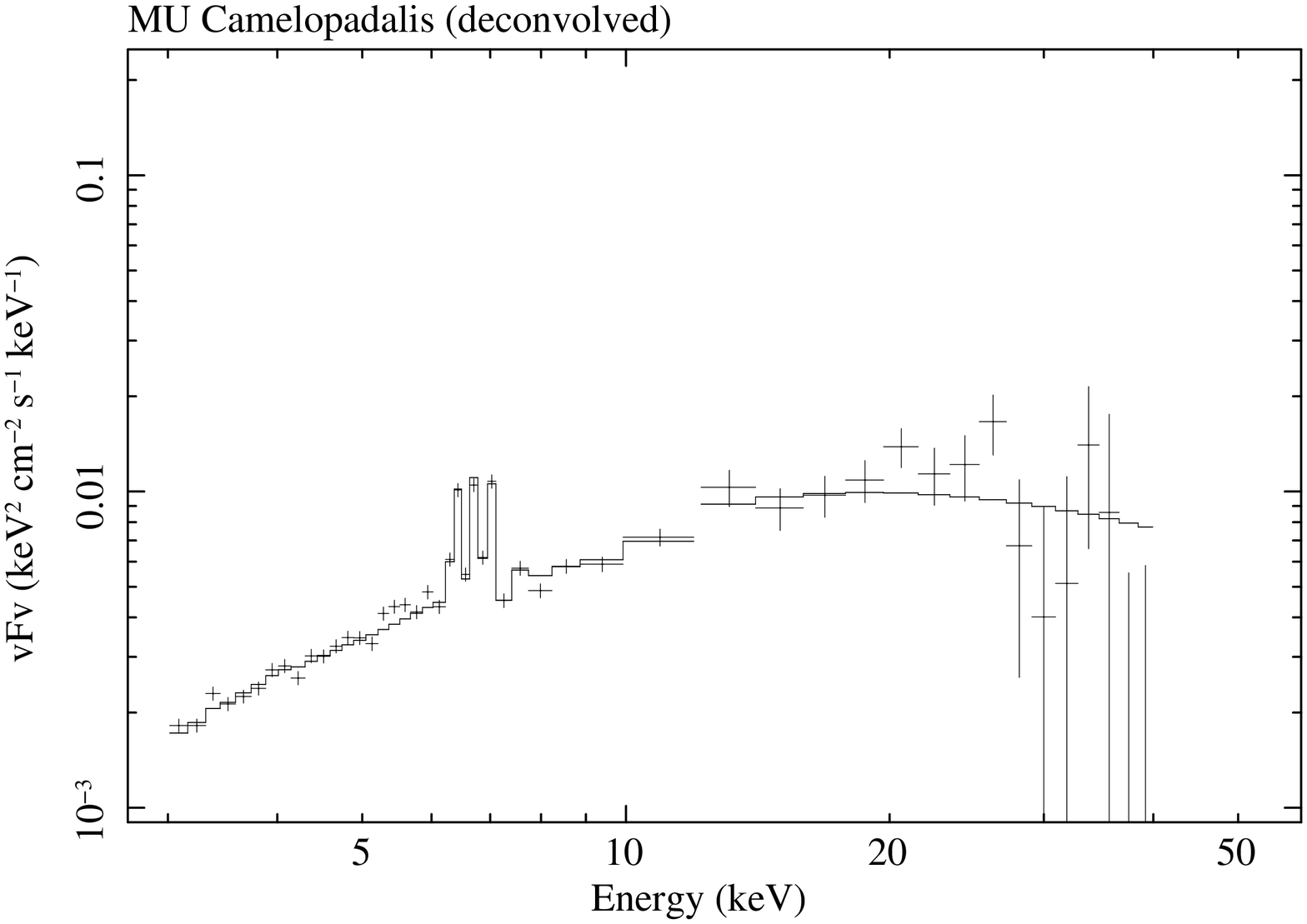}
}
\subfigure{
\includegraphics[height=6cm]{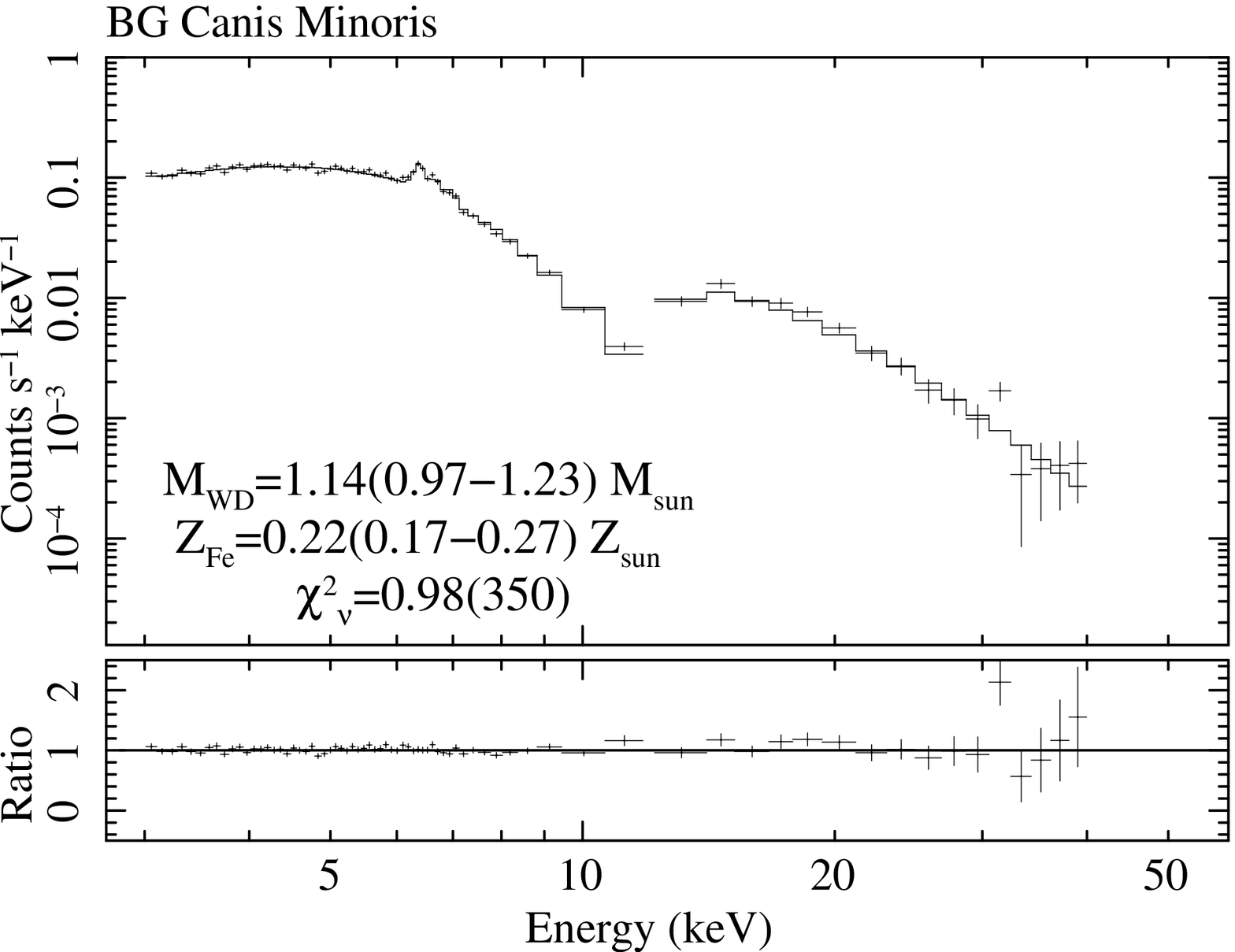}
\includegraphics[height=6cm]{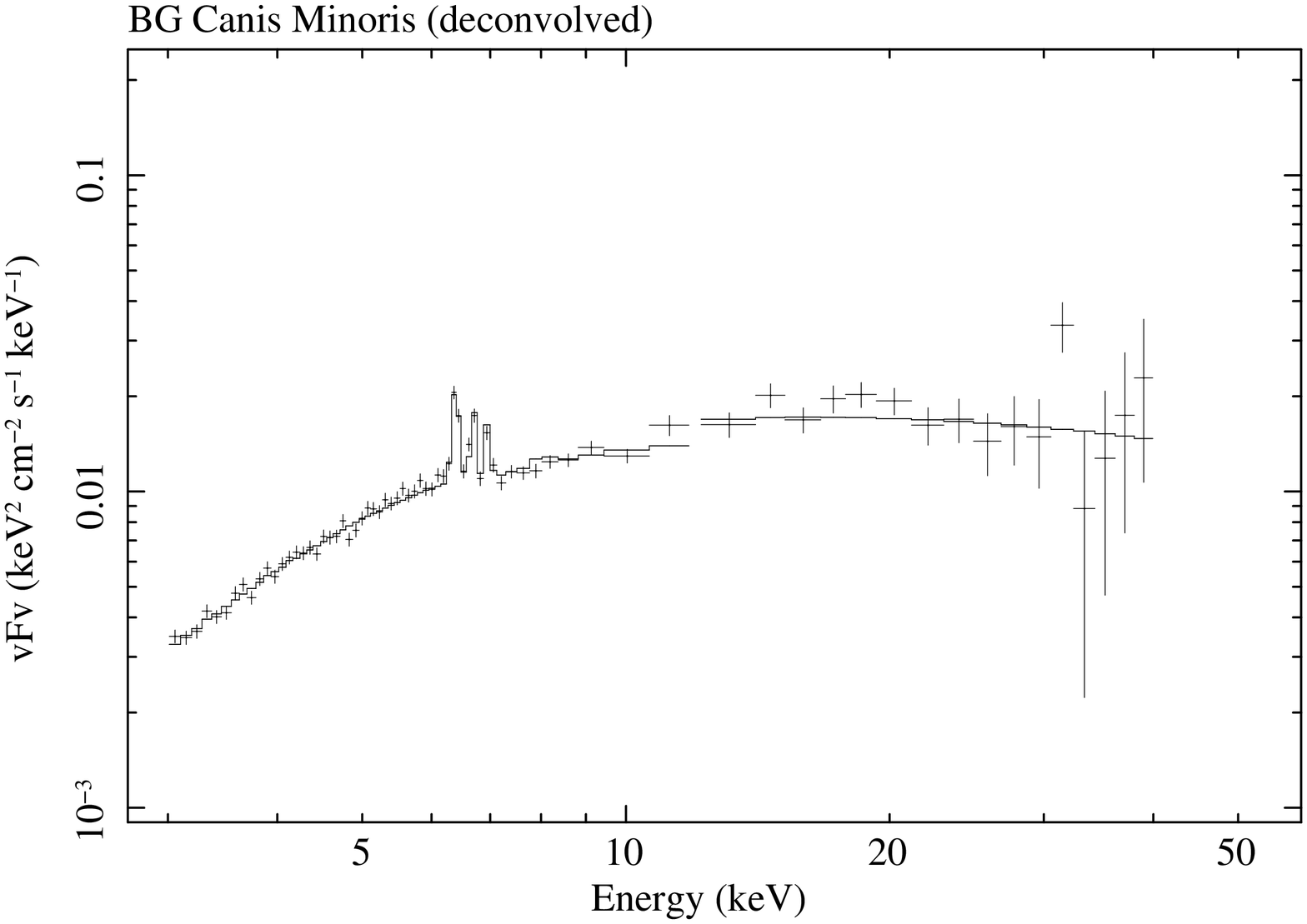}
}
\subfigure{
\includegraphics[height=6cm]{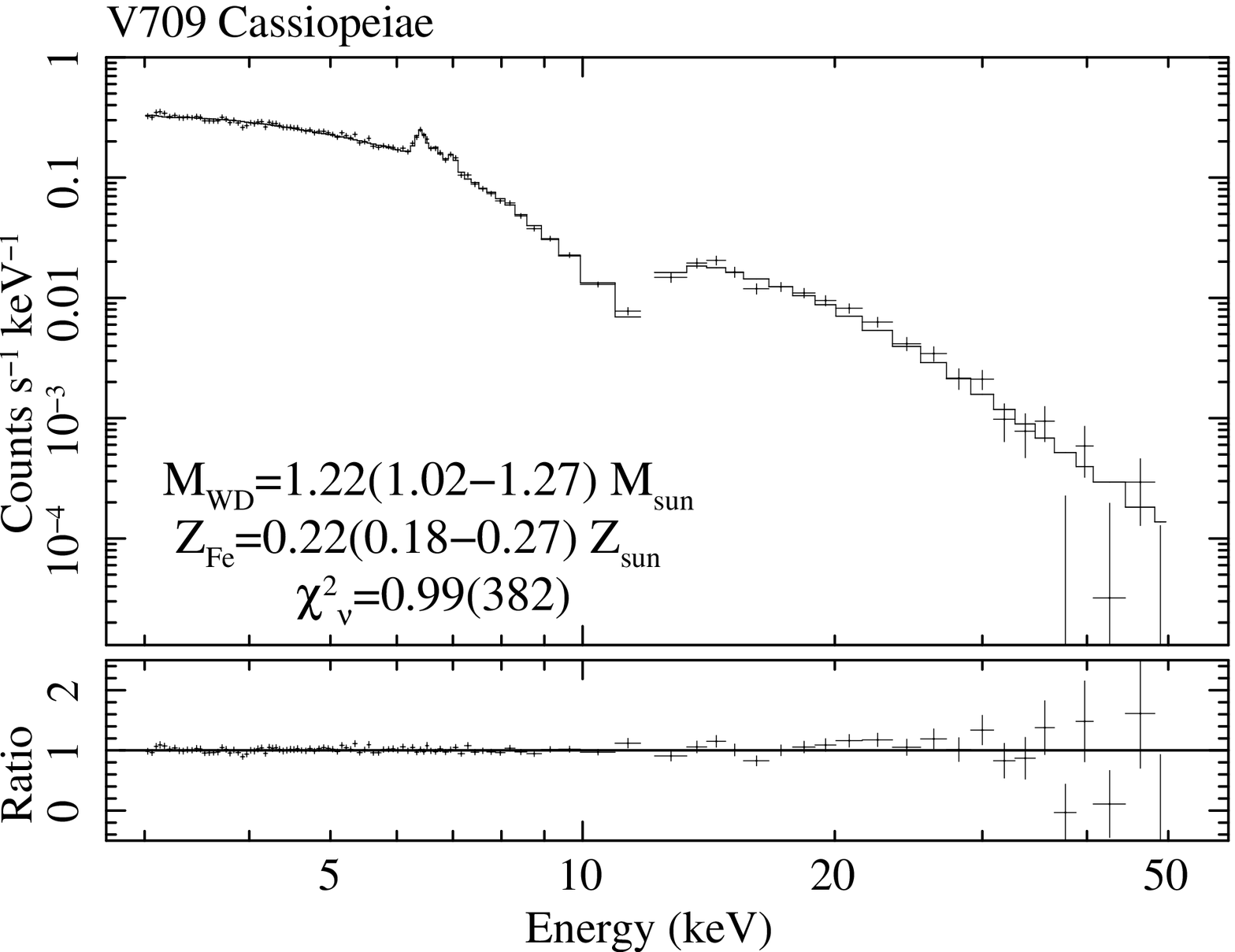}
\includegraphics[height=6cm]{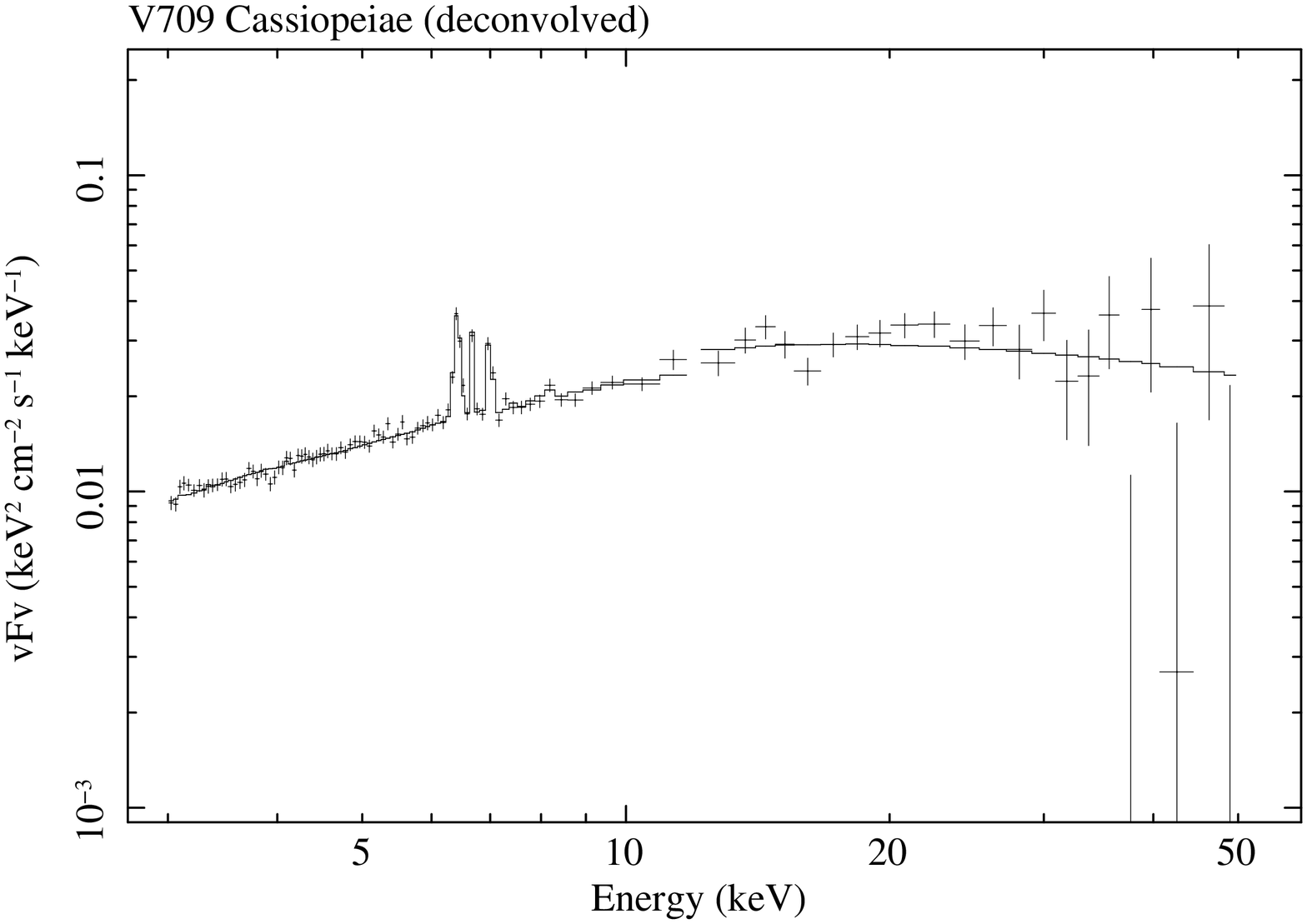}
}
\subfigure{
\includegraphics[height=6cm]{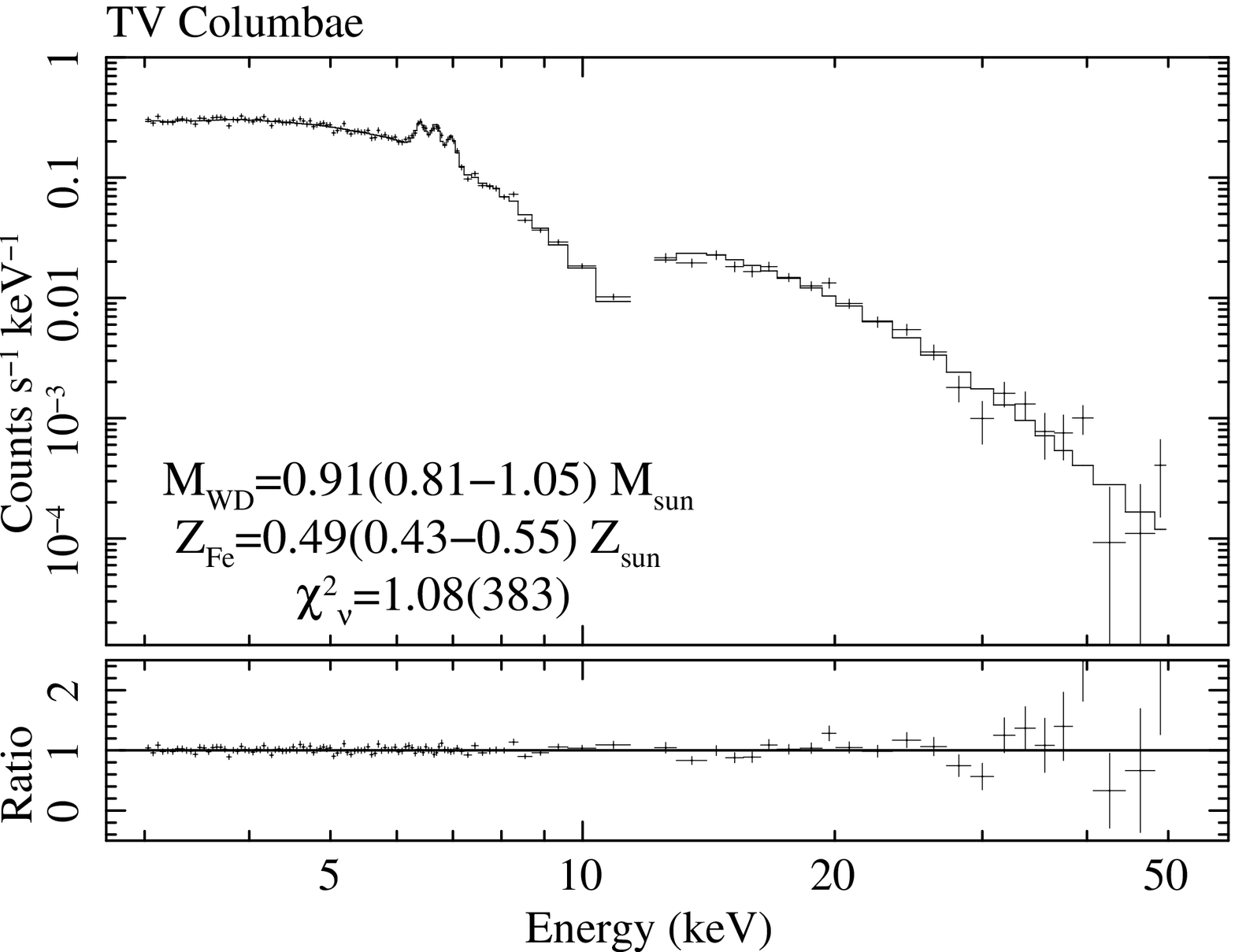}
\includegraphics[height=6cm]{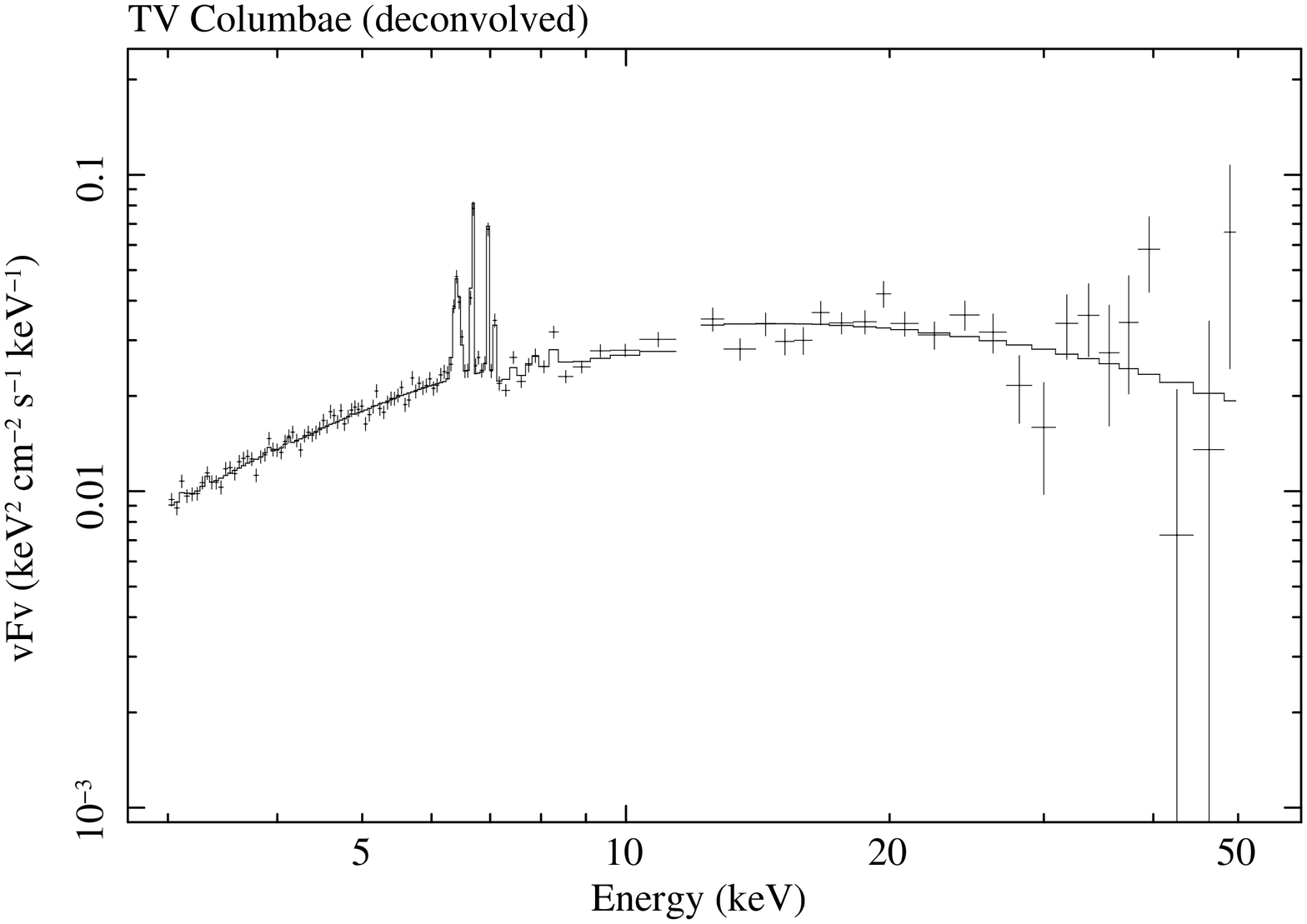}
}
\caption{
	{\it Continued from Fig. \ref{figure:pc_result}.}
}
\label{figure:pc_result:b}
\end{figure*}

\addtocounter{figure}{-1}
\begin{figure*}[htb]
\addtocounter{subfigure}{1}
\centering
\subfigure{
\includegraphics[height=6cm]{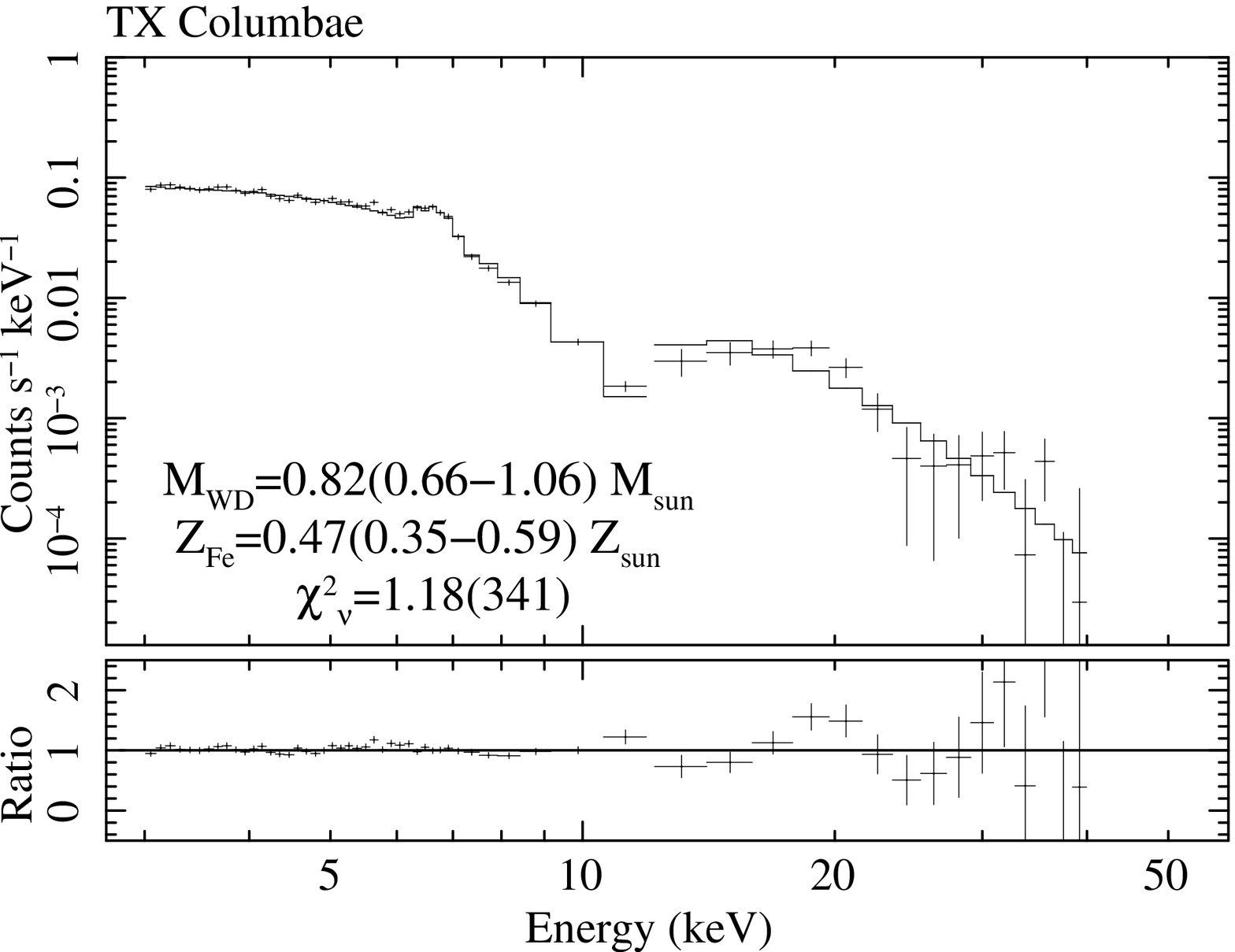}
\includegraphics[height=6cm]{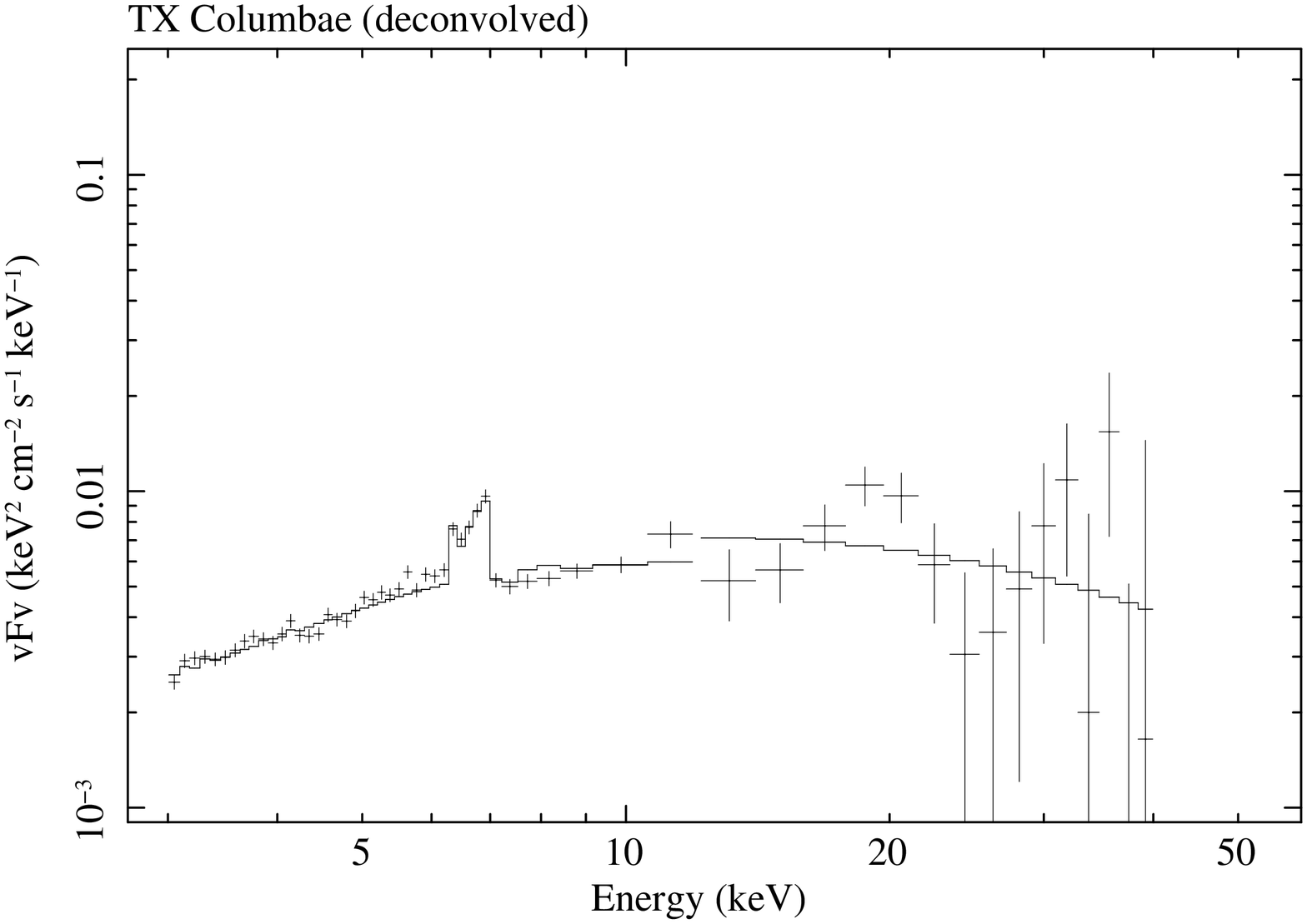}
}
\subfigure{
\includegraphics[height=6cm]{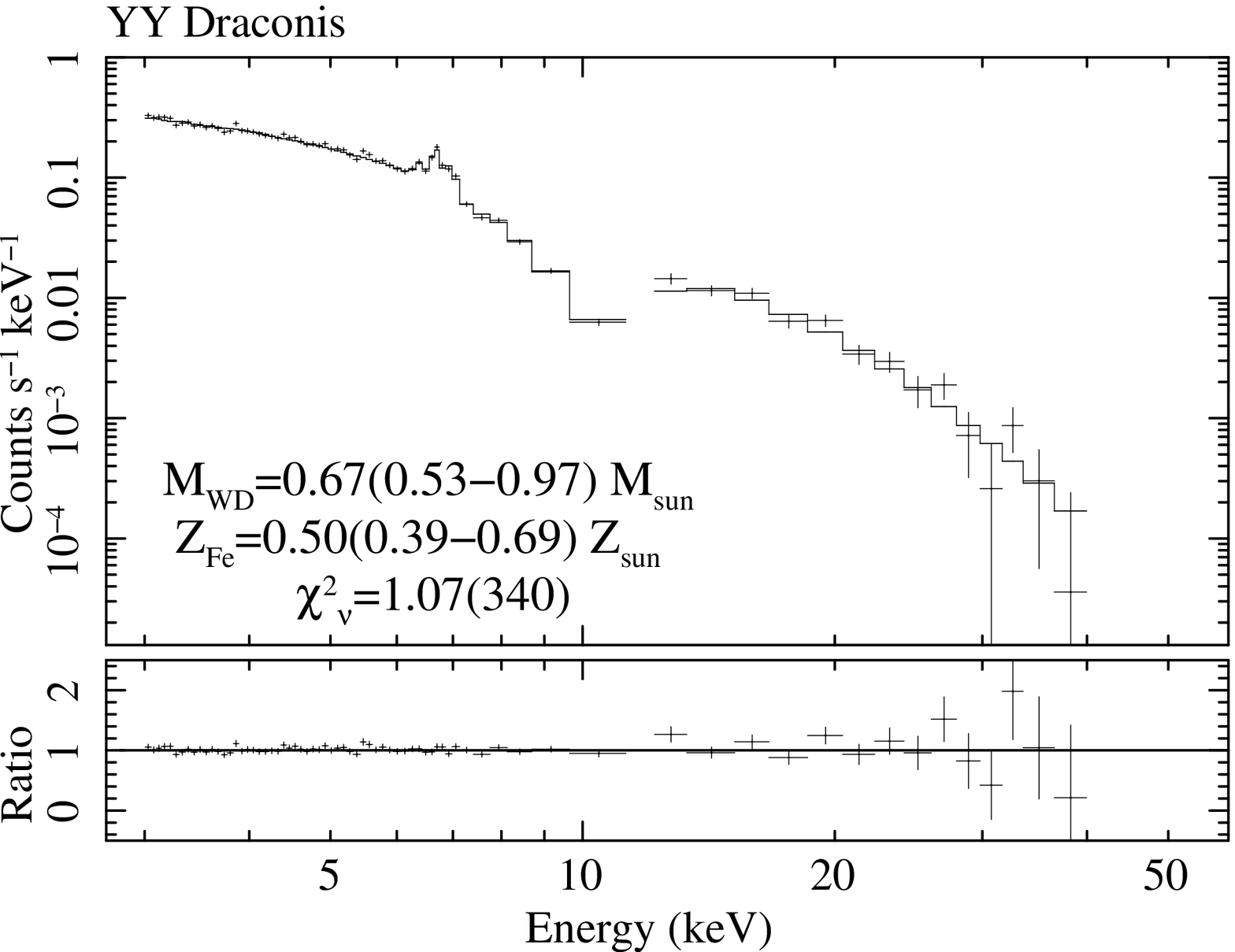}
\includegraphics[height=6cm]{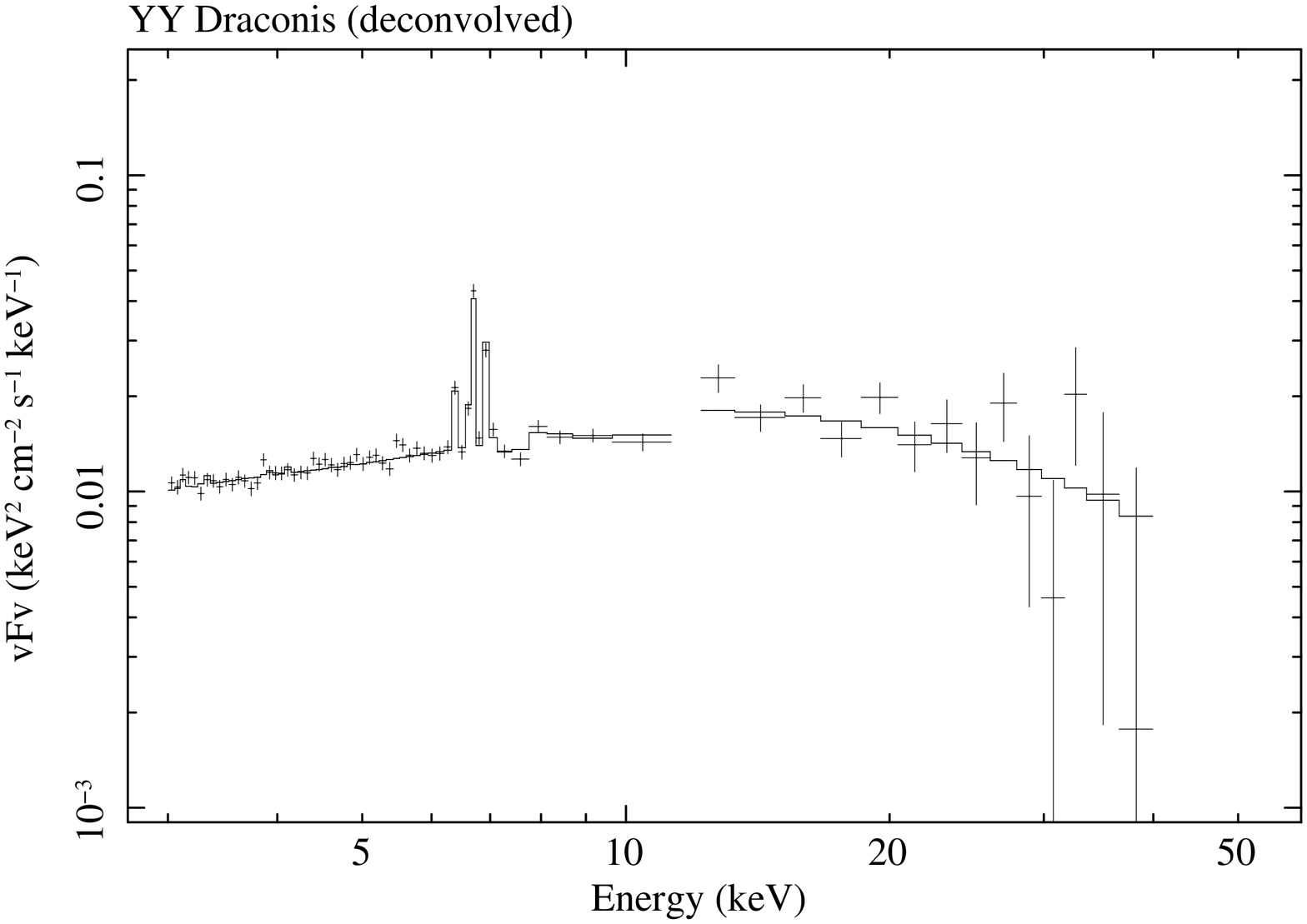}
}
\subfigure{
\includegraphics[height=6cm]{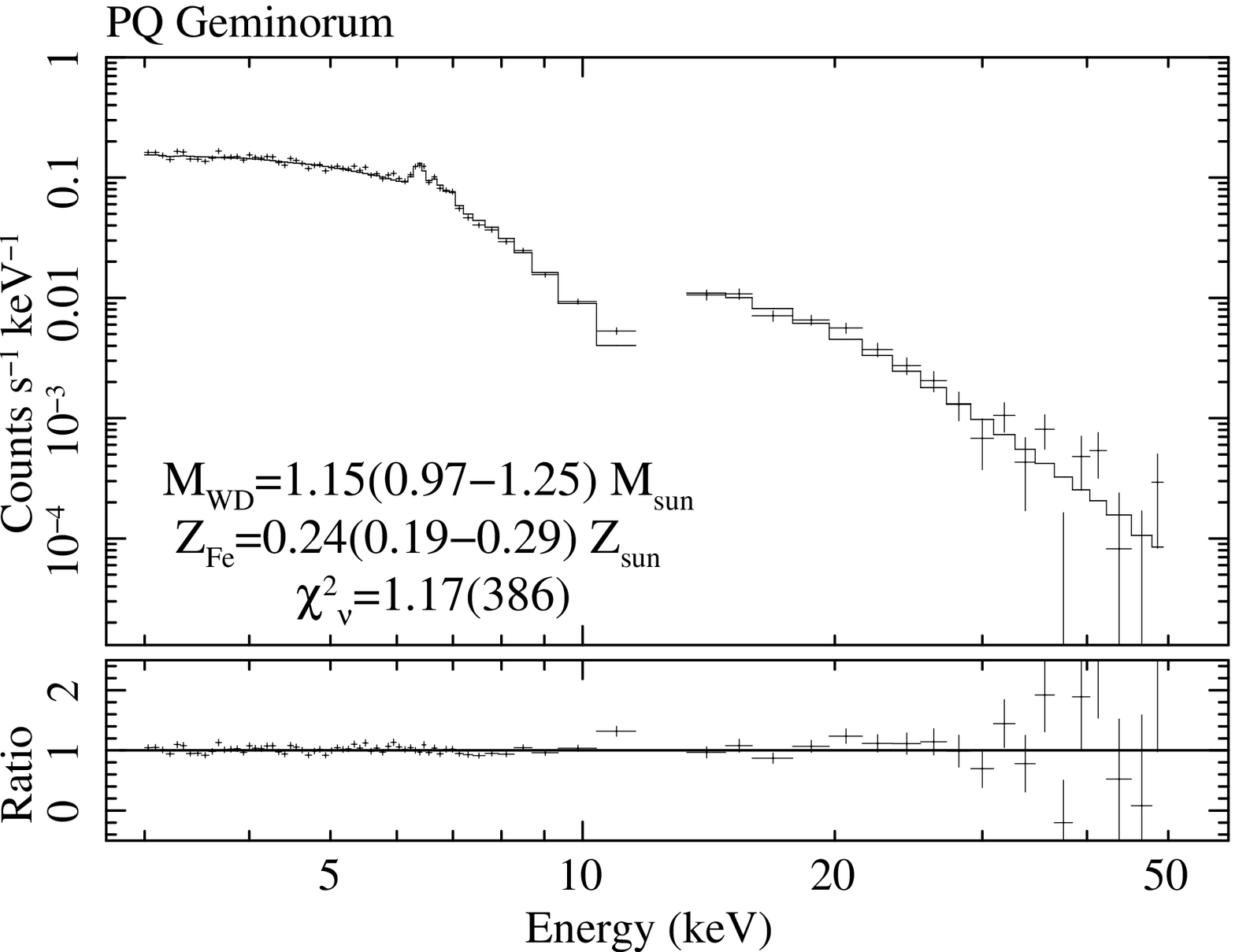}
\includegraphics[height=6cm]{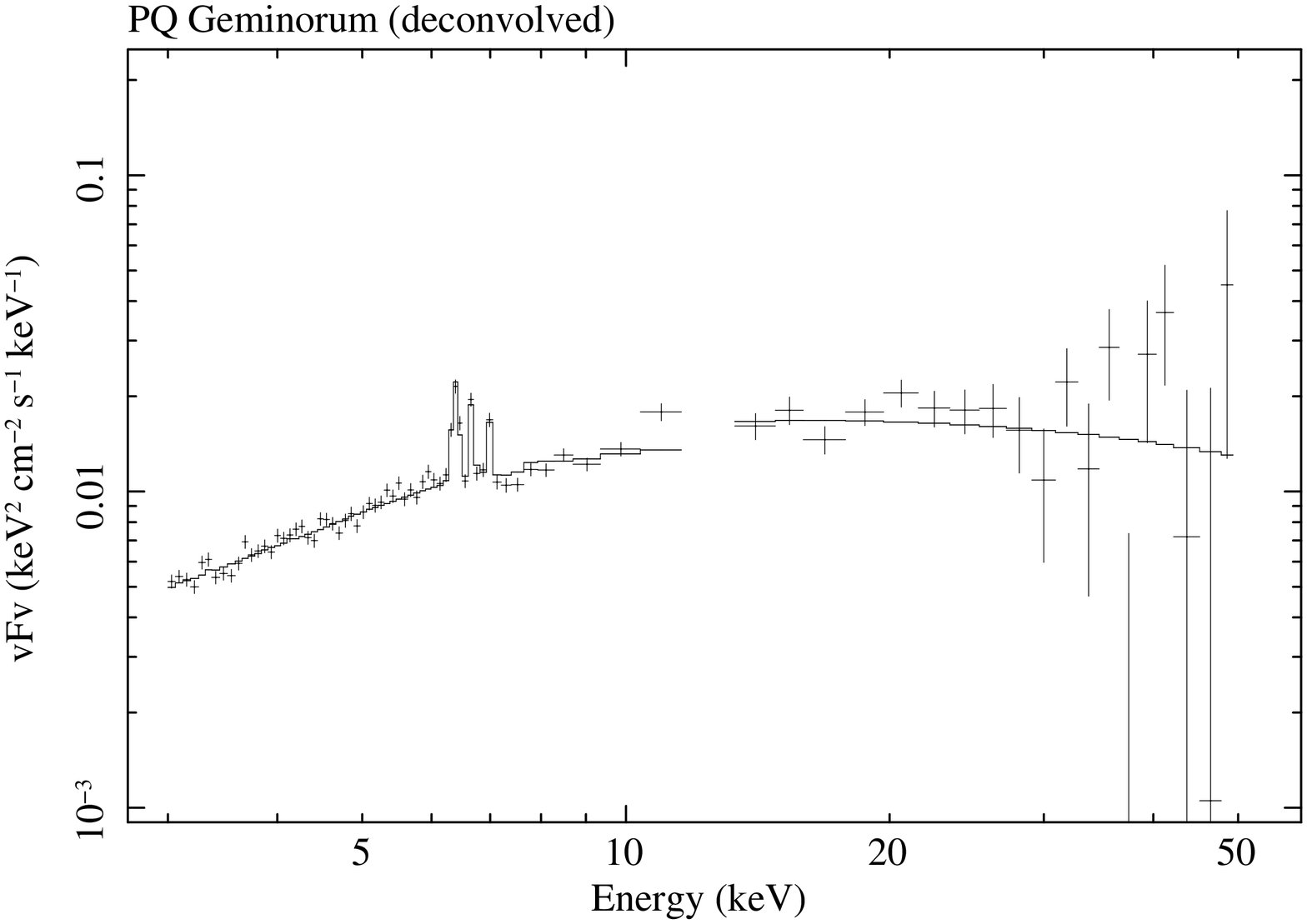}
}
\subfigure{
\includegraphics[height=6cm]{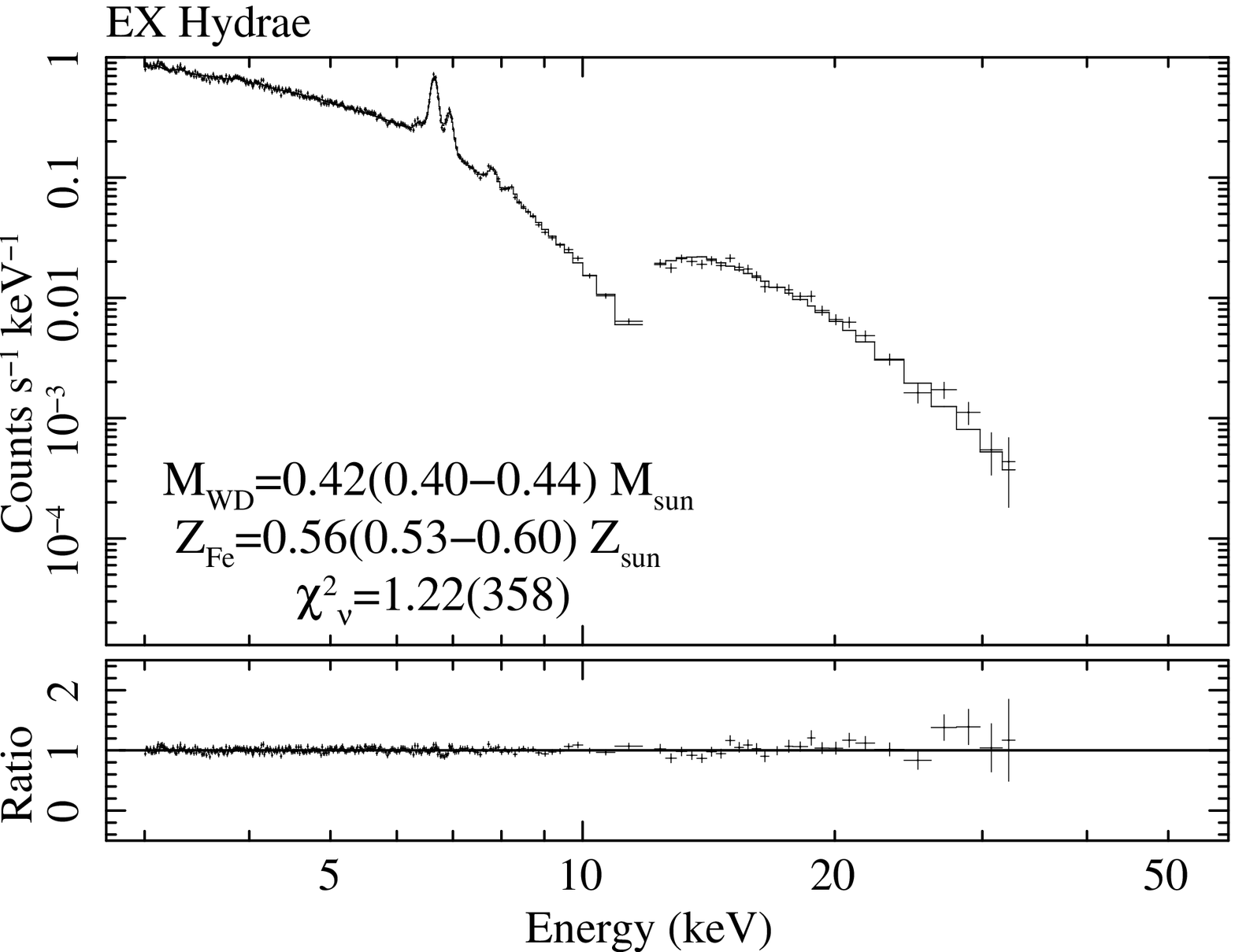}
\includegraphics[height=6cm]{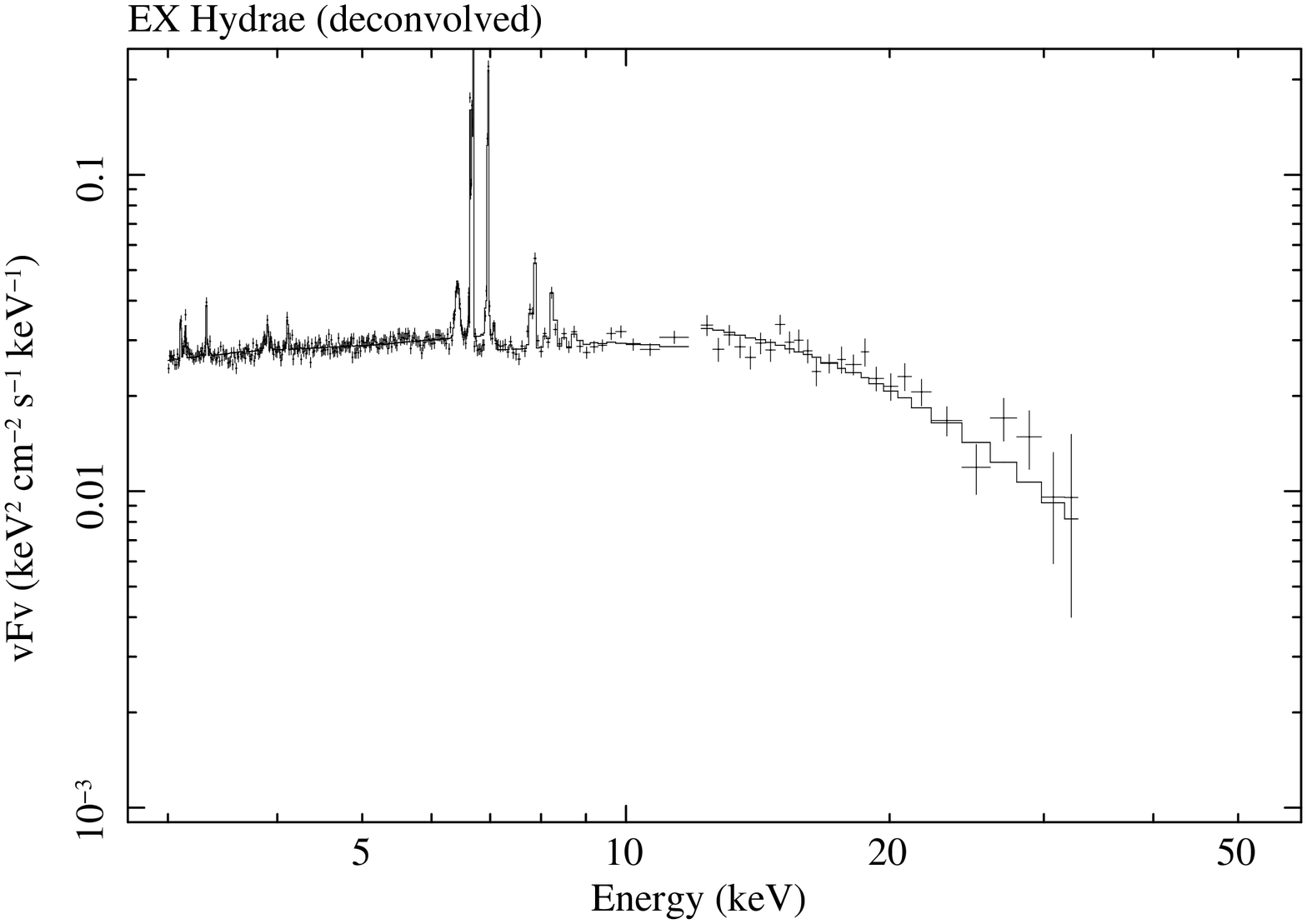}
}
\caption{
	{\it Continued.}
}
\label{figure:pc_result:c}
\end{figure*}

\addtocounter{figure}{-1}
\begin{figure*}[htb]
\addtocounter{subfigure}{1}
\centering
\subfigure{
\includegraphics[height=6cm]{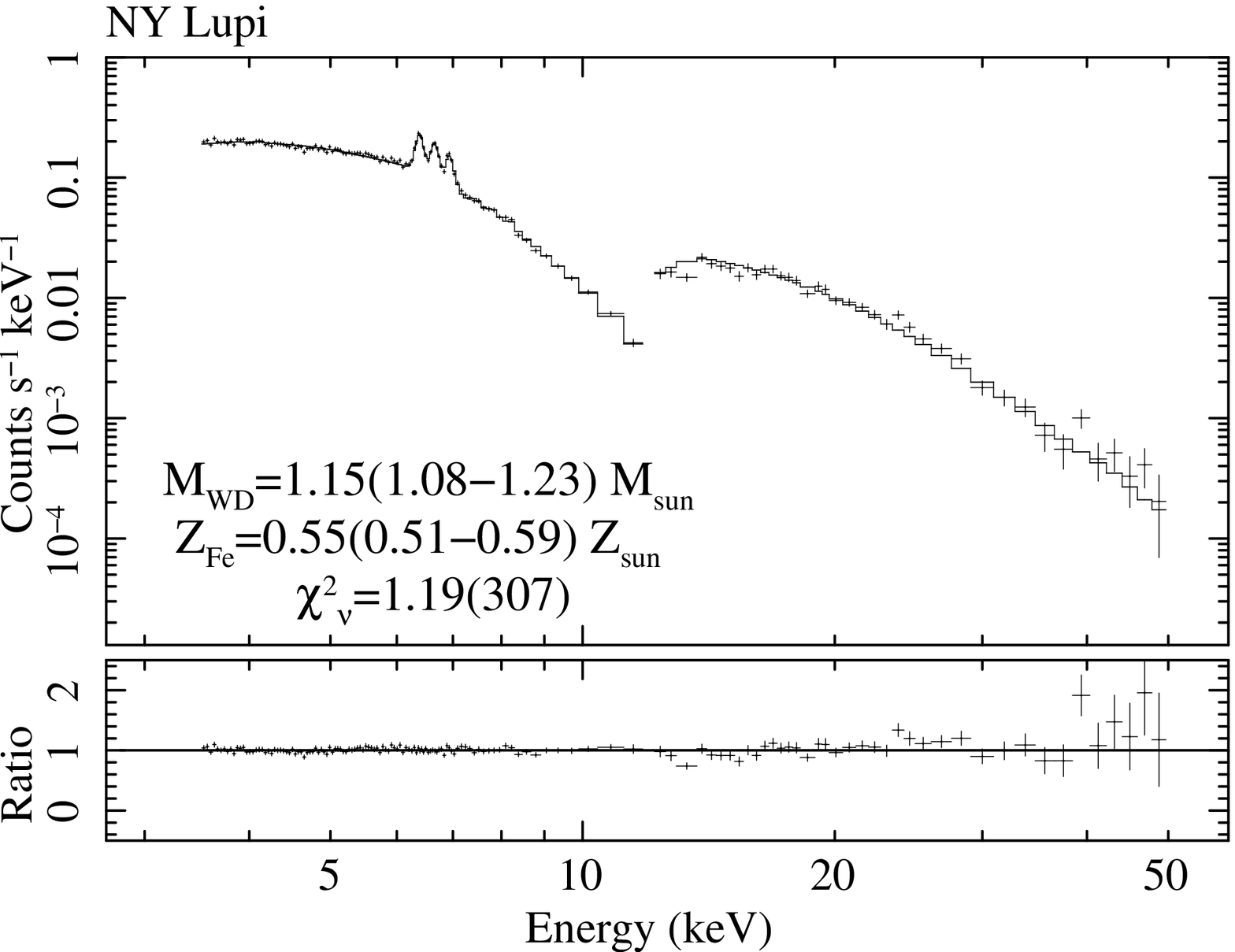}
\includegraphics[height=6cm]{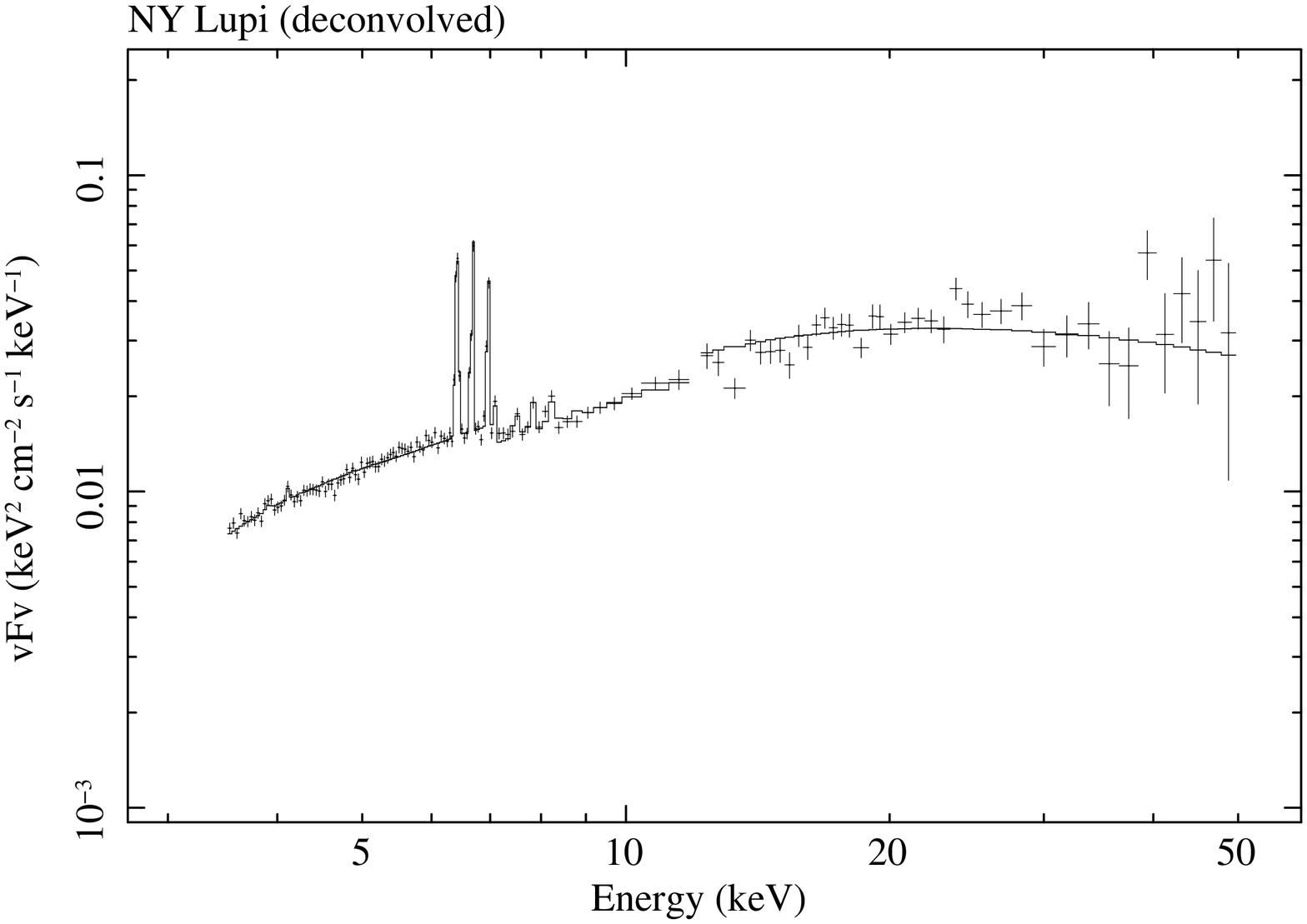}
}
\subfigure{
\includegraphics[height=6cm]{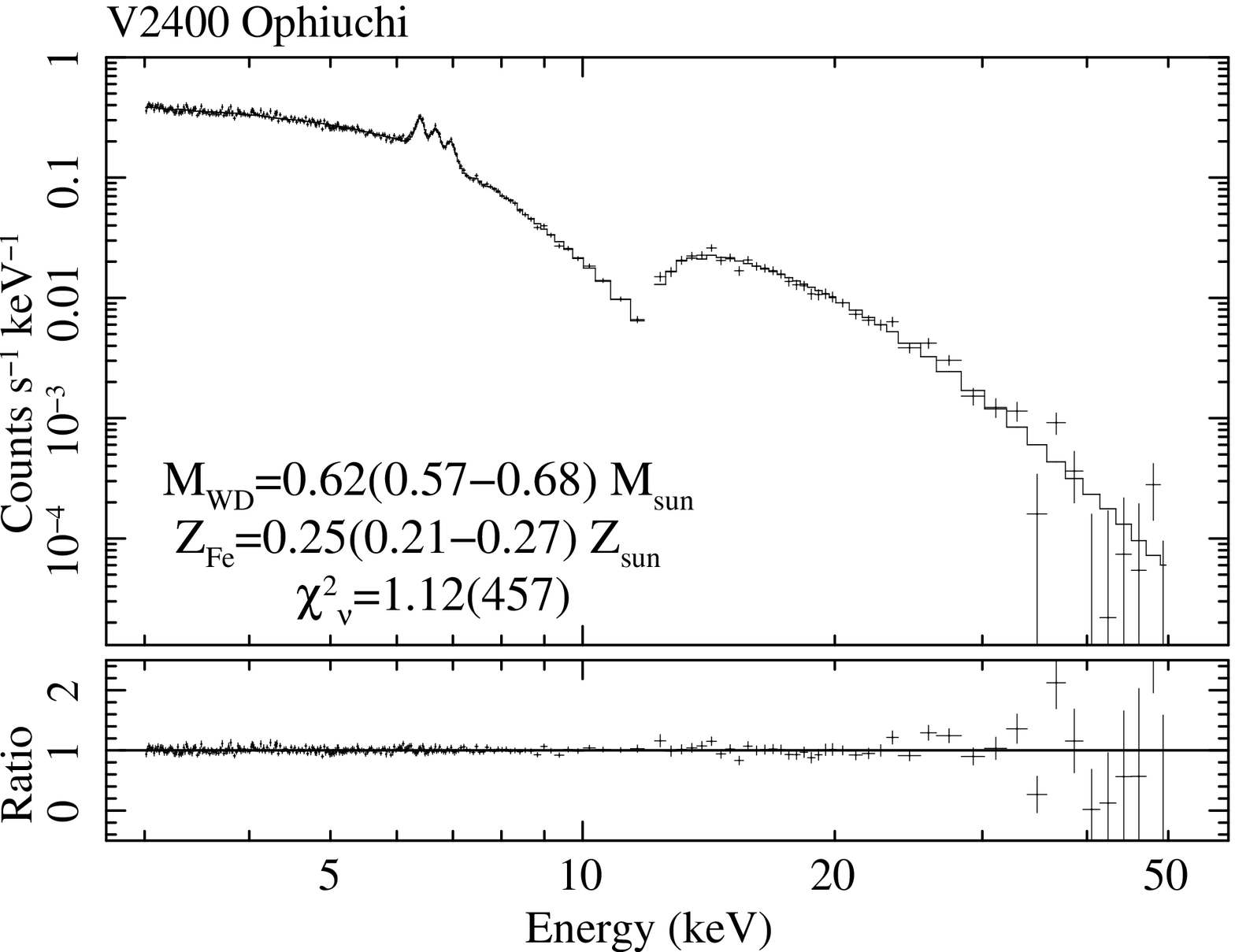}
\includegraphics[height=6cm]{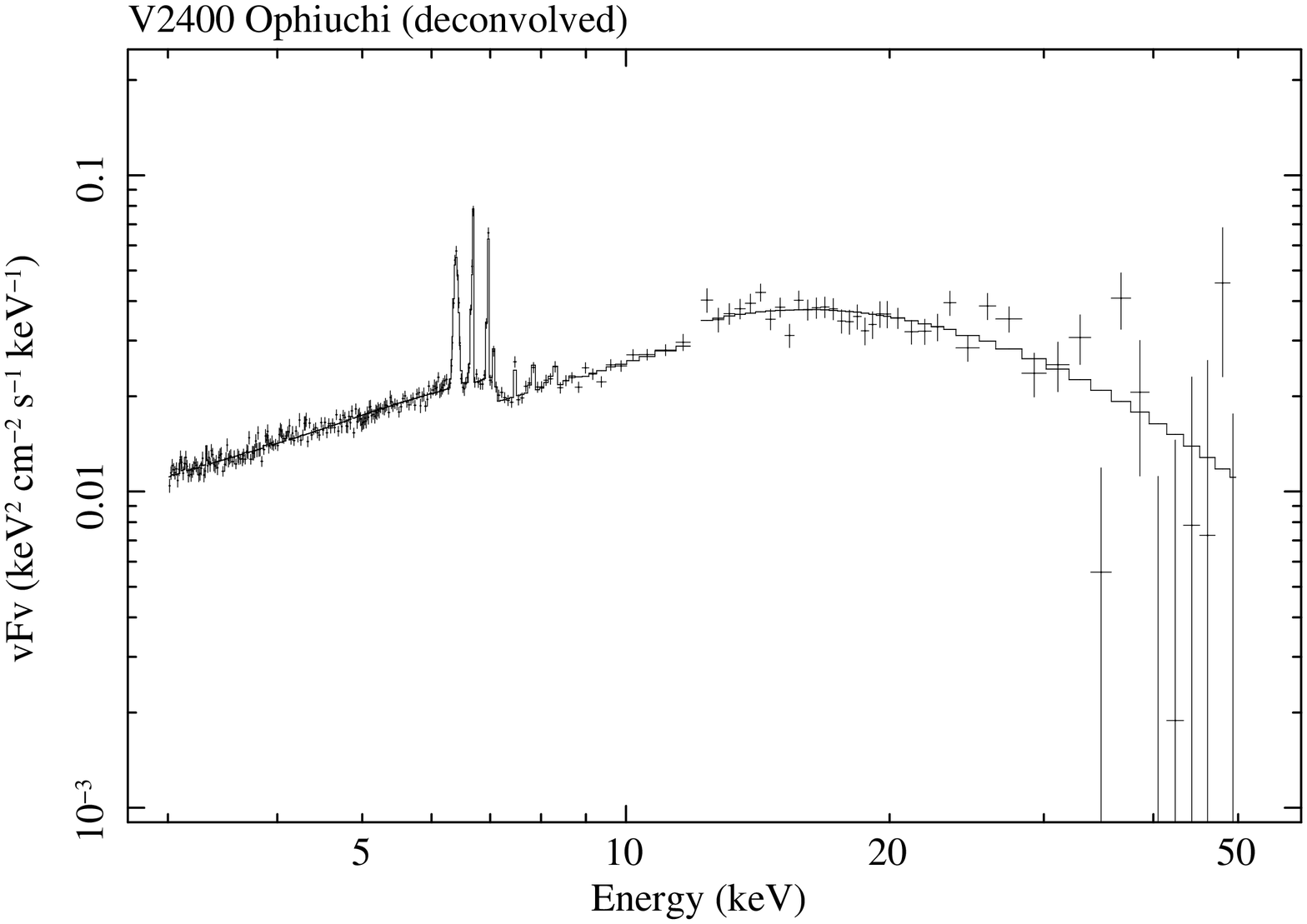}
}
\subfigure{
\includegraphics[height=6cm]{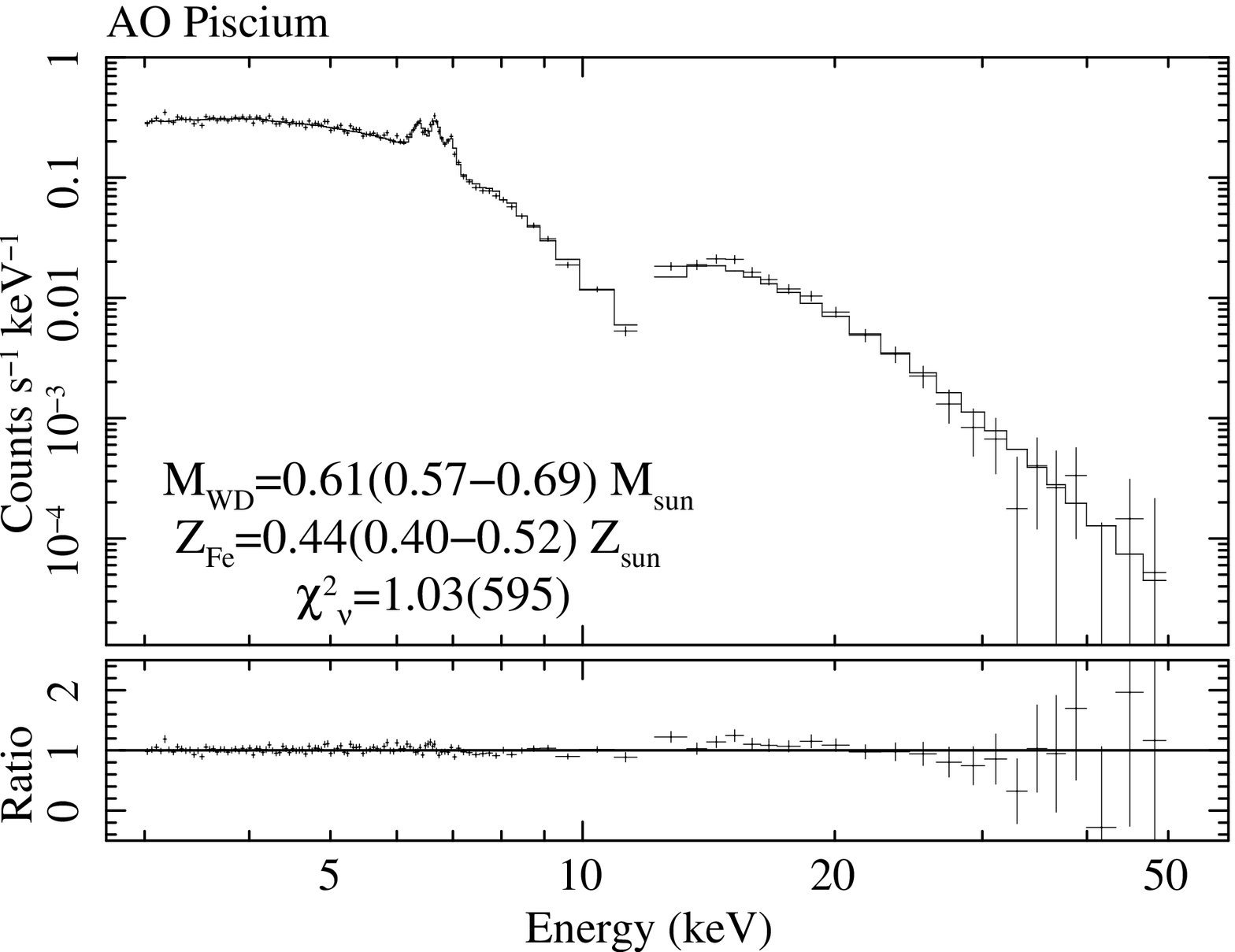}
\includegraphics[height=6cm]{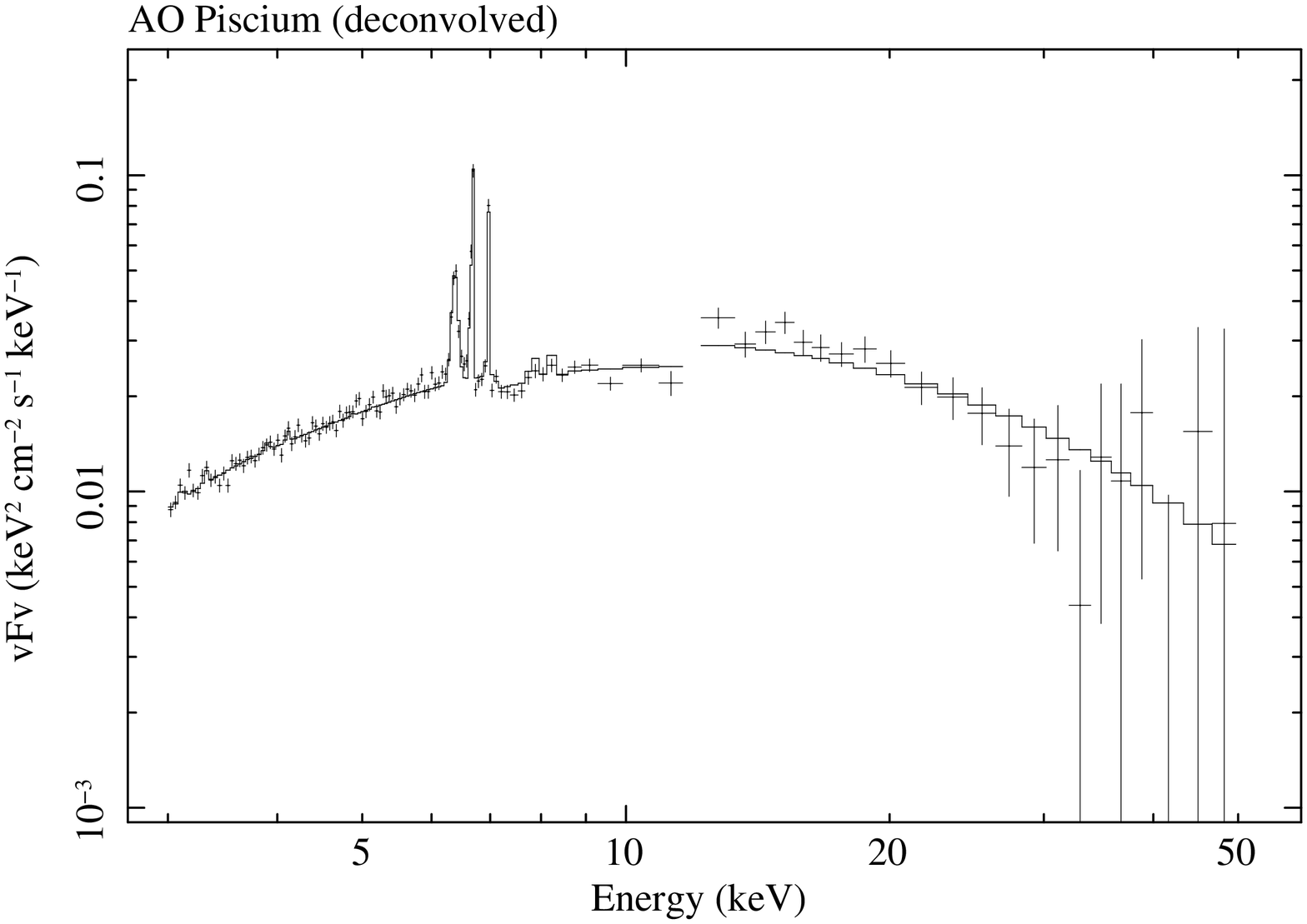}
}
\subfigure{
\includegraphics[height=6cm]{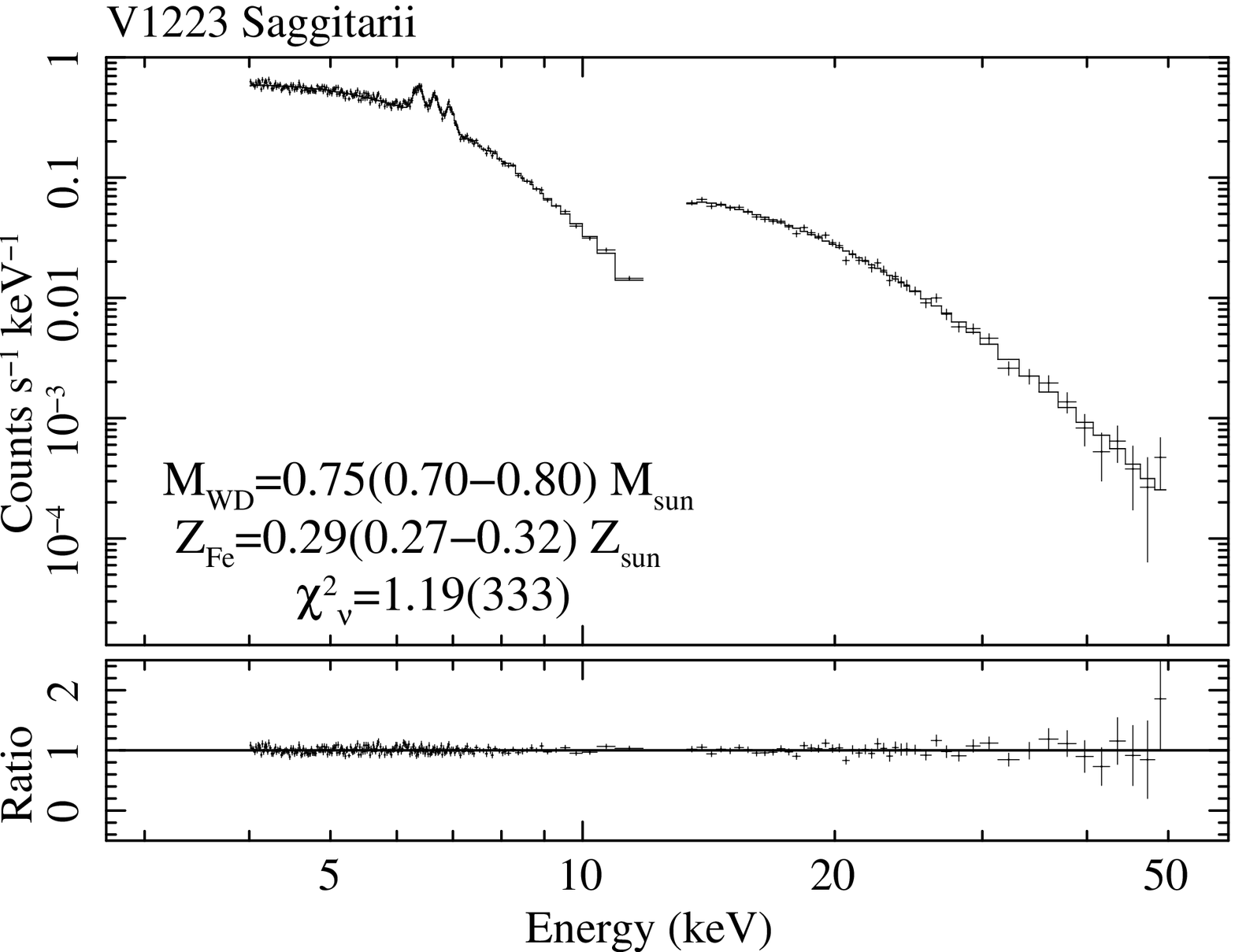}
\includegraphics[height=6cm]{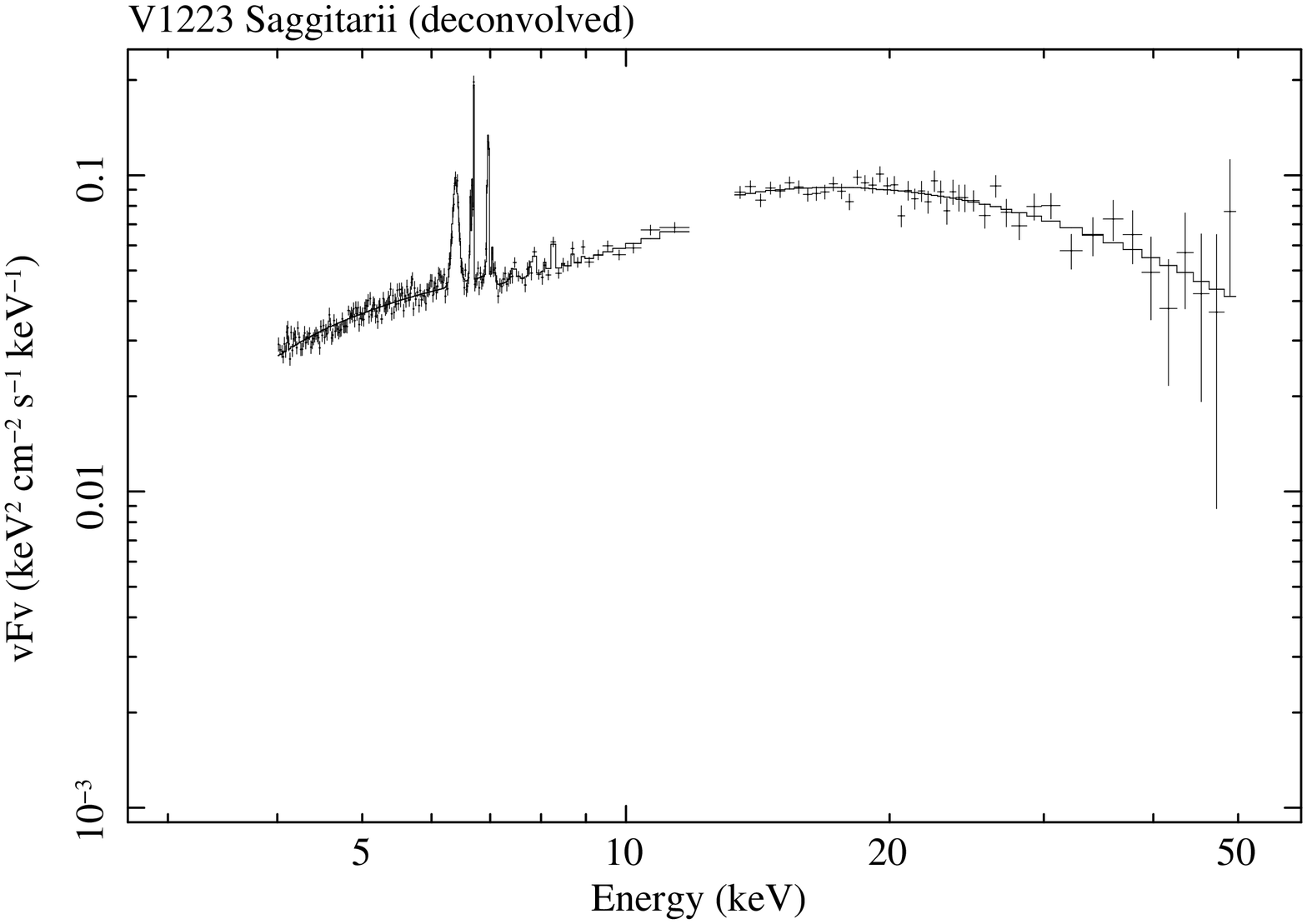}
}
\caption{
	{\it Continued.}
}
\label{figure:pc_result:c}
\end{figure*}

\addtocounter{figure}{-1}
\begin{figure*}[htb]
\addtocounter{subfigure}{1}
\centering
\subfigure{
\includegraphics[height=6cm]{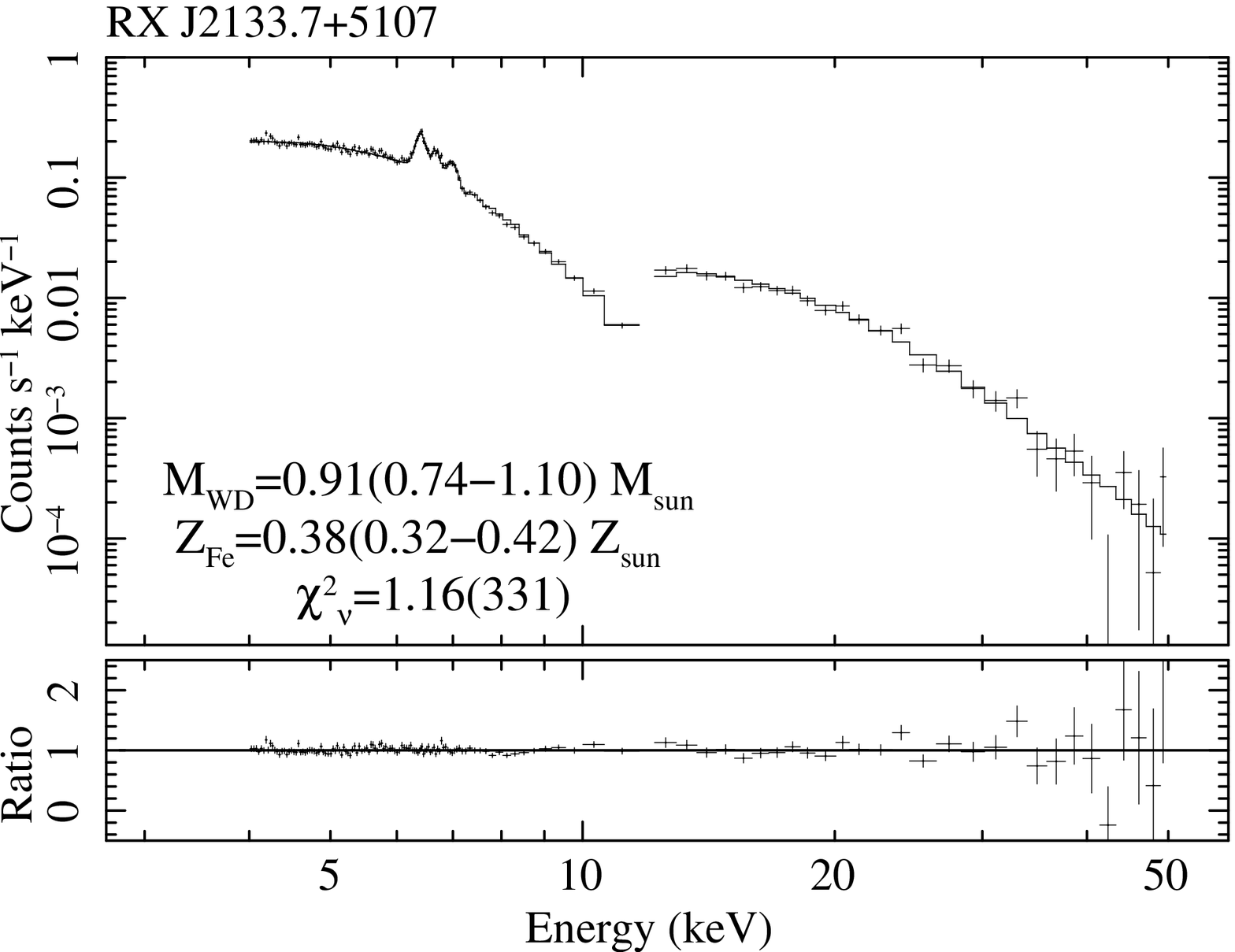}
\includegraphics[height=6cm]{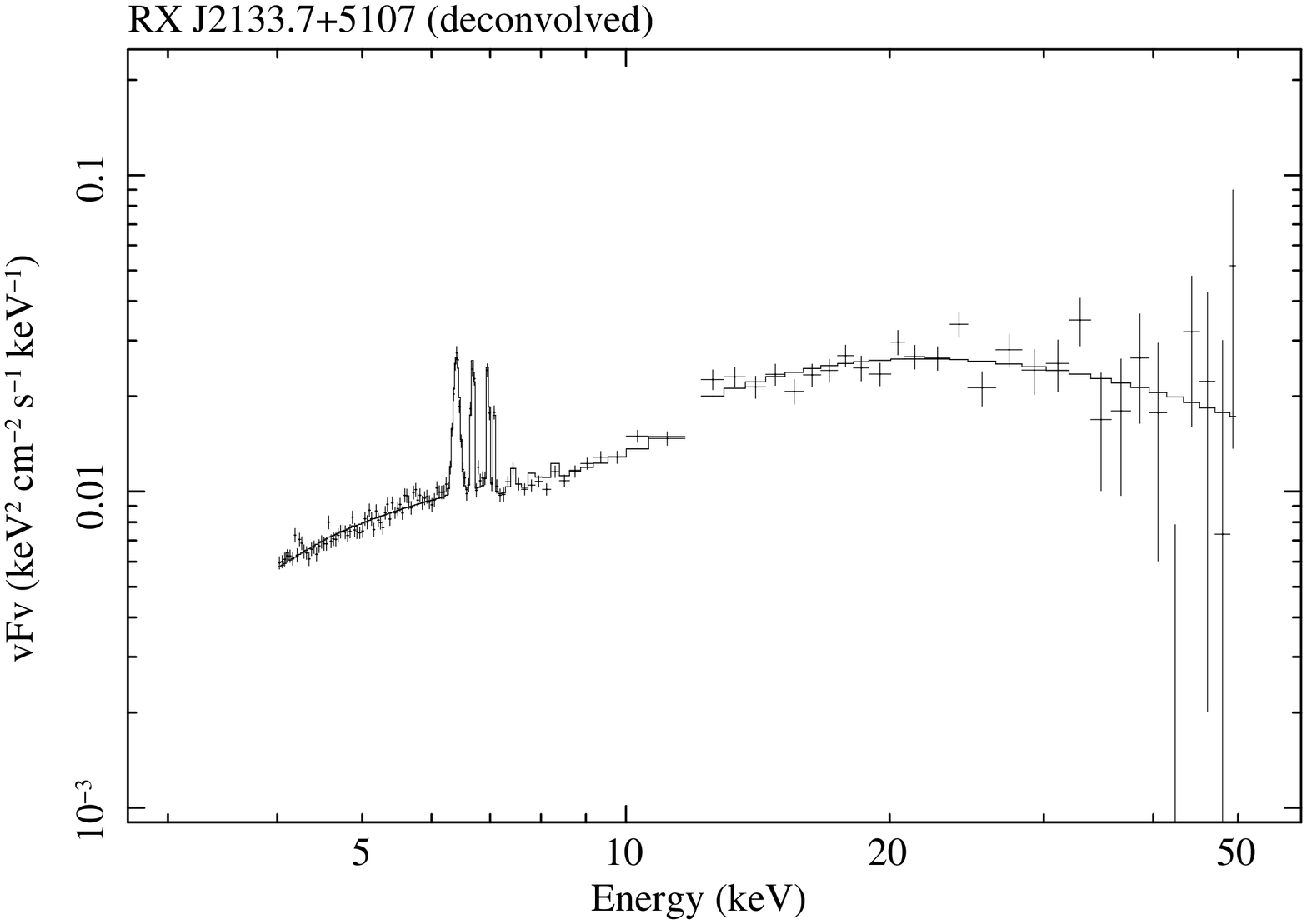}
}
\subfigure{
\includegraphics[height=6cm]{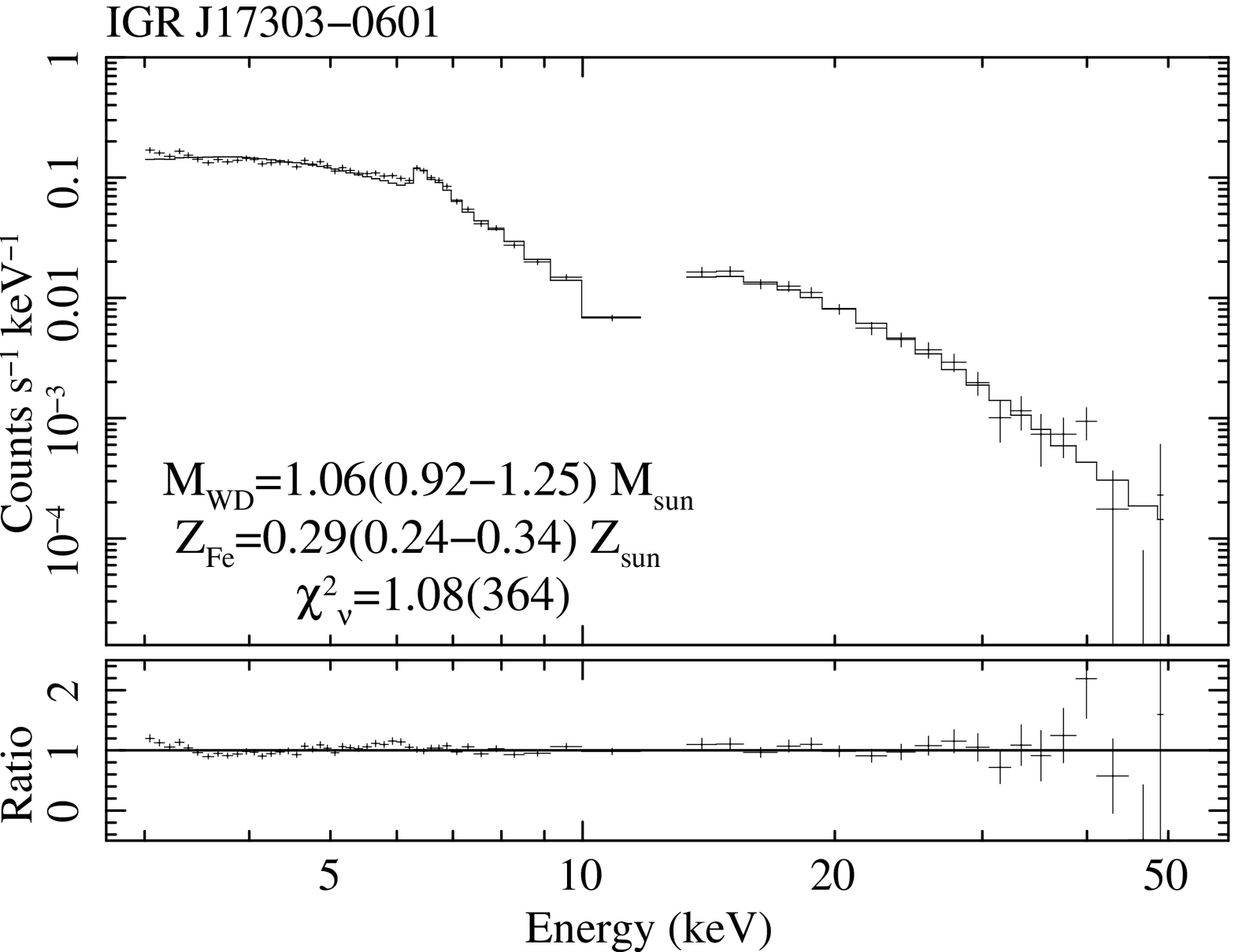}
\includegraphics[height=6cm]{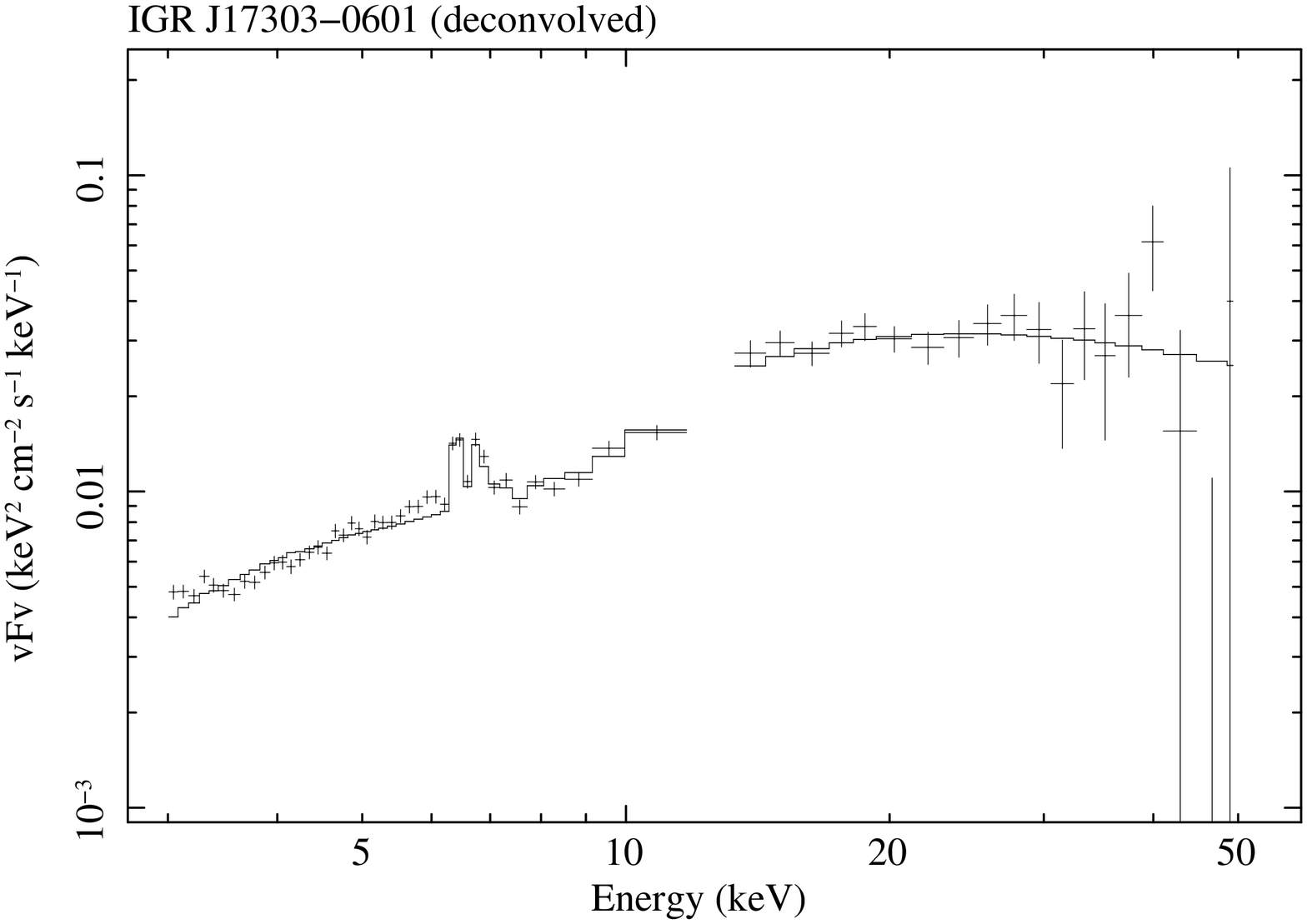}
}
\subfigure{
\includegraphics[height=6cm]{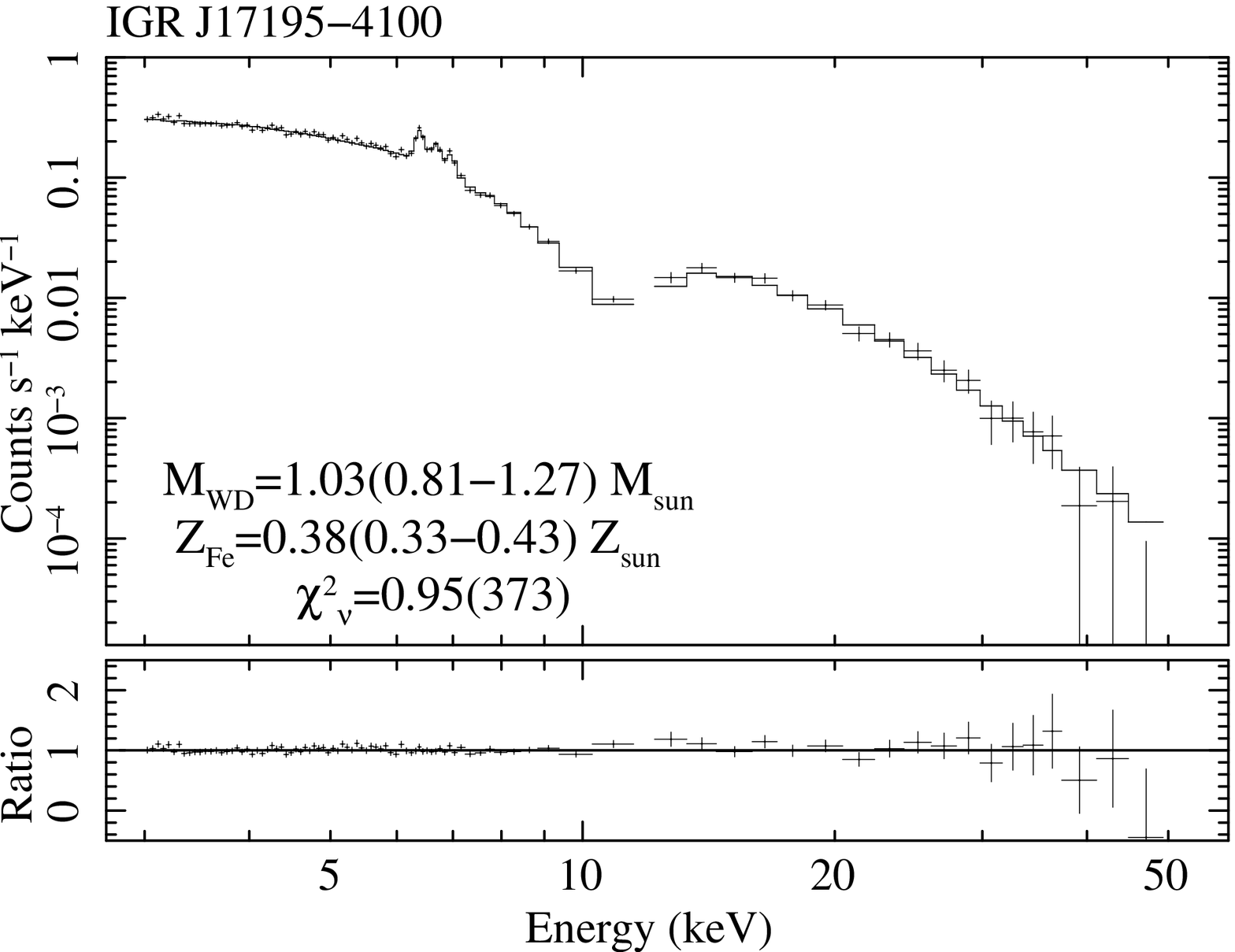}
\includegraphics[height=6cm]{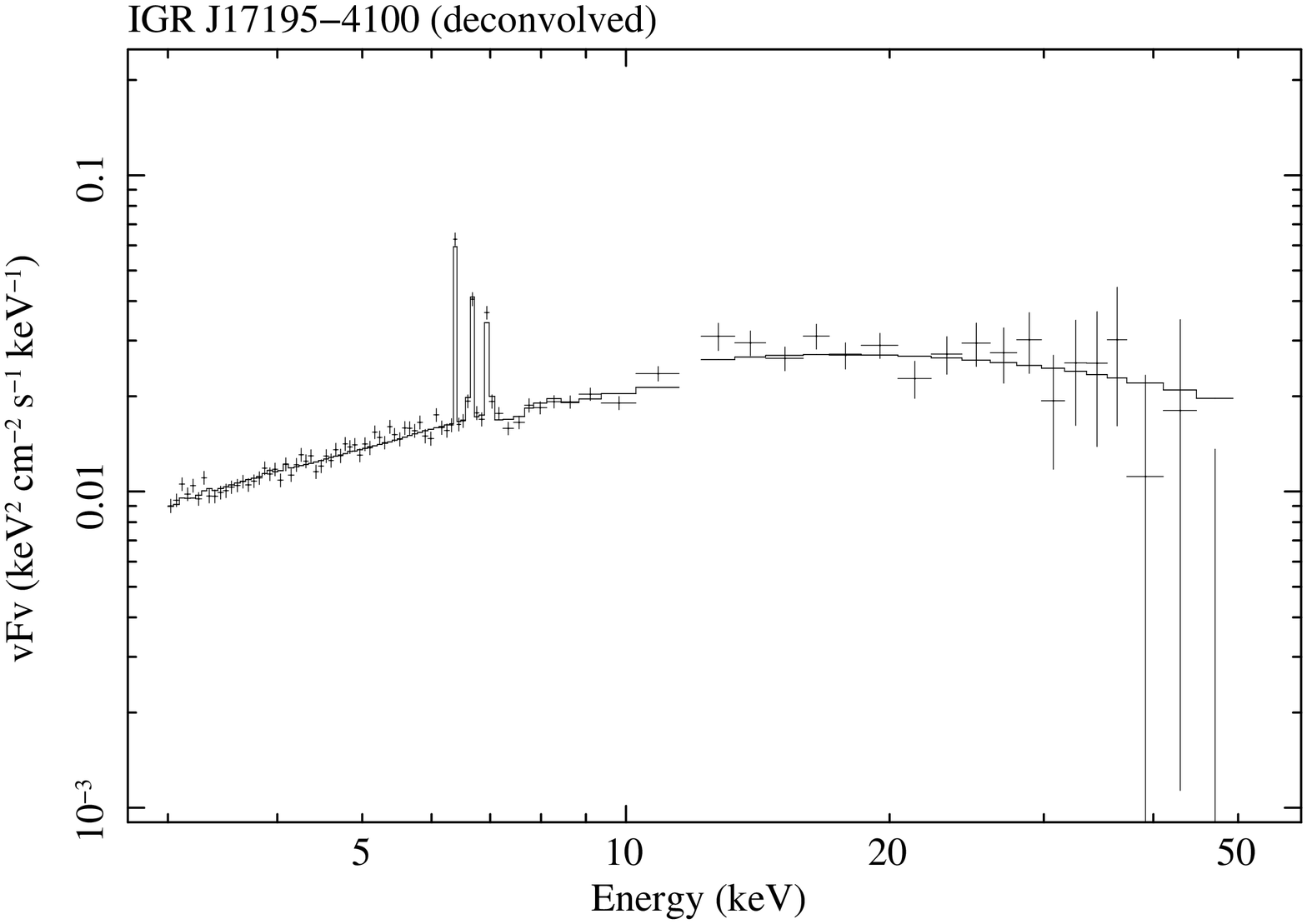}
}
\caption{
	{\it Continued.}
}
\label{figure:pc_result:c}
\end{figure*}
}

\end{document}